\documentclass{ws-rv9x6}
\usepackage{subfigure}   % required only when side-by-side / subfigures are used
\usepackage{ws-rv-thm}   % comment this line when `amsthm / theorem / ntheorem` package is used
\usepackage{ws-rv-van}   % numbered citation & references (default)
\newcommand{\deri}[2]{{\displaystyle \frac{\partial #1 }{\partial #2 }}}
\newcommand{\derri}[2]{{\displaystyle \frac{\partial^2 #1 }{\partial #2 }}}
\newcommand{\onlinecite}[1]{\hspace{-1 ex} \nocite{#1}\citenum{#1}} 

\makeindex
\begin{document}

\setcounter{chapter}{9}
\chapter[Computational simulations of glasses]{Computational simulations of\\
the vibrational properties of glasses}~\label{ra_ch10}

\author[H. Mizuno and A. Ikeda]{Hideyuki Mizuno and Atsushi Ikeda}
%\index[aindx]{Author, F.} % or \aindx{Author, F.}
%\index[aindx]{Author, S.} % or \aindx{Author, S.}

\address{Graduate School of Arts and Sciences,\\
The University of Tokyo, Tokyo 153-8902, Japan.\\
hideyuki.mizuno@phys.c.u-tokyo.ac.jp
}

%%%%%%%%%%%%%%%%%%%%%%%%%%%%%%%%%%%%%%%%%%%%%%%%%%%%%%%%%%%%%%%%%%%%%%%%%%%%%%%%%%%%%%%%%%%
\begin{abstract}
Glasses show vibrational properties that are markedly different to those of crystals which are known as phonons.
For example, excess low-frequency modes (the so-called boson peak), vibrational localization, and strong scattering of phonons have been the most discussed topics, and a theoretical understanding of these phenomena is challenging.
To address this problem, computational simulations are a powerful tool, which have been employed by many previous works.
In this chapter, we describe simulation methods for studying the vibrational properties of glasses (and any solid-state materials).
We first present a method for studying vibrational eigenmodes.
Since vibrational motions of particles are excited along eigenmodes, the eigenmodes are fundamental to descriptions of vibrational properties.
The eigenmodes in glasses are non-phonon modes in general, and some of them are even localized in space.
We next present a method of analysing phonon transport, which is also crucial for understanding vibrational properties.
Since phonons are not eigenmodes in glasses, they are decomposed into several different, non-phonon eigenmodes.
As a result, phonons in glasses are strongly scattered.
In addition, we describe how to analyse the elastic response.
The elastic response of glasses is also anomalous with respect to that of crystals.
Finally, we briefly introduce recent advances that have been achieved by means of large-scale computational simulations.
\end{abstract}

%\markright{Customized Running Head for Odd Page} % default is Chapter Title.
\body

%\tableofcontents

%%%%%%%%%%%%%%%%%%%%%%%%%%%%%%%%%%%%%%%%%%%%%%%%%%%%%%%%%%%%%%%%%%%%%%%%%%%%%%%%%%%%%%%%%%%
\section{Introduction}
Currently, there is no doubt that computational simulations play an important role in the development of fundamental science as well as engineering applications.
Molecular Dynamics (MD) simulations and Monte Carlo (MC) simulations have been established to simulate the behaviours of materials in dense gas, liquid, and solid states at the microscopic, molecular level~\cite{Allen_1986,Frenkel_2002}.
The importance of these molecular simulations comes from the fact that they provide exact, quasi-experimental data on well-defined simulation models of materials.
The usefulness of simulations is also based on the fact that they can access data that cannot be obtained through experiments.
From the theoretical point of view, exact data on prototypical models are valuable for understanding the fundamental mechanisms of phenomena as well as for testing the validity of proposed theories.
In the past, molecular simulations have been employed to solve many important problems.
For example, in statistical physics, phase transitions, such as the gas-liquid transition and the paramagnetic-ferromagnetic transition, are among the topics that are most widely studied by means of computer simulations~\cite{Newman_1999}.
Additionally, to address the problem of glass transition, many previous works have relied on simulations of, e.g., the drastically slowed dynamics near the glass transition~\cite{Cavagna_2009,Berthier_2011}.

Computational simulations are also a powerful tool for studying the vibrational properties of glasses.
For crystals, thanks to their periodicity and symmetry, analytical formulations can be obtained for the vibrational motions of the molecules, which give the concept of phonons (lattice waves)~\cite{Ashcroft_1976,Kittel_1996}
\footnote{
Phonons are quantized lattice waves, but here we use the term ``phonons" for general lattice waves.
We can also consider phonons in glasses, which are sinusoidal waves propagating in disordered structures.
Such phonons in glasses will be discussed in Section~\ref{sec.phonon}.
}.
Particularly, the Debye theory has been established to explain the behaviour of the vibrational density of states (vDOS) in a crystal.
In contrast, for glasses, due to the lack of periodicity and symmetry, analytical calculations are much difficult to perform.
Although some mean-field theories, such as replica theory~\cite{Franz_2015} and effective medium theory~\cite{schirmacher_2006,schirmacher_2007,Wyart_2010,DeGiuli_2014}, have been proposed, it is crucial to test their validity and extend them to the 3-dimensional case.
Also, in many past works, e.g., Refs.~\onlinecite{Zeller_1971,Phillips_1981,Elliott_1990,buchenau_1984,Yamamuro_1996,Masciovecchio_2006,Monaco_2009,Baldi_2010,Ferrante_2013,Baldi_2014,Perez-Castaneda_2014,Perez-Castaneda2_2014,Kabeya_2016}, experiments have been performed to study the thermal and vibrational properties of glasses.
Although the vDOS and dynamic structure factor can be measured in experiments by monitoring the responses of various probes, such as light, X-rays, and neutrons, it is generally difficult to directly observe the vibrational motions of molecules.
Considering this situation of separate purely theoretical and purely experimental works, computational simulations take advantage of the ability to directly observe and understand the vibrational motions of particles in well-defined simulation models.
There are many relevant problems regarding the vibrational properties of glasses, e.g., excess low-frequency modes (the so-called boson peak), vibrational localization, and the strong scattering of phonons.
A theoretical understanding of these phenomena remains to be developed, and doing so will be challenging.
To address these problems, computer simulations have been employed in many previous works, e.g., Refs.~\onlinecite{Schober_1991,mazzacurati_1996,Taraskin_1999,taraskin_2000,Ruocco_2000,Tanguy_2002,Leonforte_2005,Leonforte_2006,Wyart_2006,Silbert_2005,Silbert_2009,shintani_2008,Monaco2_2009,Marruzzo_2013,Mizuno2_2013,Mizuno_2014,Mizuno2_2016,Ikeda_2013,Crespo_2016,Beltukov_2016,Gelin_2016,Milkus_2016,Krausser_2017,Baumgarten_2017,Lerner_2016,Mizuno_2017,Shimada_2017,Lerner_2017,Lerner2_2017,Angelani_2018,Bouchbinder_2018,Mizuno_2018,Shimada_2018,Wang_2019,Beltukov_2018,Saitoh_2019,Tong_2019}.

In this chapter, we review simulation methods for analysing the vibrational properties of glasses; these methods can also be applied for any solid-state materials.
In Section~\ref{sec.model}, we describe simulation models of glasses.
We first present a brief description of MD simulations, and we next explain how to obtain glass systems through simulations.
Section~\ref{sec.modeanalysis} introduces a method of analysing vibrational eigenmodes.
Since the vibrational motions of particles are excited along eigenmodes, the eigenmodes are fundamental to descriptions of vibrational properties.
Section~\ref{sec.phonon} provides a method of analysing phonon transport, which is also a fundamental property.
In addition, we also present a method of analysing elastic moduli in Section~\ref{sec.moduli}.
Finally, Section~\ref{sec.recent} briefly introduces recent advances that have been achieved by means of large-scale computational simulations.

%%%%%%%%%%%%%%%%%%%%%%%%%%%%%%%%%%%%%%%%%%%%%%%%%%%%%%%%%%%%%%%%%%%%%%%%%%%%%%%%%%%%%%%%%%%
\section{Simulation models}~\label{sec.model}
Several excellent books~(e.g., Refs.~\onlinecite{Allen_1986,Frenkel_2002}) have already been published that describe molecular simulations, i.e., MD and MC simulations.
Here, we provide only a brief description of MD simulations and explain how we can obtain glass systems through MD simulations (we can also use MC simulations to obtain glasses in the same way).
In any solid-state material, including glasses and crystals, the constituent particles vibrate around the so-called inherent structure
\footnote{
The inherent structure of a crystal is a periodic lattice structure.
}.
We also explain how to obtain this inherent structure through computational simulations.

%%%%%%%%%%%%%%%%%%%%%%%%%%%%%%%%%%%%%%%%%%%%%%%%%%%%%%%%%%%%%%%%%%%%%%%%%%%%%%%%%%%%%%%%%%%
\subsection{Molecular Dynamics~(MD) simulations}
In an MD simulation, we define a system composed of an enormous number of particles and numerically solve the equations of motion to evolve the dynamics of those constituent particles.
Here, let us consider a 3-dimensional system composed of $N$ particles in a cubic box of length $L$ and volume $V \equiv L^3$.
The mass of particle $i$~($i=1,2,\cdots,N$) is denoted by $m_i$, and the position of particle $i$ is $\mathbf{r}_i = [r_{ix},r_{iy},r_{iz}]^T$~(a 3-dimensional vector),~where $T$ denotes transposition.
To represent the configuration of all the constituent particles, we introduce a $3N$-dimensional vector $\mathbf{r} \equiv \left[ \mathbf{r}_1^T, \mathbf{r}_2^T, \cdots, \mathbf{r}_N^T \right]^T$.
In this chapter, we treat vectorial quantities, such as $\mathbf{r}_i$ and $\mathbf{r}$, as \textit{vertical} vectors.
We suppose that the total potential energy of the system, $\Phi$, is a function of the positions of the particles: $\Phi(\mathbf{r}) \equiv \Phi(\mathbf{r}_1,\mathbf{r}_2,\cdots,\mathbf{r}_N)$.
For the case of a pair-wise potential, where particles $i$ and $j$ interact through the potential $\phi_{ij}$, $\Phi(\mathbf{r}) = \sum_{i<j} \phi_{ij}$.
The equation of motion is then
\begin{equation} 
m_i \frac{d^2 \mathbf{r}_i}{dt^2} = - \frac{\partial \Phi}{\partial \mathbf{r}_i} \quad (i=1,2,\cdots,N), \label{eoma}
\end{equation}
or, in terms of the $3N$-dimensional vector $\mathbf{r}$,
\begin{equation}
\mathcal{M} \frac{d^2 \mathbf{r}}{dt^2} = - \frac{\partial \Phi}{\partial \mathbf{r}}, \label{eom}
\end{equation}
where $t$ is time and we introduce the mass matrix $\mathcal{M}$, which is a $3N \times 3N$ diagonal matrix:
\begin{equation}
\mathcal{M} \equiv \text{diag}(m_1,m_1,m_1,m_2,m_2,m_2,\cdots,m_N,m_N,m_N).
\end{equation}

We start with some initial configuration of particles, $\mathbf{r}(t=0)$, and velocity, $d\mathbf{r}(t=0)/dt$, and solve Eq.~(\ref{eoma})~(or Eq.~(\ref{eom})).
Here, we need to implement appropriate boundary conditions in the space.
In this chapter, we consider periodic boundary conditions in all directions.
Thermodynamic quantities such as temperature $T$ and pressure $p$ can be calculated from the trajectory $\mathbf{r}(t)$: for example, the temperature is calculated from the total kinetic energy of the particles.
We may use a heat bath and/or a pressure bath to control the temperature and/or the pressure, respectively.
As the simulation runs, the system approaches a steady, equilibrium state.

Several potentials $\Phi(\mathbf{r})$ have been proposed for modelling different types of glass systems.
Here, we briefly introduce the potentials for modelling atomic glasses~(packed glasses) and covalently bonded glasses~(network glasses).
\begin{enumerate}
\renewcommand{\labelenumi}{(\Roman{enumi})}
\item Atomic glasses~(packed glasses): \\
To model an atomic glass composed of rare gas atoms such as argons, we can employ the Lennard-Jones (LJ) potential~\cite{Hansen_2006}:
\begin{equation} 
\Phi_\text{LJ}(\mathbf{r}) = \sum_{i<j} 4\epsilon_{ij} \left[ \left( \frac{\sigma_{ij}}{r_{ij}} \right)^{12} - \left( \frac{\sigma_{ij}}{r_{ij}} \right)^{6} \right], \label{lj}
\end{equation}
where $r_{ij} \equiv | \mathbf{r}_i -\mathbf{r}_j |$ is the distance between particles $i$ and $j$ and $\sigma_{ij}$ and $\epsilon_{ij}$ represent the length and energy scales, respectively.
We also often employ the soft-core (SC) potential, which is the repulsive part of the LJ potential:
\begin{equation} 
\Phi_\text{SC}(\mathbf{r}) = \sum_{i<j} \epsilon_{ij} \left( \frac{\sigma_{ij}}{r_{ij}} \right)^{12}. \label{sc}
\end{equation}
In addition, as the simplest model of a glass, a finite-range harmonic potential has been employed:
\begin{equation} 
\Phi_\text{HA}(\mathbf{r}) = \sum_{i<j} \frac{\epsilon_{ij}}{2} \left( 1- \frac{r_{ij}}{\sigma_{ij}} \right)^{2} H(\sigma_{ij}-r_{ij}), \label{ha}
\end{equation}
where $H(x)$ is the Heaviside step function: $H(x) = 1$ for $x \ge 0$ and $H(x) = 0$ for $x<0$.
This harmonic potential was originally proposed for modelling granular materials, emulsions, foams, etc.~\cite{van_Hecke_2009}.

\smallskip
\item Covalently bonded glasses~(network glasses): \\
To model silica~(SiO$_2$) glass, which is one example of a covalently bonded glass, the Beest-Kramer-Santen potential~\cite{Beest_1990} has been proposed:
\begin{equation} 
\Phi_\text{BKS}(\mathbf{r}) = \sum_{i<j} \left[ \frac{q_i q_j e^2}{r_{ij}} +A_{ij} \exp(-B_{ij} r_{ij}) - \frac{C_{ij}}{r_{ij}^6} \right], \label{bks}
\end{equation}
where particles $i$ and $j$ are Si or O atoms, $q_i e$ is the electric charge ($e$ is elementary charge), and $A_{ij}$, $B_{ij}$, and $C_{ij}$ are constants.
The first term in Eq.~(\ref{bks}) represents the Coulomb interaction.
In addition, the Stillinger-Weber potential~\cite{Stillinger_1985}, $\Phi_\text{SW}(\mathbf{r})$, has been proposed for modelling amorphous silicon.
$\Phi_\text{SW}(\mathbf{r})$ includes three-body interactions to represent the bending rigidity.
For the explicit equation for $\Phi_\text{SW}$, please see Ref.~\onlinecite{Stillinger_1985}.
\end{enumerate}
In addition, for metallic glasses, Daw and Baskes have proposed the embedded-atom method~(EAM) based on density functional theory~\cite{Daw_1984}.
Additionally, for polymer glasses, we can perform coarse-grained simulations by using the Kremer--Grest model~\cite{Kremer_1990}, which treats polymer chains as
linear series of monomer beads (particles).
In the following section, we take a system with the SC potential~[Eq.~(\ref{sc})] as an example.
We present numerical results not only for the glass but also for the crystal, which are obtained from Refs.~\onlinecite{Mizuno2_2013,Mizuno_2014,Mizuno2_2016}.
By comparing the glass and crystal, we discuss the characteristic features of glasses.
The values of quantities are presented in units of the mass ($m$), length ($\sigma$), and energy ($\epsilon$) scales that are typical of the constituent particles.

%%%%%%%%%%%%%%%%%%%%%%%%%%%%%%%%%%%%%%%%%%%%%%%%%%%%%%%%%%%%%%%
\begin{figure}[t]
\centerline{
\subfigure[Glass.]
{\includegraphics[width=0.5\textwidth]{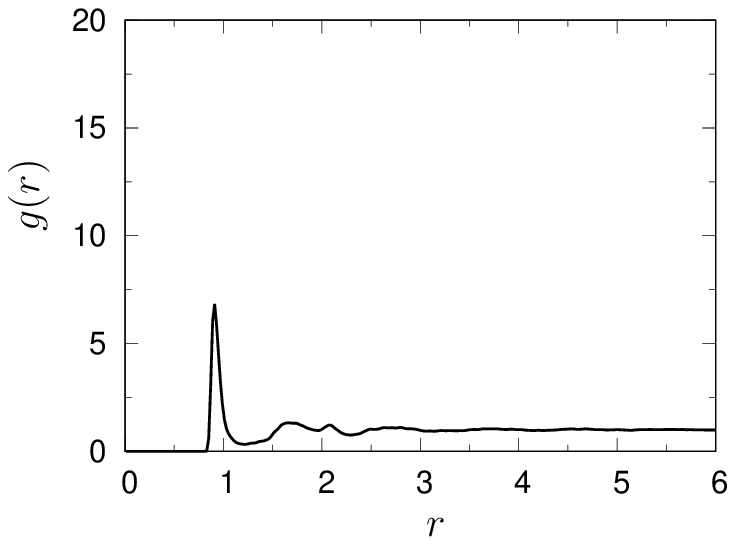}\label{fig1a}}
\hspace*{0mm}
\subfigure[Crystal.]
{\includegraphics[width=0.5\textwidth]{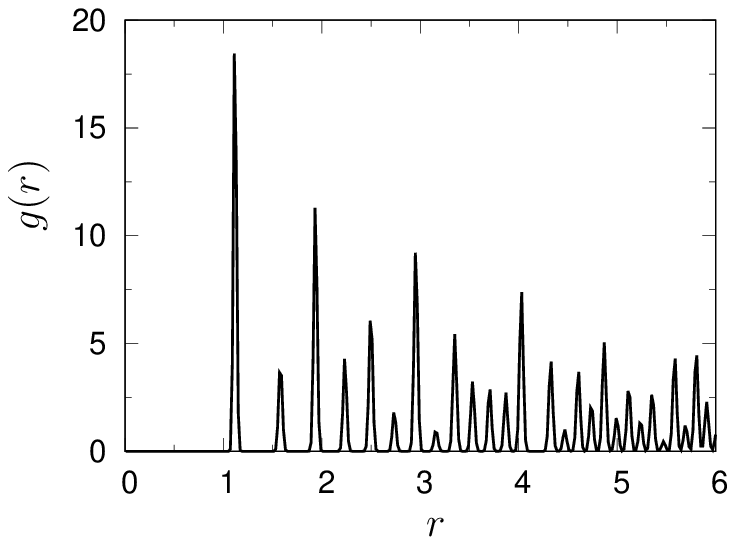}\label{fig1b}}
}
\caption{
Radial distribution function $g(r)$ of a soft-core (SC) system at a low temperature of $T=10^{-2}$, well below $T_g$ and $T_m$.
(a) Glass.
(b) Face-centred cubic (FCC) crystal.
}
\label{fig1}
\end{figure}
%%%%%%%%%%%%%%%%%%%%%%%%%%%%%%%%%%%%%%%%%%%%%%%%%%%%%%%%%%%%%%%

%%%%%%%%%%%%%%%%%%%%%%%%%%%%%%%%%%%%%%%%%%%%%%%%%%%%%%%%%%%%%%%%%%%%%%%%%%%%%%%%%%%%%%%%%%%
\subsection{Glass and inherent structure}
By means of an MD simulation, we obtain the glass as follows~\cite{Cavagna_2009,Berthier_2011}.
We first equilibrate the system in the liquid phase above the melting temperature, $T_m$.
We next rapidly quench the system below $T_m$.
If we avoid crystallization
\footnote{
To avoid crystallization, we may employ a poly-disperse system composed of a number of different kinds of particles with different mass, length, and energy scales.
},
we can maintain the liquid phase even below $T_m$; such a phase is called a supercooled liquid phase.
When we further quench the system below the glass transition temperature, $T_g$, the system is finally frozen in a disordered configuration, without crystallization.
To study the structural properties of this system, we can measure the radial distribution function $g(r)$ or the static structure factor $S(q)$~\cite{Hansen_2006}.
Figure~\ref{fig1} presents the $g(r)$ functions for two SC systems, (a) glass and (b) crystal, for comparison.
We can see long-range order in the crystal, while the glass shows only short-range order.
Although glasses are solids like crystals (i.e., they exhibit rigidity), they also possess disordered structures as liquids do~\cite{Cavagna_2009,Berthier_2011}.

In the glass phase below $T_g$, the particles vibrate around the inherent structure, which is denoted by $\mathbf{R} \equiv \left[ \mathbf{R}_1^T, \mathbf{R}_2^T, \cdots, \mathbf{R}_N^T \right]^T$
\footnote{
Glasses may show ageing phenomena at a time scale that is much longer than that of vibrations.
Ageing induces rearrangements of some particles and alters the inherent structure.
In this chapter, we will not consider this effect.
}.
In the inherent structure, the system is in a state of mechanical equilibrium, and the potential energy $\Phi(\mathbf{r})$ takes its minimum value in the $3N$-dimensional space:
\begin{equation}
\left. \frac{\partial \Phi}{\partial \mathbf{r}} \right|_{\mathbf{r}=\mathbf{R}} = 0. \label{firstderivative}
\end{equation}
Thus, we can numerically obtain $\mathbf{R}$ by minimizing $\Phi(\mathbf{r})$.
For this purpose, several numerical techniques have been established, e.g., the steepest descent method~\cite{Press_2007}, the conjugate gradient method~\cite{Press_2007}, and the fire algorithm~\cite{Bitzek_2006}.
We note that minimizing the potential corresponds to quenching the system to zero temperature, $T=0$, and the inherent structure is the configuration of the particles at $T=0$.

%%%%%%%%%%%%%%%%%%%%%%%%%%%%%%%%%%%%%%%%%%%%%%%%%%%%%%%%%%%%%%%%%%%%%%%%%%%%%%%%%%%%%%%%%%%
\section{Vibrational eigenmodes}\label{sec.modeanalysis}
In a glass (or any solid-state material), the constituent particles vibrate around the inherent structure.
These vibrational motions of particles are excited along vibrational eigenmodes.
The eigenmodes are therefore fundamental to understanding the vibrational properties.
In this section, we explain how to obtain the eigenmodes by means of computational simulations.
We then describe the vDOS, which represents the statistics of the eigenmodes.
We also explain the order parameters used to characterize each vibrational eigenmode.
In particular, we introduce the participation ratio, which measures the extent of localization, and the phonon order parameter, which measures the extent of phonon-like vibrations.

%%%%%%%%%%%%%%%%%%%%%%%%%%%%%%%%%%%%%%%%%%%%%%%%%%%%%%%%%%%%%%%%%%%%%%%%%%%%%%%%%%%%%%%%%%%
\subsection{General description}
Let us introduce the displacement vector ($3N$-dimensional vector) of the particles relative to the inherent structure $\mathbf{R}$ (multiplied by the mass factor $\sqrt{\mathcal{M}}$) as follows:
\begin{equation}
\mathbf{u} = \sqrt{\mathcal{M}}(\mathbf{r} - \mathbf{R}).
\end{equation}
We then expand the potential $\Phi(\mathbf{r})$ around $\mathbf{R}$ in a power series of $\mathbf{u}$ as follows~($T$ denotes transposition):
\begin{equation}
\Phi(\mathbf{r}) = \Phi(\mathbf{R}) + \frac{1}{2} \mathbf{u}^T \mathcal{D} \mathbf{u} + \mathcal{O}(|\mathbf{u}|^3). \label{expand}
\end{equation}
The first derivative of $\Phi(\mathbf{r})$ is zero~(mechanical equilibrium), as expressed in Eq.~(\ref{firstderivative}).
$\mathcal{D}$ is the so-called dynamical matrix (a $3N \times 3N$ matrix) which is the second derivative of $\Phi(\mathbf{r})$~\cite{Ashcroft_1976,Kittel_1996}:
\begin{equation}
\mathcal{D} = \left. \frac{1}{\sqrt{\mathcal{M}}} \left[ \frac{\partial^2 \Phi}{\partial \mathbf{r} \partial \mathbf{r}^T} \right|_{\mathbf{r}=\mathbf{R}} \right] \frac{1}{\sqrt{\mathcal{M}}}.
\end{equation}
$\mathcal{D}$ contains $3\times 3$ matrix elements, $(\mathcal{D})_{ij}$, corresponding to particles $i$ and $j$:
\begin{equation}
(\mathcal{D})_{ij} = \left. \frac{1}{\sqrt{m_i m_j}} \left[ \frac{\partial^2 \Phi}{\partial \mathbf{r}_i \partial \mathbf{r}_j^T} \right|_{\mathbf{r}=\mathbf{R}} \right].
\end{equation}
In the case of a pair-wise potential, $\phi_{ij}(r_{ij})$, which depends only on the distance $r_{ij}$ between particles $i$ and $j$, $(\mathcal{D})_{ij}$ is formulated as
\begin{equation}
(\mathcal{D})_{ij} = 
\left\{
\begin{aligned}
& \frac{1}{m_i} \sum_{k=1,k\neq i}^{N} \left[ \frac{d^2\phi_{ik}}{dr_{ik}^2} \frac{\mathbf{r}_{ik} \mathbf{r}^T_{ik}}{r_{ik}^2} + \frac{1}{r_{ik}} \frac{d\phi_{ik}}{dr_{ik}} \left(  I_3 -  \frac{\mathbf{r}_{ik} \mathbf{r}^T_{ik}}{r_{ik}^2} \right) \right] & (i = j), \\
&  - \frac{1}{\sqrt{m_i m_j}} \left[ \frac{d^2\phi_{ij}}{dr_{ij}^2} \frac{\mathbf{r}_{ij} \mathbf{r}^T_{ij}}{r_{ij}^2} + \frac{1}{r_{ij}} \frac{d\phi_{ij}}{dr_{ij}} \left( I_3 - \frac{\mathbf{r}_{ij} \mathbf{r}^T_{ij}}{r_{ij}^2} \right) \right] & (i \neq j),
\end{aligned} 
\right.
\end{equation}
where $I_3$ is the $3 \times 3$ unit matrix, $\mathbf{r}_{ij} \equiv \mathbf{r}_{i}-\mathbf{r}_{j}$, and $r_{ij} \equiv \left| \mathbf{r}_{ij} \right|$.
Since $|\mathbf{u}| \ll 1$ at low temperatures, we neglect higher-order terms of $\mathcal{O}(|\mathbf{u}|^3)$ in $\Phi(\mathbf{r})$ as expressed in Eq.~(\ref{expand}) and substitute this expression into Eq.~(\ref{eom}) to obtain the linearized equation of motion~(harmonic approximation):
\begin{equation} 
\frac{d^2 \mathbf{u}(t)}{dt^2} = - \mathcal{D} \mathbf{u}(t). \label{eom2}
\end{equation}

To solve Eq.~(\ref{eom2}), we perform a Fourier transform and obtain
\begin{equation} 
\omega^2 \tilde{\mathbf{u}}(\omega) = \mathcal{D} \tilde{\mathbf{u}}(\omega), \label{eom2a}
\end{equation}
where $\omega$ is the frequency and $\tilde{\mathbf{u}}(\omega)$ is the Fourier transform of $\mathbf{u}(t)$:
\begin{equation}
\tilde{\mathbf{u}}(\omega) = \int \mathbf{u}(t) \exp \left(-\text{i} \omega t \right) dt.
\end{equation}
$u(t)$ is then obtained by taking the inverse Fourier transform of $\tilde{\mathbf{u}}(\omega)$:
\begin{equation}
{\mathbf{u}}(t) = \frac{1}{2\pi} \int \tilde{\mathbf{u}}(\omega) \exp \left(\text{i} \omega t \right) d\omega.
\end{equation}
We thus encounter the eigenvalue problem of the matrix $\mathcal{D}$~[Eq.~(\ref{eom2a})].

Since $\mathcal{D}$ is a symmetric matrix, it can be diagonalized by an orthonormal matrix ${X}$, and its eigenvalues $\lambda^k$ ($k=1,2,\cdots,3N$) are all real numbers:
\begin{equation}
{X}^T \mathcal{D} {X}=  \text{diag}(\lambda^1,\lambda^2,\cdots,\lambda^{3N}), \label{diagonal}
\end{equation}
\begin{equation}
X^T X = I_{3N}, \label{diagonal2}
\end{equation}
where $I_{3N}$ is the $3N \times 3N$ unit matrix.
We write the matrix in the form of $X \equiv \left[ \mathbf{e}^{1}, \mathbf{e}^2, \cdots, \mathbf{e}^{3N} \right]$, thereby introducing $3N$-dimensional eigenvectors, $\mathbf{e}^k \equiv \left[ \mathbf{e}_1^{kT}, \mathbf{e}_2^{kT}, \cdots, \mathbf{e}_N^{kT} \right]^T$ ($k=1,2,\cdots,3N$).
The orthonormality condition, Eq.~(\ref{diagonal2}), then gives
\begin{equation}
\mathbf{e}^{k} \cdot \mathbf{e}^{l} = \sum_{i=1}^N \mathbf{e}^{k}_i \cdot \mathbf{e}^{l}_i = \delta_{kl}, \label{orthnormal}
\end{equation}
where $\delta_{kl}$ is the Kronecker delta
\footnote{
The set of $3N$ eigenvectors $\left\{ \mathbf{e}^{1}, \mathbf{e}^2, \cdots, \mathbf{e}^{3N} \right\}$ can be treated as an orthonormal basis in $3N$-dimensional space for $\mathbf{u}$.
}.
We thus obtain $3N$ sets of eigenvalues $\lambda^k$ and eigenvectors $\mathbf{e}^k$ for the matrix $\mathcal{D}$:
\begin{equation}
\lambda^k \mathbf{e}^k = \mathcal{D} \mathbf{e}^k \quad (k=1,2,\cdots,3N), \label{eigenequation}
\end{equation}
and $\mathcal{D}$ can be described as
\begin{equation}
\mathcal{D} = \sum_{k=1}^{3N} \lambda^k \mathbf{e}^{k} \mathbf{e}^{kT}.
\end{equation}
If the system is stable, the eigenvalues are positive, $\lambda^k >0$, and the eigenfrequencies, $\omega^k \equiv \sqrt{\lambda^k}$, are real numbers.
A general solution for $\tilde{\mathbf{u}}(\omega)$ in Eq.~(\ref{eom2a}) is thus obtained as a superposition of the $\mathbf{e}^{k}$:
\begin{equation}
\tilde{\mathbf{u}}(\omega) = \sum_{k=1}^{3N} A^k \delta(\omega-\omega^k) \mathbf{e}^{k}, \label{mode1}
\end{equation}
where the $A^k$ are complex constants, and $\mathbf{u}(t)$ is
\begin{equation}
\mathbf{u}(t) = \frac{1}{2\pi} \sum_{k=1}^{3N} A^k \exp\left( \text{i} \omega^k t \right) \mathbf{e}^{k}. \label{mode1a}
\end{equation}

The values of the $A^k$ are determined from the initial conditions $\mathbf{u}=\mathbf{u}_0$ and $d \mathbf{u}/dt=\dot{\mathbf{u}}_0$ at $t=0$, as $\text{Re}(A^k) = 2\pi\left( \mathbf{e}^k \cdot \mathbf{u}_0 \right)$ and $\text{Im}(A^k) = -2\pi\left( \mathbf{e}^k \cdot \dot{\mathbf{u}}_0 \right)/\omega^k$ (where Re and Im denote the real and imaginary parts, respectively), and we finally obtain $\mathbf{u}(t)$ as follows:
\begin{equation}
\begin{aligned}
\mathbf{u}(t) &= \sum_{k=1}^{3N} u^k(t) \mathbf{e}^{k}, \\
u^k(t) &= \left( \mathbf{e}^k \cdot \mathbf{u}_0 \right) \cos\left( \omega^k t \right) + \left( \mathbf{e}^k \cdot \dot{\mathbf{u}}_0\right) \frac{ \sin \left( \omega^k t \right)}{\omega^k}, \label{mode1b}
\end{aligned}
\end{equation}
which is a superposition of terms of the form $u^k(t) \mathbf{e}^{k}$, which represents the vibrations of particles along the eigenvector $\mathbf{e}^{k}$ (vibration of particle $i$ along $\mathbf{e}_i^k$) with frequency $\omega^k$ and which we call the vibrational eigenmodes $k$.

The kinetic energy, $K$, and the potential energy, $\Delta \Phi \equiv \Phi - \Phi(\mathbf{R})$, are described as sums of the energies of the eigenmodes $k$:
\begin{equation}
\begin{aligned}
K(t) &= \frac{1}{2} \frac{d \mathbf{u}(t)}{dt} \cdot \frac{d \mathbf{u}(t)}{dt} = \frac{1}{2} \sum_{k=1}^{3N} \left( \frac{d {u}^k(t)}{dt} \right)^2, \\
\Delta \Phi (t) &= \frac{1}{2} \mathbf{u}(t)^T \mathcal{D} \mathbf{u}(t) = \frac{1}{2} \sum_{k=1}^{3N} \left( \omega^k {u}^k(t) \right)^2.
\end{aligned}
\end{equation}
The total energy, $K(t) + \Delta \Phi(t) = \left( 1/2 \right) \sum_{k=1}^{3N} \left[ {\omega^k}^2 \left( \mathbf{e}^k \cdot \mathbf{u}_0 \right)^2 + \left( \mathbf{e}^k \cdot \dot{\mathbf{u}}_0 \right)^2 \right]$, is constant, independent of time~(energy conservation).

%%%%%%%%%%%%%%%%%%%%%%%%%%%%%%%%%%%%%%%%%%%%%%%%%%%%%%%%%%%%%%%%%%%%%%%%%%%%%%%%%%%%%%%%%%%
\subsection{Vibrational eigenmodes in elastic media (elastic waves)}
As the simplest case, let us consider the eigenmodes in an elastic medium, i.e., elastic waves.
Elastic media are continuum systems (i.e., not particulate systems), and their eigenfrequencies and eigenvectors are determined by continuum mechanics~\cite{Ashcroft_1976,Kittel_1996}.

An elastic wave is specified by the wavevector $\mathbf{q}$ and the polarization $\alpha$.
In the 3-dimensional case, there are three polarizations: two transverse ($\alpha = T_1,\ T_2$) and one longitudinal ($\alpha =L$).
An eigenvector of elastic waves is a continuous function of the position $\mathbf{r} \in V$ (where $V$ is the volume of the system)
\footnote{
The continuous variable $\mathbf{r}$ for an elastic medium corresponds to the particle index ``$i$'' in a particulate system.
}:
\begin{equation}
\mathbf{e}^{\mathbf{q},\alpha}_\text{el} \equiv \left[ \mathbf{e}^{\mathbf{q},\alpha}_\text{el}(\mathbf{r});\ \mathbf{r} \in V \right] \equiv  \left[ \mathbf{s}_\alpha( \hat{\mathbf{q}} ) \frac{ \exp(\text{i}\mathbf{q}\cdot\mathbf{r}) }{\sqrt{V}};\ \mathbf{r} \in V \right]. \label{elasticeigen}
\end{equation}
$\mathbf{s}_\alpha(\hat{\mathbf{q}})$ is the polarization vector, which depends on the direction of $\mathbf{q}$, i.e., $\hat{\mathbf{q}} \equiv \mathbf{q}/q$ (where $q \equiv |\mathbf{q}|$ is the wavenumber), and can be determined from the elastic equation of motion that corresponds to the eigen equation~(\ref{eigenequation}):
\begin{equation}
\omega^2 \mathbf{s}_\alpha(\hat{\mathbf{q}}) = \left( \frac{ \mathcal{C}(\hat{\mathbf{q}}) }{\rho} \right) q^2 \mathbf{s}_\alpha(\hat{\mathbf{q}}), \label{elasticeigeneq}
\end{equation}
where $\rho$ is the mass density and $\mathcal{C}(\hat{\mathbf{q}})$~(a $3\times 3$ tensor) is the elastic modulus tensor in the $\hat{\mathbf{q}}$ direction
\footnote{
Given the elastic modulus tensor $C_{\alpha \beta \gamma \delta}$~(a $3\times 3 \times 3 \times 3$ tensor), where $\alpha, \beta, \gamma, \delta = x,y,z$, $\mathcal{C}(\hat{\mathbf{q}})$ is defined as $\left(\mathcal{C}(\hat{\mathbf{q}}) \right)_{\alpha \delta} = \sum_{\beta,\gamma} C_{\alpha \beta \gamma \delta} \hat{q}_\beta \hat{q}_\gamma$, where $\hat{\mathbf{q}} \equiv (\hat{q}_x, \hat{q}_y, \hat{q}_z)$.
We will describe the measurement of $C_{\alpha \beta \gamma \delta}$ in Section~\ref{sec.moduli}.
}.

Since $\left( \mathcal{C}(\hat{\mathbf{q}})/\rho \right) q^2$~(a $3 \times 3$ matrix) is positive and symmetric, its eigenvalues $\omega^2$ are real, positive numbers, and the eigenvectors $\mathbf{s}_\alpha(\hat{\mathbf{q}})$ are orthonormalized as $\mathbf{s}_\alpha(\hat{\mathbf{q}}) \cdot \mathbf{s}_{\alpha'}(\hat{\mathbf{q}}) = \delta_{\alpha \alpha'}$.
The eigenvectors $\mathbf{e}^{\mathbf{q},\alpha}_\text{el}(\mathbf{r})$ are then orthonormalized as
\footnote{
For an elastic medium, we need to replace $\sum_{i=1}^N$ in Eq.~(\ref{orthnormal}) with $\int_V d^3\mathbf{r}$.
}
\begin{equation}
\mathbf{e}^{\mathbf{q},\alpha}_\text{el} \cdot \mathbf{e}^{\mathbf{q}',\alpha' \ast}_\text{el} = \int_V d^3\mathbf{r}\ \mathbf{e}^{\mathbf{q},\alpha}_\text{el}(\mathbf{r}) \cdot \mathbf{e}^{\mathbf{q}',\alpha' \ast}_\text{el}(\mathbf{r}) = \delta_{\mathbf{q}\mathbf{q}'}\delta_{\alpha \alpha'},
\end{equation}
where $\ast$ denotes the complex conjugate.
Eq.~(\ref{elasticeigeneq}) determines the eigenfrequency $\omega_\alpha (\mathbf{q})$ as a linear function of $q$:
\begin{equation}
\omega_\alpha (\mathbf{q}) = c_\alpha (\hat{\mathbf{q}}) q, \label{lineardispersion}
\end{equation}
where $c_\alpha (\hat{\mathbf{q}})$ is the sound speed, which is the square root of the eigenvalue of $\mathcal{C}(\hat{\mathbf{q}})/\rho$.
We note that Eq.~(\ref{elasticeigeneq}) is the so-called Christoffel equation, which can be solved analytically following the Every algorithm~\cite{Every_1980}.

In the case of an isotropic medium, the elastic modulus tensor contains two independent moduli: the shear modulus $G$ and the bulk modulus $K$.
$\mathbf{s}_{T_1}(\hat{\mathbf{q}})$ and $\mathbf{s}_{T_2}(\hat{\mathbf{q}})$ are perpendicular to $\mathbf{q}$, while $\mathbf{s}_{L}(\hat{\mathbf{q}})$ is parallel to $\mathbf{q}$: $\hat{\mathbf{q}} \cdot \mathbf{s}_{T_1}(\hat{\mathbf{q}}) = \hat{\mathbf{q}} \cdot \mathbf{s}_{T_2}(\hat{\mathbf{q}})=0$ and $\mathbf{s}_{L}(\hat{\mathbf{q}}) = \hat{\mathbf{q}}$.
$c_\alpha (\hat{\mathbf{q}})$ does not depend on $\hat{\mathbf{q}}$, and it is calculated as
\begin{equation}
c_{T_1} = c_{T_2} = \sqrt{\frac{G}{\rho}}, \qquad c_{L}=\sqrt{\frac{K + 4G/3}{\rho}}.
\label{fccsoundspeediso}
\end{equation}
%

%%%%%%%%%%%%%%%%%%%%%%%%%%%%%%%%%%%%%%%%%%%%%%%%%%%%%%%%%%%%%%%
\begin{figure}[t]
\centerline{
\includegraphics[width=0.95\textwidth]{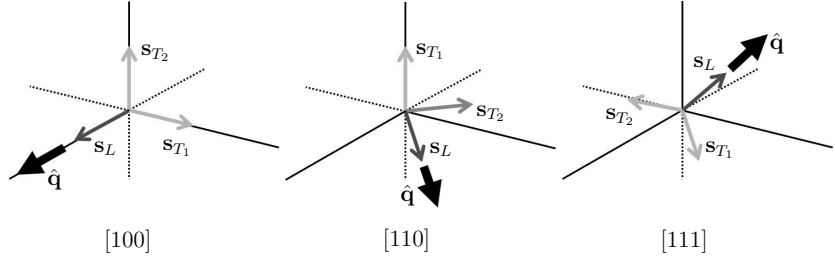}
}
\caption{
Schematic illustration of the polarization vector $\mathbf{s}_\alpha(\hat{\mathbf{q}})$ in a medium with cubic symmetry.
We show $\mathbf{s}_{T_1}(\hat{\mathbf{q}})$, $\mathbf{s}_{T_2}(\hat{\mathbf{q}})$, and $\mathbf{s}_L(\hat{\mathbf{q}})$ in three directions of $\hat{\mathbf{q}}$: $[100]$, $[110]$, and $[111]$~(Miller indices).
}
\label{fig2}
\end{figure}
%%%%%%%%%%%%%%%%%%%%%%%%%%%%%%%%%%%%%%%%%%%%%%%%%%%%%%%%%%%%%%%

On the other hand, in the case of an anisotropic medium, $\mathbf{s}_\alpha(\hat{\mathbf{q}})$ and $c_\alpha (\hat{\mathbf{q}})$ depend on $\hat{\mathbf{q}}$.
The simplest example is a medium with cubic symmetry, which has three independent moduli: the pure shear modulus $G_p$, the simple shear modulus $G_s$, and the bulk modulus $K$
\footnote{
$G_p$, $G_s$, and $K$ correspond to the elastic moduli for pure shear, simple shear, and bulk deformations, respectively.
These three deformations are illustrated in Fig.~\ref{fig11} of Section~\ref{sec.moduli}.
}.
Figure~\ref{fig2} shows schematic illustrations of $\mathbf{s}_\alpha(\hat{\mathbf{q}})$ in three directions, namely, $[100]$, $[110]$, and $[111]$, where we use Miller indices
\footnote{
$\hat{\mathbf{q}} = (1,0,0)$ and all equivalent vectors for $[100]$, $\hat{\mathbf{q}} = (1/\sqrt{2},1/\sqrt{2},0)$ and all equivalent vectors for $[110]$, and $\hat{\mathbf{q}} = (1/\sqrt{3},1/\sqrt{3},1/\sqrt{3})$ and all equivalent vectors for $[111]$.
}.
Additionally, $c_\alpha (\hat{\mathbf{q}})$ is calculated as
\begin{equation}
c_{T_1} = c_{T_2} = \sqrt{\frac{G_s}{\rho}},\quad c_L=\sqrt{\frac{K + 4G_p/3}{\rho}} \quad \text{for $[100]$}, \label{fccsoundspeed1}
\end{equation}
\begin{equation}
c_{T_1}= \sqrt{\frac{G_s}{\rho}},\quad c_{T_2} = \sqrt{\frac{G_p}{\rho}},\quad c_L=\sqrt{\frac{K + G_p/3 + G_s}{\rho}} \quad \text{for $[110]$}, \label{fccsoundspeed2}
\end{equation}
\begin{equation}
c_{T_1} = c_{T_2} = \sqrt{\frac{2G_p + G_s}{3\rho}},\quad c_L=\sqrt{\frac{K + 4G_s/3}{\rho}} \quad \text{for $[111]$}. \label{fccsoundspeed3}
\end{equation}
A general solution for $c_\alpha (\hat{\mathbf{q}})$ as a function of $\hat{\mathbf{q}}$ is given in Ref.~\onlinecite{Jasiukiewicz_2003}.

%%%%%%%%%%%%%%%%%%%%%%%%%%%%%%%%%%%%%%%%%%%%%%%%%%%%%%%%%%%%%%%
\begin{figure}[t]
\centerline{
\subfigure[Glass.]
{\includegraphics[width=0.4\textwidth]{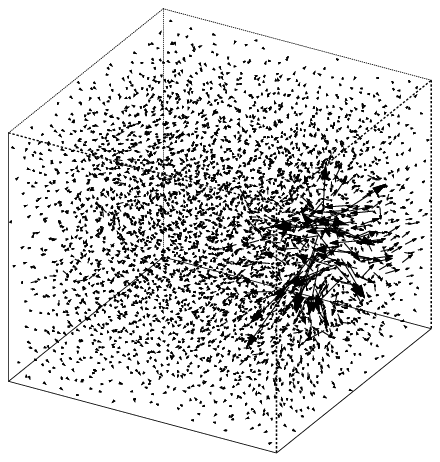}\label{fig3a}}
\hspace*{4mm}
\subfigure[Crystal.]
{\includegraphics[width=0.4\textwidth]{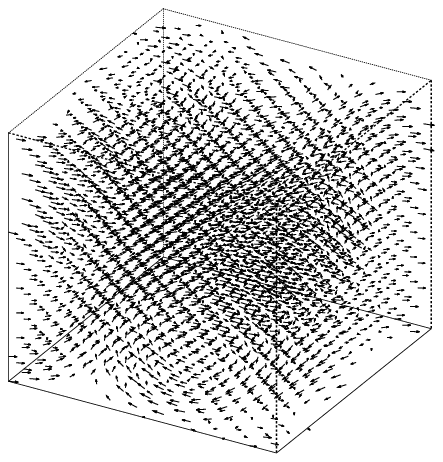}\label{fig3b}}
}

\vspace*{2mm}
\centerline{
\subfigure[Glass.]
{\includegraphics[width=0.4\textwidth]{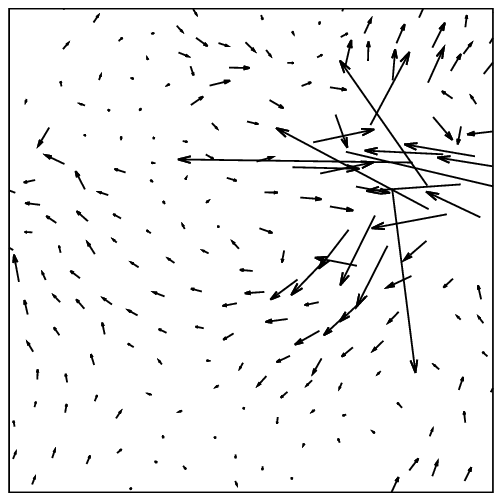}\label{fig3c}}
\hspace*{4mm}
\subfigure[Crystal.]
{\includegraphics[width=0.4\textwidth]{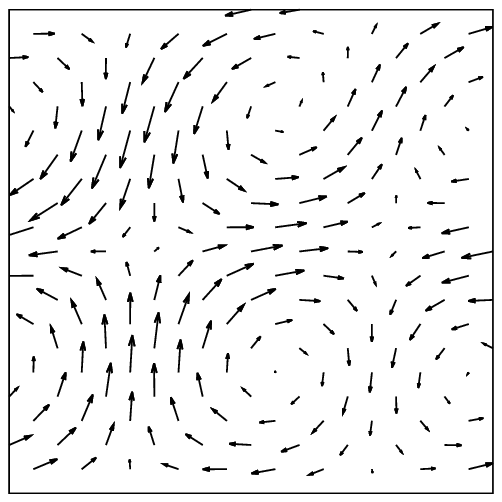}\label{fig3d}}
}
\caption{Visualization of the eigenmodes in SC systems.
(a),(c) Glass.
(b),(d) FCC crystal.
We show the eigenvector field, $\mathbf{e}^k \equiv \left[ \mathbf{e}_1^{kT}, \mathbf{e}_2^{kT},\cdots,\mathbf{e}_N^{kT} \right]^T$, as arrows.
In (c),(d), the eigenvectors on some plane are presented in a two-dimensional plot.
}
\label{fig3}
\end{figure}
%%%%%%%%%%%%%%%%%%%%%%%%%%%%%%%%%%%%%%%%%%%%%%%%%%%%%%%%%%%%%%%

%%%%%%%%%%%%%%%%%%%%%%%%%%%%%%%%%%%%%%%%%%%%%%%%%%%%%%%%%%%%%%%%%%%%%%%%%%%%%%%%%%%%%%%%%%%
\subsection{Vibrational eigenmodes in crystals (phonons)}
We next consider the eigenmodes in crystals.
The inherent structure $\mathbf{R}$ of a crystal is a periodic lattice structure (we show the radial distribution function $g(r)$ in Fig.~\ref{fig1b}).
In this case, the eigenmodes have been established to be phonons (lattice waves)~\cite{Ashcroft_1976,Kittel_1996}.
Here, we consider single-component crystals, for which the eigenmodes are phonons of the acoustic type.
For multi-component crystals, phonons of the optical type also appear, which are not considered below.

A phonon is specified by the wavevector $\mathbf{q}$ and the polarization $\alpha$, as is an elastic wave.
There are $N$ wavevectors in the first Brillouin zone, $\mathbf{q} = \mathbf{q}_1, \mathbf{q}_2,\cdots, \mathbf{q}_N$, and three polarizations, $\alpha = T_1,\ T_2,\ L$; therefore, there are $3N$ phonons in total.
The eigenvector $\mathbf{e}_\text{ph}^{\mathbf{q},\alpha}$ is described as
\begin{equation}
\mathbf{e}^{\mathbf{q},\alpha}_\text{ph} \equiv \left[ \mathbf{s}_\alpha^T(\mathbf{q}) \frac{ \exp(\text{i}\mathbf{q}\cdot\mathbf{R}_1) }{\sqrt{N}},\mathbf{s}_\alpha^T(\mathbf{q}) \frac{ \exp(\text{i}\mathbf{q}\cdot\mathbf{R}_2) }{\sqrt{N}},\cdots,\mathbf{s}_\alpha^T(\mathbf{q})\frac{ \exp(\text{i}\mathbf{q}\cdot\mathbf{R}_N) }{\sqrt{N}} \right]^T, \label{phononeigen}
\end{equation}
where the factor $1/\sqrt{N}$ is necessary for the orthonormality condition, $\mathbf{e}^{\mathbf{q},\alpha}_\text{ph} \cdot \mathbf{e}^{\mathbf{q}',\alpha' \ast}_\text{ph} = \delta_{\mathbf{q}\mathbf{q}'}\delta_{\alpha \alpha'}$.
The eigen equation~(\ref{eigenequation}) then reduces to
\begin{equation}
\omega^2 \mathbf{s}_\alpha({\mathbf{q}}) = {D}({\mathbf{q}}) \mathbf{s}_\alpha({\mathbf{q}}), \label{eomcrystal}
\end{equation}
where ${D}({\mathbf{q}})$ (a $3\times 3$ matrix) is the dynamical matrix of the unit cell of the lattice structure:
\begin{equation}
{D}({\mathbf{q}}) = \frac{1}{N} \sum_{i,j=1}^N \left(\mathcal{D}\right)_{ij} \exp(-\text{i}\mathbf{q}\cdot (\mathbf{R}_i-\mathbf{R}_j)).
\end{equation}
Since ${D}({\mathbf{q}})$ is positive and symmetric, Eq.~(\ref{eomcrystal}) gives the real number of eigenfrequencies $\omega_\alpha(\mathbf{q})$ and polarization vectors $\mathbf{s}_\alpha(\mathbf{q})$ with the orthonormality condition, $\mathbf{s}_\alpha({\mathbf{q}}) \cdot \mathbf{s}_{\alpha'}({\mathbf{q}}) = \delta_{\alpha \alpha'}$.
The eigenvectors of the phonons in Eq.~(\ref{phononeigen}) are of the same form as those of the elastic waves in Eq.~(\ref{elasticeigen}); however, we remark that they are discretized waves in a particulate system.
As an example, we visualize a transverse phonon in Figs.~\ref{fig3b} and~\ref{fig3d}.
We can see the vortex structure that is characteristic of transverse phonons.

An important feature is that the dynamical matrix, ${D}({\mathbf{q}})$, converges to that of elastic waves, $(\mathcal{C}(\hat{\mathbf{q}})/\rho)q^2$, at low wavenumbers $q \ll 1$ with
\footnote{
We can show that $\left(\mathcal{C}(\hat{\mathbf{q}}) \right)_{\alpha \delta} = \sum_{\beta,\gamma} \left( C^B_{\alpha \beta \gamma \delta} + C^C_{\alpha \beta \gamma \delta} \right)\hat{q}_\beta \hat{q}_\gamma$ with $\hat{\mathbf{q}} \equiv (\hat{q}_x, \hat{q}_y, \hat{q}_z)$, where $C^B_{\alpha \beta \gamma \delta}$ and $C^C_{\alpha \beta \gamma \delta}$ are respectively the Born term and the correction term of elastic modulus tensor, given in Eq.~(\ref{fformuation2}) of Section~\ref{sec.moduli}.
}
\begin{equation}
\frac{\mathcal{C}(\hat{\mathbf{q}})}{\rho} = -\frac{1}{2N} \sum_{i,j=1}^N \left(\mathcal{D}\right)_{ij} \left[ \hat{\mathbf{q}} \cdot \left(\mathbf{R}_i-\mathbf{R}_j\right) \right]^2,
\end{equation}
and thus, the phonons smoothly converge to elastic waves~\cite{Ashcroft_1976,Kittel_1996}
\footnote{
Acoustic phonons (the type considered here) converge to elastic waves at $q \ll 1$; however, optical phonons (not considered here) do not.
Optical phonons represent vibrations of particles within a unit cell (vibrations at microscopic scales).
}.
In particular, the phonons in cubic crystals, such as FCC crystals, converge to the elastic waves in media with cubic symmetry.
The eigenfrequency $\omega_\alpha(\mathbf{q})$ of the phonons converges to a linear function of $q$ as shown in Eq.~({\ref{lineardispersion}}).

%%%%%%%%%%%%%%%%%%%%%%%%%%%%%%%%%%%%%%%%%%%%%%%%%%%%%%%%%%%%%%%%%%%%%%%%%%%%%%%%%%%%%%%%%%%
\subsection{Vibrational eigenmodes in glasses}
Unlike for the phonons in crystals, we do not have any established formulation for the eigenmodes in glasses.
In this case, computational simulations are a powerful tool.
We can numerically diagonalize the dynamical matrix $\mathcal{D}$~(solve the eigen equation~(\ref{eigenequation})) to obtain the eigenfrequencies, $\omega^k \equiv \sqrt{\lambda^k}$, and the eigenvectors, $\mathbf{e}^k \equiv \left[ \mathbf{e}_1^{kT}, \mathbf{e}_2^{kT},\cdots,\mathbf{e}_N^{kT} \right]^T$ ($k=1,2,\cdots,3N$).
Note that when we diagonalize the $\mathcal{D}$ of a crystal, we obtain phonons, as presented in Figs.~\ref{fig3b} and~\ref{fig3d}.
We also note that for a 3-dimensional system under periodic boundary conditions, there appear three zero-frequency modes that represent uniform translations of particles. These modes are known as Goldstone modes and
are usually discarded.

Figures~\ref{fig3a} and~\ref{fig3c} present an example of eigenmodes in an SC glass.
We can see localized vibration, in which some particles vibrate considerably while other particles vibrate much less.
This localization is a characteristic feature of glasses (disordered systems)~\cite{Schober_1991,mazzacurati_1996,Taraskin_1999}.
Although phonon-like modes exist in glasses, the eigenmodes are non-phonon modes in general.
We may expect that at low frequencies, glasses will behave as uniform elastic media and their eigenmodes will converge to elastic waves, as in the case of crystals and phonons.
However, recent simulations~\cite{Lerner_2016,Mizuno_2017,Shimada_2017,Lerner_2017,Lerner2_2017,Angelani_2018,Bouchbinder_2018,Mizuno_2018,Shimada_2018,Wang_2019} have demonstrated that this is not the case, as will be discussed in Section~\ref{sec.recent}.

%%%%%%%%%%%%%%%%%%%%%%%%%%%%%%%%%%%%%%%%%%%%%%%%%%%%%%%%%%%%%%%
\begin{figure}[t]
\centerline{
\subfigure[Glass.]
{\includegraphics[width=0.5\textwidth]{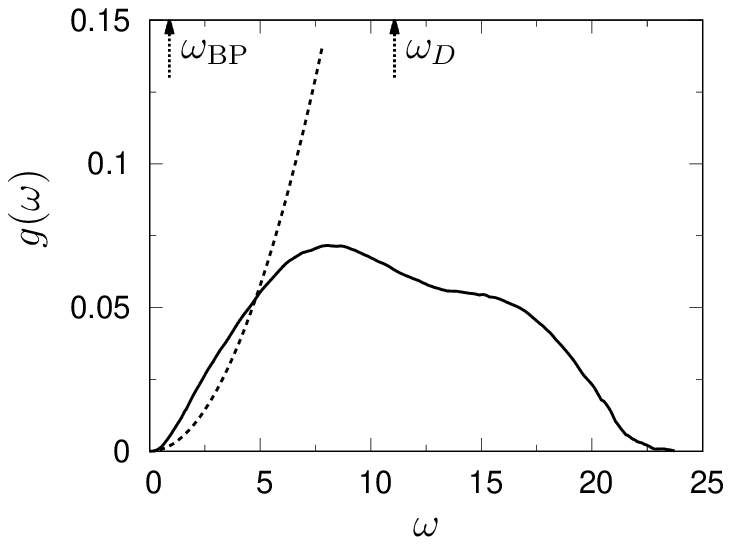}\label{fig4a}}
\hspace*{0mm}
\subfigure[Crystal.]
{\includegraphics[width=0.5\textwidth]{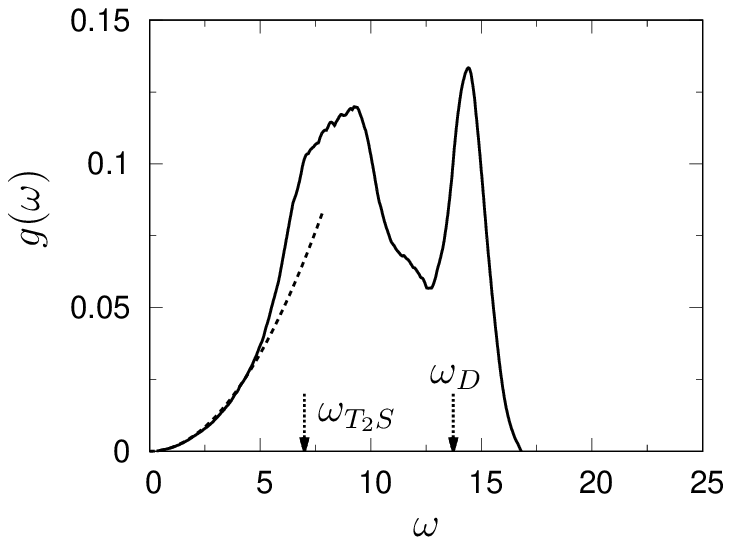}\label{fig4b}}
}
\caption{
The vDOS $g(\omega)$ of an SC system.
(a) Glass.
(b) FCC crystal.
The dashed line shows the Debye vDOS, $g_D(\omega) = A_D \omega^2$.
The Debye frequency $\omega_D$, the boson peak~(BP) frequency $\omega_\text{BP}$, and the position of the lowest-frequency, van Hove singularity $\omega_{T_2S}$ are indicated by arrows.
}
\label{fig4}
\end{figure}
%%%%%%%%%%%%%%%%%%%%%%%%%%%%%%%%%%%%%%%%%%%%%%%%%%%%%%%%%%%%%%%

%%%%%%%%%%%%%%%%%%%%%%%%%%%%%%%%%%%%%%%%%%%%%%%%%%%%%%%%%%%%%%%
\begin{figure}[t]
\centerline{
\subfigure[Glass.]
{\includegraphics[width=0.5\textwidth]{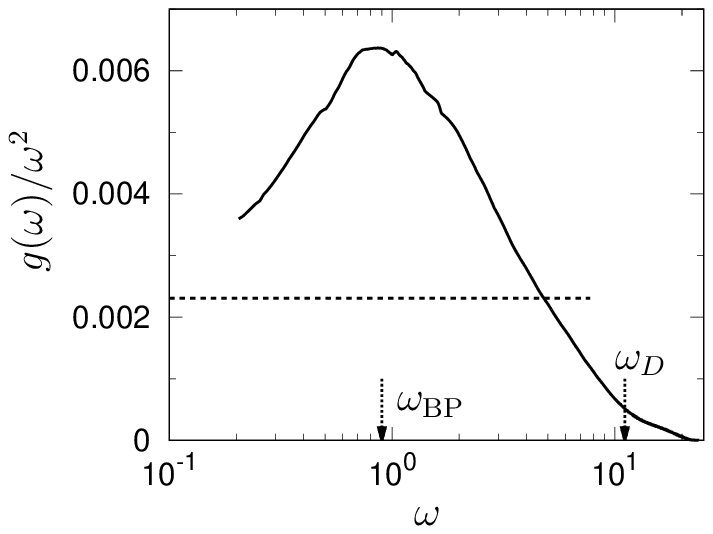}\label{fig5a}}
\hspace*{0mm}
\subfigure[Crystal.]
{\includegraphics[width=0.5\textwidth]{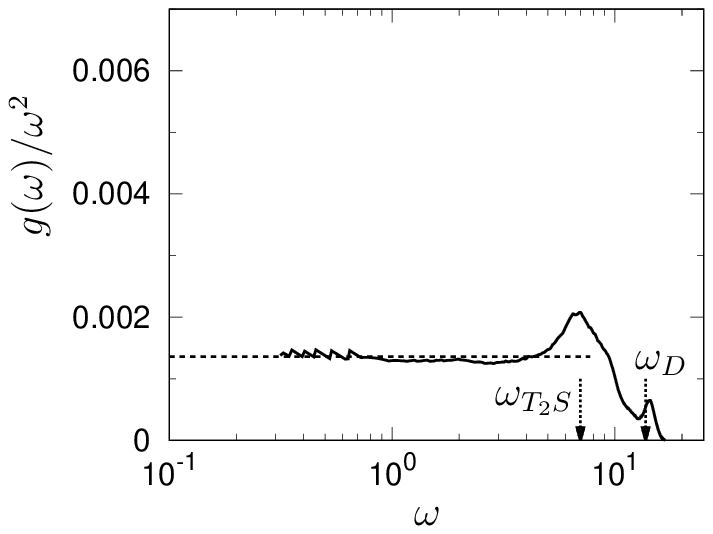}\label{fig5b}}
}
\caption{
The reduced vDOS $g(\omega)/\omega^2$ of an SC system.
(a) Glass.
(b) FCC crystal.
The dashed line shows the Debye level $A_D$.
See also the caption of Fig.~\ref{fig4}.
}
\label{fig5}
\end{figure}
%%%%%%%%%%%%%%%%%%%%%%%%%%%%%%%%%%%%%%%%%%%%%%%%%%%%%%%%%%%%%%%

%%%%%%%%%%%%%%%%%%%%%%%%%%%%%%%%%%%%%%%%%%%%%%%%%%%%%%%%%%%%%%%%%%%%%%%%%%%%%%%%%%%%%%%%%%%
\subsection{Vibrational density of states}
Now that we have obtained information on the eigenfrequencies $\omega^k$ and eigenvectors $\mathbf{e}^k$, we can next analyse them to understand the vibrational properties of glasses (or any solids).
One of the most important quantities is the vDOS $g(\omega)$.
$g(\omega)$ measures how many eigenmodes exist around the frequency $\omega$:
\begin{equation}
g(\omega) = \frac{1}{3N} \sum_{k=1}^{3N} \delta (\omega-\omega^k), \label{vdos}
\end{equation}
where $\delta(x)$ is the Dirac delta function.
Figure~\ref{fig4a} presents $g(\omega)$ for an SC glass.
For comparison, we also plot $g(\omega)$ for an SC crystal in Fig.~\ref{fig4b}.
Before looking at the results in these figures, let us define the Debye vDOS $g_D(\omega)$~\cite{Ashcroft_1976,Kittel_1996}, which is the vDOS calculated by assuming glasses or crystals to be elastic media and is a useful reference for capturing the features of the vDOS.

The eigenfrequencies of elastic waves are given by $\omega_\alpha(\hat{\mathbf{q}}) = c_\alpha (\hat{\mathbf{q}})q$ as in Eq.~(\ref{lineardispersion}).
We can then calculate the vDOS by using Eq.~(\ref{vdos}) as follows (where we replace the summation $\sum_{k=1}^{3N}$ with the integral $(L/2\pi)^3 \sum_{\alpha=T_1,T_2,L} \int d\mathbf{q}$):
\begin{equation}
g_D(\omega) = A_D \omega^2 = \frac{3}{\omega_D^3} \omega^2, \label{debyevdos}
\end{equation}
where $A_D \equiv {3}/{\omega_D^3}$ is the Debye level and $\omega_D$ is the Debye frequency.
$\omega_D \equiv \bar{c} k_D$, with $k_D \equiv \left(6\pi^2 {N}/{V} \right)^{1/3}$ being the Debye wavenumber and $\bar{c}$ being the average sound speed:
\begin{equation}
\bar{c} = \left(\frac{1}{3}\sum_{\alpha =T_1,T_2,L} \int \frac{d\Omega(\hat{\mathbf{q}})}{4\pi} \frac{1}{c_\alpha(\hat{\mathbf{q}})^3} \right)^{-1/3},
\end{equation}
where $d\Omega$ denotes the solid angle.
In the case of an isotropic system such as a glass, $\omega_D$ is given in the following simple form:
\begin{equation}
\omega_D = \left[ \frac{18 \pi^2 (N/V)}{2{c_T}^{-3} + {c_L}^{-3}} \right]^{1/3}.
\end{equation}
For an anisotropic system, we need to solve Eq.~(\ref{elasticeigeneq}) to obtain $c_\alpha (\hat{\mathbf{q}})$ as a function of $\hat{\mathbf{q}}$~\cite{Every_1980}.
An analytical solution $c_\alpha (\hat{\mathbf{q}})$ for cubic crystals is given in Ref.~\onlinecite{Jasiukiewicz_2003}.
Note that to calculate the Debye vDOS $g_D(\omega)$ (with Debye level $A_D$ and Debye frequency $\omega_D$), we need to measure the elastic moduli of the system, as described later in Section~\ref{sec.moduli}.

In Fig.~\ref{fig4}, the Debye vDOS $g_D(\omega)$ is plotted as a dashed line.
Since the phonons in a crystal converge to elastic waves at low $\omega$, the $g(\omega)$ function of the crystal converges to $g_D(\omega)$, as demonstrated in Fig.~\ref{fig4b}.
We can also clearly observe van Hove singularities~\cite{Ashcroft_1976,Kittel_1996}.
In Fig.~\ref{fig4b}, we use $\omega_{T_2S}$ to indicate the position of the lowest-frequency, van Hove singularity which corresponds to that of $T_2$ transverse phonons.

Compared to the crystal, the glass shows smoother variations and broader distributions of the eigenmodes.
Remarkably, the low-$\omega$ portion of $g(\omega)$ shows an enhancement in eigenmodes over the Debye vDOS.
To demonstrate this point more clearly, Figure~\ref{fig5} presents the reduced vDOS, which is $g(\omega)$ divided by $\omega^2$, i.e., $g(\omega)/\omega^2$.
Note that $g_D(\omega)/\omega^2$ is equal to the Debye level $A_D$, i.e., a constant value, as shown by the dashed line in Fig.~\ref{fig5}.
For the crystal, $g(\omega)/\omega^2$ coincides with $A_D$ at low $\omega$, whereas for the glass, $g(\omega)/\omega^2$ clearly shows an excess peak over $A_D$, which we call the boson peak~(BP)~\cite{buchenau_1984,Yamamuro_1996,Kabeya_2016}
\footnote{
The excess eigenmodes result in enhancement of the specific heat of a glass.
This excess specific heat is also called the boson peak~\cite{Zeller_1971,Phillips_1981,Elliott_1990}.
}.
The frequency at which $g(\omega)/\omega^2$ reaches a maximum is called the BP frequency and is denoted by $\omega_\text{BP}$ ($\omega_\text{BP} \approx 0.9$ in Fig.~\ref{fig5a}).
We remark that $g(\omega)/\omega^2$ also shows a peak for the crystal, as shown in Fig.~\ref{fig5b}; however, this peak corresponds to a van Hove singularity and appears at a frequency much higher than $\omega_\text{BP}$ ($\omega_{T_2S} \approx 7 \gg \omega_\text{BP} \approx 0.9$).
The van Hove singularity picks up vibrations at a microscopic length scale on the order of the lattice constant~\cite{Ashcroft_1976,Kittel_1996}, whereas the boson peak exhibits vibrations at a much longer, mesoscopic length scale~\cite{Leonforte_2005,Leonforte_2006}.

As $\omega$ decreases below the BP frequency $\omega_\text{BP}$, the $g(\omega)/\omega^2$ of the glass approaches the Debye level $A_0$, as shown in Fig.~\ref{fig5a}.
We might expect that $g(\omega)/\omega^2$ will converge to $A_0$ at some frequency, where the eigenmodes will converge to elastic waves.
However, this is not the case, as will be discussed in detail in Section~\ref{sec.recent}.

%%%%%%%%%%%%%%%%%%%%%%%%%%%%%%%%%%%%%%%%%%%%%%%%%%%%%%%%%%%%%%%%%%%%%%%%%%%%%%%%%%%%%%%%%%%
\subsection{Characterization of vibrational eigenmodes}
We employ order parameters to characterize the vibrations of particles in each eigenmode $k$.
Here, we introduce two order parameters.
One is the participation ratio, which measures the extent of localization~\cite{Schober_1991,mazzacurati_1996,Taraskin_1999}.
The other is the phonon order parameter, which measures the extent of phonon-like vibrations~\cite{Mizuno_2017,Shimada_2017}.

%%%%%%%%%%%%%%%%%%%%%%%%%%%%%%%%%%%%%%%%%%%%%%%%%%%%%%%%%%%%%%%
\begin{figure}[t]
\centerline{
\subfigure[Glass.]
{\includegraphics[width=0.5\textwidth]{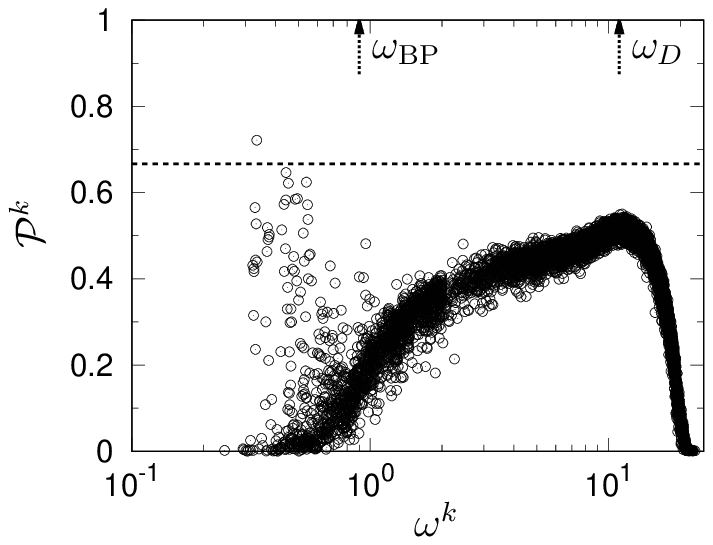}\label{fig6a}}
\hspace*{0mm}
\subfigure[Crystal.]
{\includegraphics[width=0.5\textwidth]{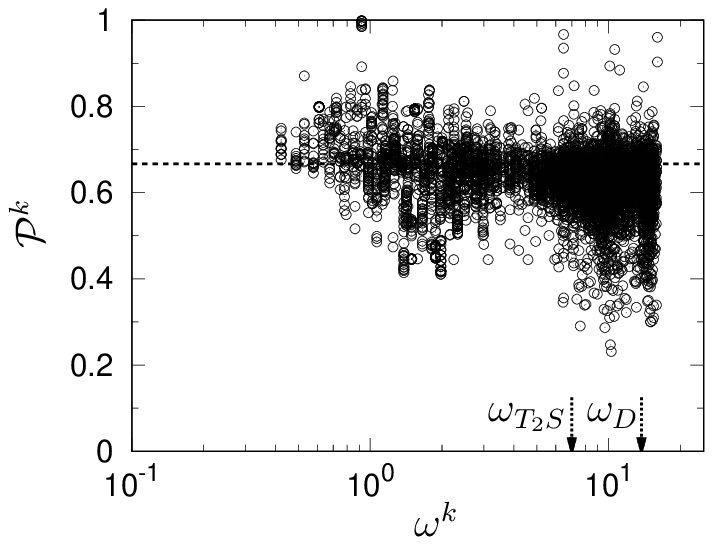}\label{fig6b}}
}
\caption{
The participation ratio $\mathcal{P}^k$ of an SC system.
(a) Glass.
(b) FCC crystal.
The dashed line corresponds to $\mathcal{P}^k = 2/3$, the value for elastic waves.
See also the caption of Fig.~\ref{fig4}.
}
\label{fig6}
\end{figure}
%%%%%%%%%%%%%%%%%%%%%%%%%%%%%%%%%%%%%%%%%%%%%%%%%%%%%%%%%%%%%%%

%%%%%%%%%%%%%%%%%%%%%%%%%%%%%%%%%%%%%%%%%%%%%%%%%%%%%%%%%%%%%%%%%%%%%%%%%%%%%%%%%%%%%%%%%%%
\subsubsection{Participation ratio}
As already seen in Fig.~\ref{fig3}, vibrational localization occurs in glasses.
The participation ratio quantitatively measures the extent of localization~\cite{Schober_1991,mazzacurati_1996,Taraskin_1999}.
Given an eigenvector, $\mathbf{e}^k \equiv \left[ \mathbf{e}_1^{kT}, \mathbf{e}_2^{kT}, \cdots, \mathbf{e}_N^{kT} \right]^T$, for eigenmode $k$, its participation ratio $\mathcal{P}^k$ is calculated as
\begin{equation}
\mathcal{P}^k \equiv \frac{1}{N} \left[ \sum_{i=1}^{N} \mathbf{e}^k_i \cdot \mathbf{e}^k_i \right]^2 \left[ \sum_{i=1}^{N} \left( \mathbf{e}^k_i \cdot \mathbf{e}^k_i \right)^2 \right]^{-1} = \frac{1}{N} \left[ \sum_{i=1}^{N} \left( \mathbf{e}^k_i \cdot \mathbf{e}^k_i \right)^2 \right]^{-1},
\label{participation}
\end{equation}
where the last equality comes from the orthonormality condition in Eq.~(\ref{orthnormal}).
$\mathcal{P}^k$ measures the fraction of particles that participate in the vibrations.
As extreme cases, $\mathcal{P}^k = \left(\sum_{i=1}^N m_i \right)^2/\left(N \sum_{i=1}^N m_i^2 \right)$ for an ideal mode in which all constituent particles vibrate equally
\footnote{
For the case of identical mass $m_i \equiv m$, $\mathcal{P}^k = \left(\sum_{i=1}^N m_i \right)^2/\left(N \sum_{i=1}^N m_i^2 \right) = 1$.
}, $\mathcal{P}^k = 1/N \ll 1$ for an ideal mode involving only one particle, and $\mathcal{P}^k = 2/3$ for elastic waves, $\mathbf{e}^{\mathbf{q},\alpha}_\text{el}$, as given in Eq.~(\ref{elasticeigen})
\footnote{
For elastic waves, we can calculate their participation ratio as $\mathcal{P}^k \equiv (1/V) \left[ \int_V d\mathbf{r}^3\ \text{Re}\left( \mathbf{e}^{\mathbf{q},\alpha}_\text{el}(\mathbf{r}) \right) \cdot \text{Re}\left( \mathbf{e}^{\mathbf{q},\alpha}_\text{el}(\mathbf{r}) \right) \right]^2 \left[ \int_V d\mathbf{r}^3\ \left\{ \text{Re}\left( \mathbf{e}^{\mathbf{q},\alpha}_\text{el}(\mathbf{r}) \right) \cdot \text{Re}\left( \mathbf{e}^{\mathbf{q},\alpha}_\text{el}(\mathbf{r}) \right) \right\}^2 \right]^{-1} =2/3$.
}.

Figure~\ref{fig6} plots $\mathcal{P}^k$ as a function of the frequency $\omega^k$ for (a) an SC glass and (b) an SC crystal.
We observe that the values of $\mathcal{P}^k$ in the crystal fluctuate around $\mathcal{P}^k=2/3$, indicating that the vibrations are similar to elastic waves.
Indeed, the eigenvectors of phonons, $\mathbf{e}^{\mathbf{q},\alpha}_\text{ph}$, have the same form as those of elastic waves, $\mathbf{e}^{\mathbf{q},\alpha}_\text{el}$~(compare Eqs.~(\ref{phononeigen}) and~(\ref{elasticeigen})).
However, phonons are not continuum waves but discrete waves, and their discrete nature causes fluctuations in $\mathcal{P}^k$ around $\mathcal{P}^k=2/3$.

On the other hand, the values of $\mathcal{P}^k$ in the glass are much different to those in the crystal.
Remarkably, the glass shows localization at the low-$\omega$ and high-$\omega$ edges, which is a characteristic feature of disordered matter.
For the low-frequency case, localization starts to occur near the BP frequency $\omega_\text{BP}$.
In addition to the participation ratio, we can also employ, for example, an effective mass~\cite{Schober_1991} to measure the extent of localization.

%%%%%%%%%%%%%%%%%%%%%%%%%%%%%%%%%%%%%%%%%%%%%%%%%%%%%%%%%%%%%%%%%%%%%%%%%%%%%%%%%%%%%%%%%%%
\subsubsection{Phonon order parameter}
As the frequency decreases, the eigenmodes in glasses tend to show phonon-like vibrations.
The phonon order parameter evaluates the extent to which an eigenmode $\mathbf{e}^k$ is similar to phonons $\mathbf{u}^{\mathbf{q},\alpha}_\text{ph}$~\cite{Mizuno_2017,Shimada_2017}:
\begin{equation}
\mathbf{u}^{\mathbf{q},\alpha}_\text{ph} \equiv \left[ \mathbf{s}_\alpha^T(\mathbf{q}) \frac{\exp(\text{i}\mathbf{q}\cdot\mathbf{R}_1)}{\sqrt{N}},\mathbf{s}_\alpha^T(\mathbf{q})\frac{\exp(\text{i}\mathbf{q}\cdot\mathbf{R}_2)}{\sqrt{N}},\cdots,\mathbf{s}_\alpha^T(\mathbf{q})\frac{\exp(\text{i}\mathbf{q}\cdot\mathbf{R}_N)}{\sqrt{N}}\right]^T. \label{phononvector}
\end{equation}
Here we define $\mathbf{u}^{\mathbf{q},\alpha}_\text{ph}$ to extend the concept of phonons to any solid-states including amorphous states.
The eigenvector $\mathbf{e}^k$ can always be expanded in terms of a series of $3N$ phonons $\mathbf{u}^{\mathbf{q},\alpha}_\text{ph}$: $\mathbf{e}^k = \sum_{\mathbf{q},\alpha} A^k_{\mathbf{q},\alpha} \mathbf{u}^{\mathbf{q},\alpha}_\text{ph}$.
In the case of a crystal where the inherent structure $\mathbf{R}$ is a lattice structure, $\mathbf{u}^{\mathbf{q},\alpha}_\text{ph}$ are exactly the same as the eigenvectors $\mathbf{e}^{\mathbf{q},\alpha}_\text{ph}$ in Eq.~(\ref{phononeigen}), which are orthonormalized as $\mathbf{u}^{\mathbf{q},\alpha}_\text{ph} \cdot \mathbf{u}^{\mathbf{q}',\alpha' \ast}_\text{ph} = \delta_{\mathbf{q}\mathbf{q}'}\delta_{\alpha \alpha'}$.
For the case of a glass where $\mathbf{R}$ is a disordered structure, the phonons $\mathbf{u}^{\mathbf{q},\alpha}_\text{ph}$ (which are not the eigenvectors in general) are not exactly identical to but can be approximately orthonormalized as $\mathbf{u}^{\mathbf{q},\alpha}_\text{ph} \cdot \mathbf{u}^{\mathbf{q}',\alpha' \ast}_\text{ph} \approx \delta_{\mathbf{q}\mathbf{q}'}\delta_{\alpha \alpha'}$.
Thus, we can calculate the projection of $\mathbf{e}^k$ onto one particular phonon, $\mathbf{u}^{\mathbf{q},\alpha}_\text{ph}$, as follows:
\begin{equation}
\begin{aligned}
O^k_{\mathbf{q},\alpha} = \left| A^k_{\mathbf{q},\alpha} \right|^2 \approx \left| \mathbf{u}^{\mathbf{q},\alpha}_\text{ph} \cdot \mathbf{e}^k \right|^2. \label{equofok2}
\end{aligned}
\end{equation}
Here, note that $\sum_{\mathbf{q},\alpha} O^k_{\mathbf{q},\alpha} \approx 1$ since $\mathbf{e}^k \cdot \mathbf{e}^k = 1$.

If eigenmode $k$ shows the phonon vibration, then $\mathbf{e}^k$ can be described as the sum of a finite number of phonons with large overlap: $\mathbf{e}^k = \sum_{\mathbf{q},\alpha;\ O^k_{\mathbf{q},\alpha} \ge N_m/3N} A^k_{\mathbf{q},\alpha} \mathbf{u}^{\mathbf{q},\alpha}_\text{ph}$.
Here, we define ``large overlap'' to mean that $O^k_{\mathbf{q},\alpha} \ge N_m/3N$, i.e., the phonons overlap by more than $N_m$ modes.
Considering this, we define the phonon order parameter $O^k$ as
\begin{equation}
\begin{aligned}
O^k & =  \sum_{\mathbf{q},\alpha;\ O^k_{\mathbf{q},\alpha} \ge N_m/3N} O^k_{\mathbf{q},\alpha}. \label{equofok}
\end{aligned}
\end{equation}
$O^k = 1$ for a phonon, whereas $O^k = 0$ for a mode that is considerably different to a phonon.
To calculate the phonon order parameter in Eq.~(\ref{equofok}), we need to set an appropriate value of $N_m$; the obtained results and conclusions should not depend on the choice of $N_m$.
$N_m = 100$ might be appropriate, but this depends on the considered system.
Simulation data for the phonon order parameter will be presented in Section~\ref{sec.recent}.

%%%%%%%%%%%%%%%%%%%%%%%%%%%%%%%%%%%%%%%%%%%%%%%%%%%%%%%%%%%%%%%%%%%%%%%%%%%%%%%%%%%%%%%%%%%
\section{Phonon transport}\label{sec.phonon}
Phonon transport is a fundamental property of solid-state materials~\cite{Ashcroft_1976,Kittel_1996}.
Phonons play a role in carrying heat energy.
In crystals, phonons are the vibrational eigenmodes, and they propagate without any attenuation in the zero-temperature, harmonic limit.
The scattering of phonons is induced through anharmonic effects at finite temperatures.
This situation is drastically changed in glasses, where the eigenmodes are non-phonon modes, as seen in Section~\ref{sec.modeanalysis}.
In this case, a phonon will be decomposed into several different eigenmodes in some frequency range~\cite{taraskin_2000}, and as a result, it will be strongly scattered.
The disordered structures therefore scatter phonons propagating in glasses.

A phonon is specified by the wavevector $\mathbf{q}$ (wavenumber $q \equiv |\mathbf{q}|$ and $\hat{\mathbf{q}} \equiv \mathbf{q}/q$) and the polarization $\alpha$, where $\alpha = T_1,\ T_2$~(transverse) or $L$~(longitudinal).
Its transport is characterized by two quantities: the propagation frequency $\Omega(\mathbf{q})$ and the attenuation rate $\Gamma_\alpha(\mathbf{q})$.~(For the frequency $\Omega(\mathbf{q})$, we do not explicitly write the polarization index $\alpha$.)
The propagation speed is given by $c_\alpha(\mathbf{q}) = \Omega(\mathbf{q})/q$
\footnote{
$c_\alpha(\mathbf{q}) = \Omega(\mathbf{q})/q$ is the phase speed, while the group speed is given by $v_\alpha(\mathbf{q}) = \left. d\Omega/dq\right|_{\hat{\mathbf{q}}}$.
In the low-wavenumber region, these two speeds coincide with each other.
}.
Below, we will describe how to measure $\Omega(\mathbf{q})$ or $c_\alpha(\mathbf{q})$, and $\Gamma_\alpha(\mathbf{q})$ through simulations.

%%%%%%%%%%%%%%%%%%%%%%%%%%%%%%%%%%%%%%%%%%%%%%%%%%%%%%%%%%%%%%%%%%%%%%%%%%%%%%%%%%%%%%%%%%%
\subsection{Dynamic structure factor}
Let us introduce the dynamic structure factor to study phonon transport.
The dynamic structure factor can be measured through scattering experiments involving light, X-rays, or neutrons.
In the computational simulation approach, we first perform an MD simulation at a finite temperature $T>0$ and obtain the corresponding trajectory data, $\mathbf{r}(t) \equiv \left[ \mathbf{r}_1^T(t), \mathbf{r}_2^T(t),\cdots,\mathbf{r}_N^T(t) \right]^T$.
From this trajectory, the dynamic structure factor, $S_\alpha (\mathbf{q},\omega)$ (where $\alpha =T$ or $L$), is calculated as the Fourier transform of the current-current correlation function~\cite{shintani_2008,Monaco2_2009,Marruzzo_2013,Mizuno_2014}:
\begin{equation}
S_\alpha (\mathbf{q},\omega) = \left( \frac{q}{\omega} \right)^2 \frac{1}{2 \pi} \int \frac{1}{N} \left< \mathbf{j}_\alpha(\mathbf{q},t) \cdot \mathbf{j}_\alpha^\ast(\mathbf{q},0) \right>_0 \exp\left( {\text{i} \omega t} \right) dt, \label{eq.sqw}
\end{equation}
where $\left< \right>_0$ denotes the ensemble average over configurations at $t=0$.
$\mathbf{j}_\alpha (\mathbf{q},t)$ is the transverse~($\alpha = T$) or longitudinal~($\alpha = L$) current:
\begin{equation}
\begin{aligned}
\mathbf{j}_T (\mathbf{q},t) &= \sum_{i=1}^N \left[ \mathbf{v}_i(t) - \left( \mathbf{v}_i(t) \cdot \hat{\mathbf{q}} \right) \hat{\mathbf{q}} \right] \exp\left( \text{i} \mathbf{q}\cdot \mathbf{r}_i(t) \right),\\
\mathbf{j}_L (\mathbf{q},t) &= \sum_{i=1}^N \left( \mathbf{v}_i(t) \cdot \hat{\mathbf{q}} \right) \hat{\mathbf{q}} \exp\left( \text{i} \mathbf{q}\cdot \mathbf{r}_i(t) \right),
\end{aligned}
\end{equation}
where $\mathbf{v}_i = d\mathbf{r}_i/dt$ is the velocity of particle $i$.
For the longitudinal case, the conservation law relates the current $\mathbf{j}_L (\mathbf{q},t)$ and the number density $\hat{\rho}(\mathbf{q},t) \equiv \sum_{i=1}^N \exp \left( \text{i} \mathbf{q}\cdot \mathbf{r}_i(t) \right)$ in the form
\begin{equation}
\frac{\partial \hat{\rho}(\mathbf{q},t)}{\partial t} = \text{i} \mathbf{q} \cdot \mathbf{j}_L (\mathbf{q},t),
\end{equation}
and $S_L (\mathbf{q},\omega)$ corresponds to the density-density correlation function
\footnote{
The factor $(q/\omega)^2$ in $S_\alpha (\mathbf{q},\omega)$ as expressed in Eq.~(\ref{eq.sqw}) is necessary for $S_L (\mathbf{q},\omega)$ to correspond to the density-density correlation function.
}.

We can understand the dynamic structure factor $S_\alpha (\mathbf{q},\omega)$ as follows.
The phonon displacement vector, $\mathbf{u}^{\mathbf{q},\alpha}_\text{ph}$, is given in Eq.~(\ref{phononvector}).
When we take the polarization vectors to be $\mathbf{s}_T(\mathbf{q}) = \hat{\mathbf{q}}^\perp$~(where $\perp$ denotes a perpendicular vector) and $\mathbf{s}_L(\mathbf{q}) = \hat{\mathbf{q}}$, the current-current correlation function $\left< \mathbf{j}_\alpha(\mathbf{q},t) \cdot \mathbf{j}^\ast_\alpha(\mathbf{q},0) \right>_0$ is formulated as
\begin{equation}
\frac{1}{N} \left< \mathbf{j}_\alpha(\mathbf{q},t) \cdot \mathbf{j}^\ast_\alpha(\mathbf{q},0) \right>_0 = \left<  \left( \mathbf{v}(t) \cdot \mathbf{u}^{\mathbf{q},\alpha}_\text{ph} \right) \left( \mathbf{v}(0) \cdot \mathbf{u}^{\mathbf{q},\alpha}_\text{ph} \right)^\ast \right>_0 \times \left[ 1 + \mathcal{O} \left( \sqrt{T} \right) \right], \label{projection}
\end{equation}
where $\mathbf{v}(t) \equiv \left[ \mathbf{v}_1^T(t), \mathbf{v}_2^T(t),\cdots,\mathbf{v}_N^T(t) \right]^T$ is the velocity vector of the system.
Therefore, $S_\alpha (\mathbf{q},\omega)$ at the low-$T$ harmonic limit converges to the correlation function of the velocity projected onto $\mathbf{u}^{\mathbf{q},\alpha}_\text{ph}$.
Thermal energy activates many different phonons with different wavevectors and different polarizations.
From among these activated phonons, $S_\alpha (\mathbf{q},\omega)$ picks up a specific phonon $\mathbf{u}^{\mathbf{q},\alpha}_\text{ph}$ of wavevector $\mathbf{q}$ and polarization $\alpha$.
We remark that the actual polarization vectors of the phonons can deviate from $\mathbf{s}_T(\mathbf{q}) = \hat{\mathbf{q}}^\perp$ and $\mathbf{s}_L(\mathbf{q}) = \hat{\mathbf{q}}$.
However, they still considerably overlap with $\mathbf{s}_T(\mathbf{q}) = \hat{\mathbf{q}}^\perp$ and $\mathbf{s}_L(\mathbf{q}) = \hat{\mathbf{q}}$, and we can properly analyse phonon transport by means of $S_\alpha (\mathbf{q},\omega)$.

%%%%%%%%%%%%%%%%%%%%%%%%%%%%%%%%%%%%%%%%%%%%%%%%%%%%%%%%%%%%%%%
\begin{figure}[t]
\centerline{
\subfigure[Glass, $\alpha =T$.]
{\includegraphics[width=0.5\textwidth]{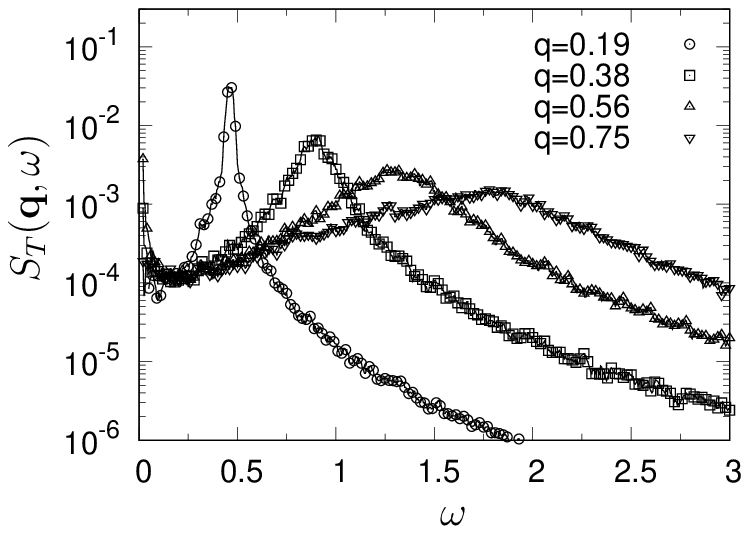}\label{fig7a}}
\hspace*{0mm}
\subfigure[Crystal, $\alpha =T$.]
{\includegraphics[width=0.5\textwidth]{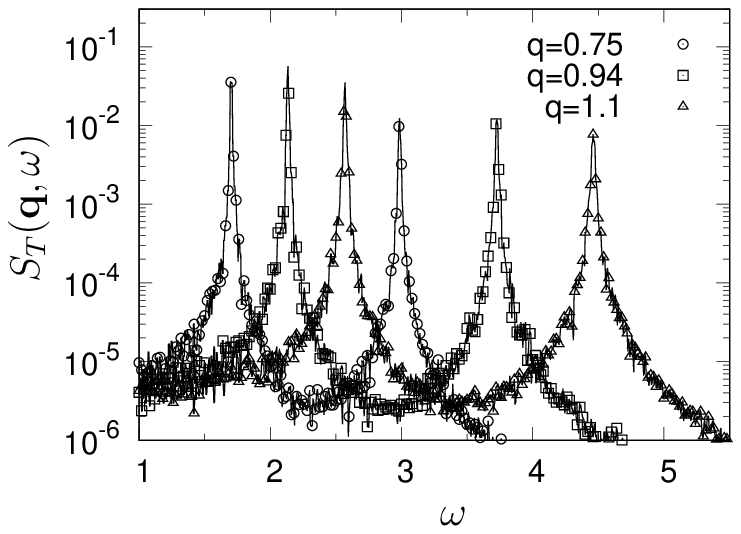}\label{fig7b}}
}

\vspace*{2mm}
\centerline{
\subfigure[Glass, $\alpha =L$.]
{\includegraphics[width=0.5\textwidth]{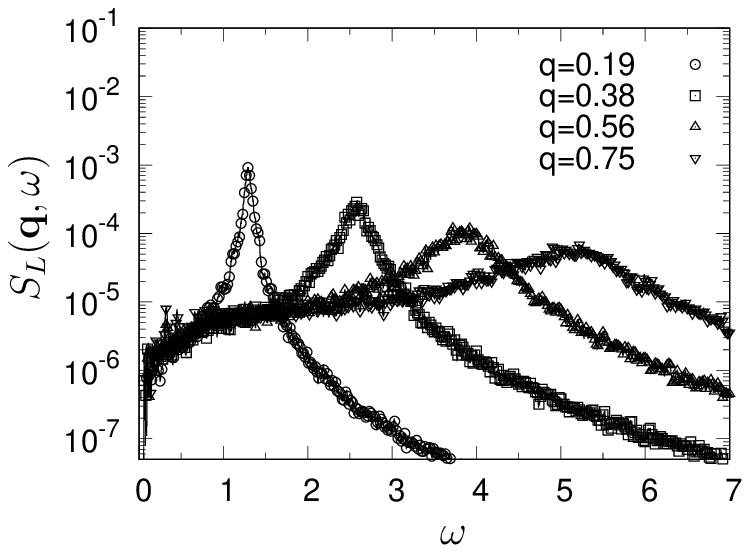}\label{fig7c}}
\hspace*{0mm}
\subfigure[Crystal, $\alpha =L$.]
{\includegraphics[width=0.5\textwidth]{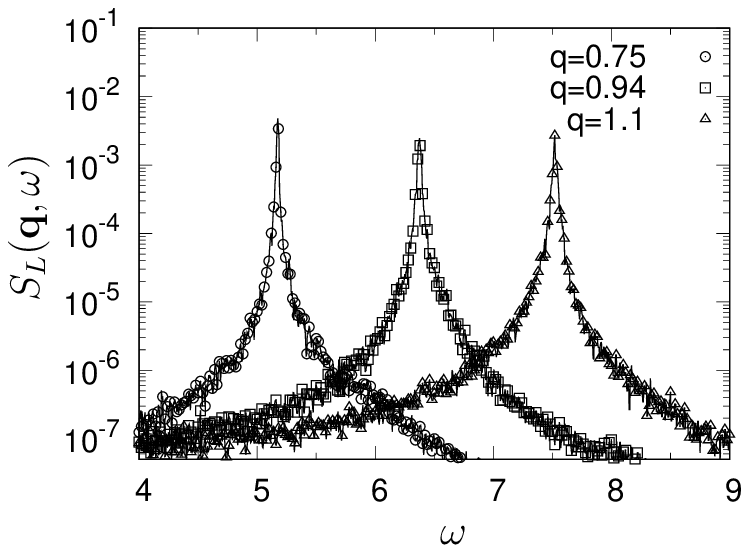}\label{fig7d}}
}
\caption{
The dynamic structure factor $S_{\alpha}(\mathbf{q},\omega)$ in the $[110]$ direction for an SC system at a low temperature of $T=10^{-2}$, well below $T_g$ and $T_m$.
(a) Glass, $\alpha =T$.
(b) FCC crystal, $\alpha =T$.
(c) Glass, $\alpha =L$.
(d) FCC crystal, $\alpha =L$.
}
\label{fig7}
\end{figure}
%%%%%%%%%%%%%%%%%%%%%%%%%%%%%%%%%%%%%%%%%%%%%%%%%%%%%%%%%%%%%%%

Figure~\ref{fig7} shows $S_{\alpha} (\mathbf{q},\omega)$ in the $[110]$ direction~(Miller indices) for an SC system.
The temperature is very low, $T=10^{-2}$, well below the glass transition temperature $T_g$ and the melting temperature $T_m$.
We plot $S_{\alpha} (\mathbf{q},\omega)$ for a glass and for an FCC crystal.
We can clearly observe the Brillouin peaks for both the glass and the crystal.
For the case of the FCC crystal, $S_{T}(\mathbf{q},\omega)$ for transverse phonons shows two peaks at a fixed $q$, as shown in Fig.~\ref{fig7b}, which correspond to a $T_2$ phonon at the lower $\omega$ and a $T_1$ phonon at the higher $\omega$ (see Fig.~\ref{fig2} for the polarization vectors $\mathbf{s}_{T_2}$ and $\mathbf{s}_{T_1}$).
On the other hand, the longitudinal $S_{L} (\mathbf{q},\omega)$ shows a single Brillouin peak, as shown in Fig.~\ref{fig7d}.

Unlike the FCC crystal, since the glass is an isotropic system, it shows a single Brillouin peak in both the transverse and longitudinal dynamic structure factors, as shown in Figs.~\ref{fig7a} and~\ref{fig7c}.
Notably, the Brillouin peaks are rather broad compared to those in the crystal.
This finding indicates that phonon attenuation is considerably enhanced in glasses, as we will see below.

Based on the data for $S_{\alpha} (\mathbf{q},\omega)$, the propagation frequency $\Omega(\mathbf{q})$ and the attenuation rate $\Gamma_\alpha(\mathbf{q})$ can be extracted by fitting the spectral region around the Brillouin peak to the damped harmonic oscillator model~\cite{shintani_2008,Monaco2_2009,Marruzzo_2013,Mizuno_2014}:
\begin{equation}
S_\alpha(\mathbf{q},\omega) \propto \frac{\Gamma_\alpha(\mathbf{q}) \Omega^2(\mathbf{q})}{\left( \omega^2-\Omega^2(\mathbf{q}) \right)^2 + \omega^2\Gamma_\alpha^2(\mathbf{q})}. \label{eq:dho}
\end{equation}
This fitting procedure in the frequency domain is equivalent to fitting the correlation function $\left< \mathbf{j}_\alpha(\mathbf{q},t) \cdot \mathbf{j}^\ast_\alpha(\mathbf{q},0) \right>_0$ to the following function in the time domain:
\begin{equation}
\frac{1}{N} \left< \mathbf{j}_\alpha(\mathbf{q},t) \cdot \mathbf{j}^\ast_\alpha(\mathbf{q},0) \right>_0 \propto \cos\left( \Omega(\mathbf{q}) t\right) \exp\left(-\frac{\Gamma_\alpha(\mathbf{q})}{2} t \right) \label{eq:dho2}.
\end{equation}
In the following, we will show data for $\Omega(\mathbf{q})$, $c_\alpha(\mathbf{q}) \equiv \Omega(\mathbf{q})/q$, and $\Gamma_\alpha(\mathbf{q})$ that are extracted from the $S_{\alpha} (\mathbf{q},\omega)$ presented in Fig.~\ref{fig7}.

%%%%%%%%%%%%%%%%%%%%%%%%%%%%%%%%%%%%%%%%%%%%%%%%%%%%%%%%%%%%%%%
\begin{figure}[t]
\centerline{
\subfigure[Glass, $\alpha =T$.]
{\includegraphics[width=0.5\textwidth]{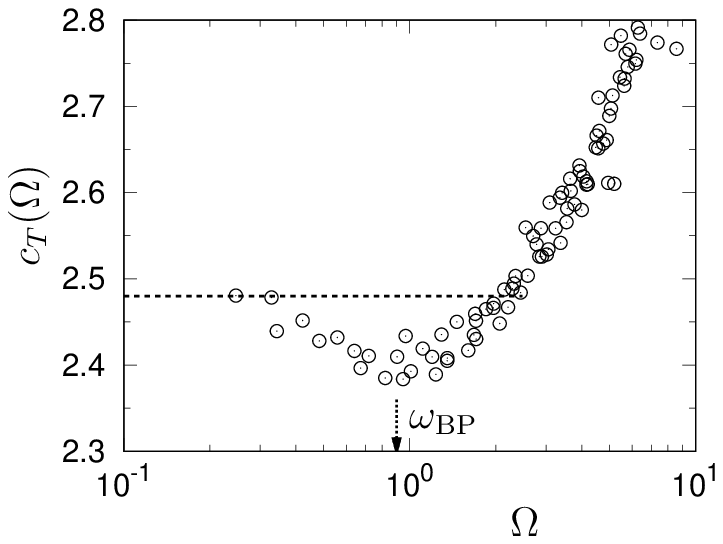}\label{fig8a}}
\hspace*{0mm}
\subfigure[Crystal, $\alpha =T_2$ in \text{$[110]$}.]
{\includegraphics[width=0.5\textwidth]{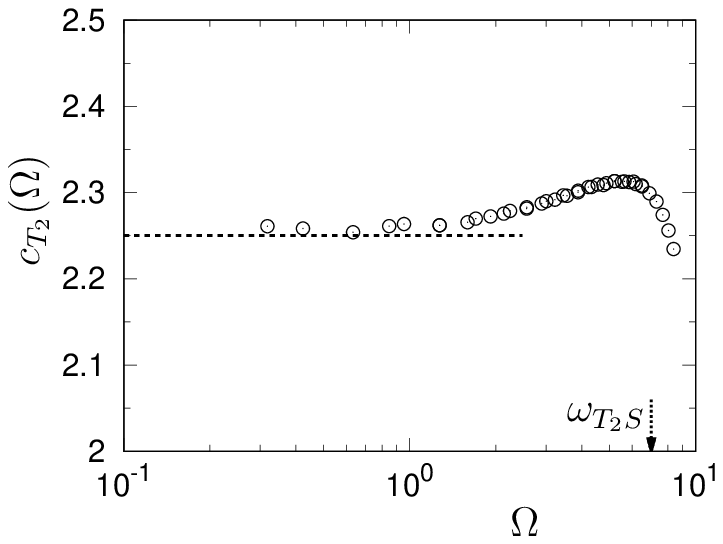}\label{fig8b}}
}

\vspace*{2mm}
\centerline{
\subfigure[Glass, $\alpha =L$.]
{\includegraphics[width=0.5\textwidth]{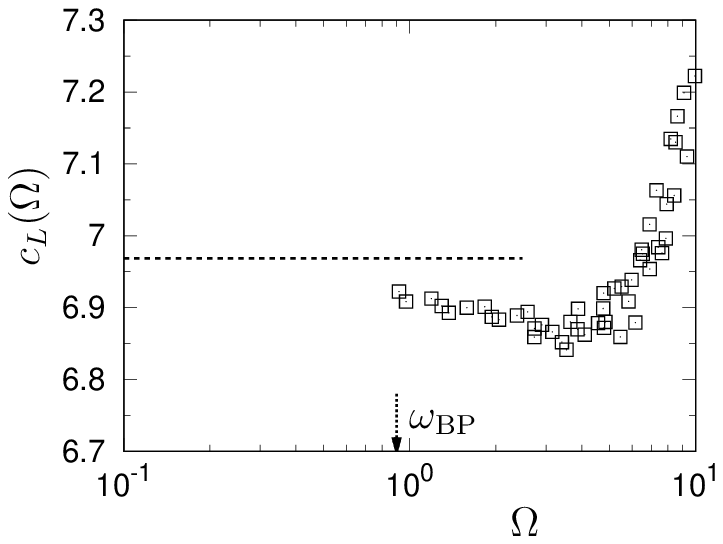}\label{fig8c}}
\hspace*{0mm}
\subfigure[Crystal, $\alpha =L$ in \text{$[110]$}.]
{\includegraphics[width=0.5\textwidth]{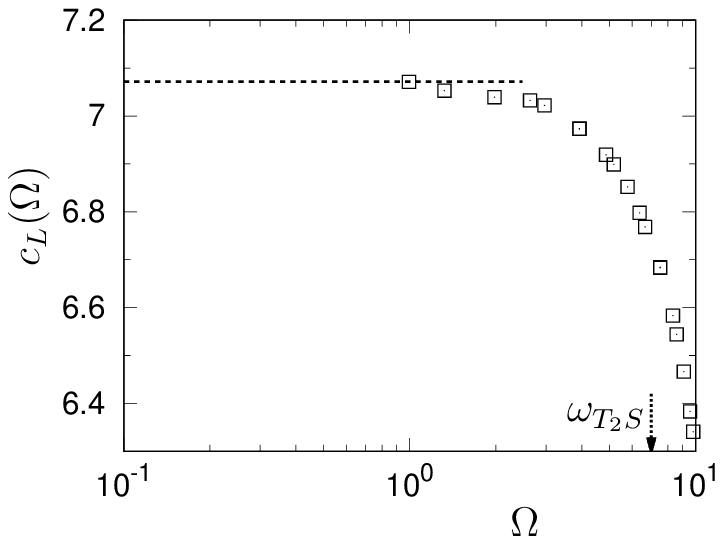}\label{fig8d}}
}
\caption{
The propagation speed $c_{\alpha}(\Omega)$ for an SC system.
(a) Glass, $\alpha =T$.
(b) FCC crystal, $\alpha =T_2$ in $[110]$.
(c) Glass, $\alpha =L$.
(d) FCC crystal, $\alpha =L$ in $[110]$.
For the glass~[(a),(c)] which is an isotropic system, we plot the data for the different directions ($[100]$, $[110]$, $[111]$) and for $\alpha = T_1,\ T_2$ all together, which coincide with each other.
The dashed line indicates the macroscopic sound speed calculated from the elastic moduli, as $c_{T0}=\sqrt{G/\rho}$ and $c_{L0}=\sqrt{(K + 4G/3)/\rho}$ as in Eq.~(\ref{fccsoundspeediso}).
For the crystal~[(b),(d)], the macroscopic sound speed for $[110]$ is plotted by the dashed line, as $c_{T_20}=\sqrt{G_p/\rho}$ or $c_{L0}=\sqrt{(K + G_p/3 + G_s)/\rho}$ as in Eq.~(\ref{fccsoundspeed2}).
The BP frequency $\omega_\text{BP}$ and the position of the lowest-frequency, van Hove singularity $\omega_{T_2S}$ are indicated by arrows.
}
\label{fig8}
\end{figure}
%%%%%%%%%%%%%%%%%%%%%%%%%%%%%%%%%%%%%%%%%%%%%%%%%%%%%%%%%%%%%%%

%%%%%%%%%%%%%%%%%%%%%%%%%%%%%%%%%%%%%%%%%%%%%%%%%%%%%%%%%%%%%%%
\begin{figure}[t]
\centerline{
\subfigure[Glass, $\alpha =T,\ L$.]
{\includegraphics[width=0.5\textwidth]{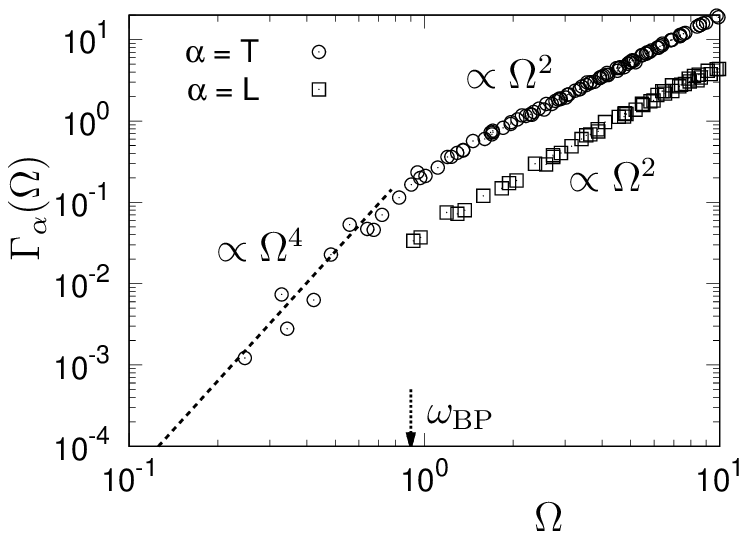}\label{fig9a}}
\hspace*{0mm}
\subfigure[Crystal, $\alpha =T_2,\ L$ in \text{$[110]$}.]
{\includegraphics[width=0.5\textwidth]{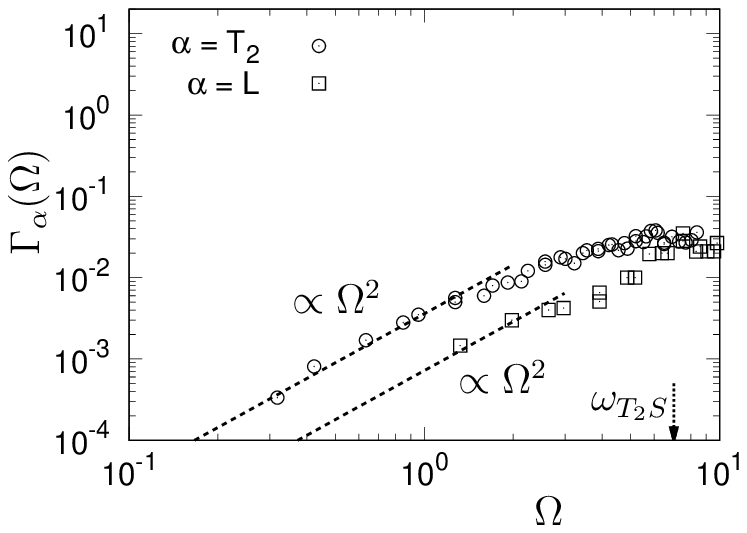}\label{fig9b}}
}
\caption{
The attenuation rate $\Gamma_{\alpha}(\Omega)$ for an SC system.
(a) Glass, $\alpha =T,\ L$.
(b) FCC crystal, $\alpha =T_2,\ L$ in $[110]$.
The dashed line indicates the power-law scaling with the frequency $\Omega$.
See also the caption of Fig.~\ref{fig8}.
}
\label{fig9}
\end{figure}
%%%%%%%%%%%%%%%%%%%%%%%%%%%%%%%%%%%%%%%%%%%%%%%%%%%%%%%%%%%%%%%

%%%%%%%%%%%%%%%%%%%%%%%%%%%%%%%%%%%%%%%%%%%%%%%%%%%%%%%%%%%%%%%%%%%%%%%%%%%%%%%%%%%%%%%%%%%
\subsection{Propagation frequency and attenuation rate}
Figures~\ref{fig8} and~\ref{fig9} plot the sound speed $c_\alpha(\Omega)$ and the attenuation rate $\Gamma_\alpha(\Omega)$, respectively, as functions of the frequency $\Omega$.
Here, we treat $c_\alpha$ and $\Gamma_\alpha$ as functions of $\Omega$ by transforming $\mathbf{q}$ into $\Omega$ via the relation $\Omega = \Omega(\mathbf{q})$.
For the FCC crystal, we focus on the lower-frequency transverse ($\alpha = T_2$) and longitudinal ($\alpha = L$) phonons in the $[110]$ direction.
On the other hand, for the glass, which is an isotropic system, there are no distinctions among the different directions or between $\alpha = T_1$ and $T_2$, so we show the results for all three different directions ($[100]$, $[110]$, $[111]$) and for $\alpha = T_1,\ T_2$ all together; these results indeed coincide with each other, as seen in Figs.~\ref{fig8} and~\ref{fig9}.

Let us first consider the data for the crystal.
Since the phonons in crystals converge to elastic waves at low $\Omega$, $c_\alpha(\Omega)$ converges to the macroscopic value $c_{\alpha 0}$, i.e., $c_{T_20}=\sqrt{G_p/\rho}$ or $c_{L0}=\sqrt{(K + G_p/3 + G_s)/\rho}$ (see Eq.~(\ref{fccsoundspeed2})), as is indeed shown in Figs.~\ref{fig8b} and~\ref{fig8d}.
In this frequency region, the dispersion curve is a straight line, $\Omega(\mathbf{q}) = c_{\alpha 0} q$, as in Eq.~(\ref{lineardispersion}).
It starts to deviate from linearity as $\Omega$ (or $q$) increases towards the edge of the first Brillouin zone, where the van Hove singularity occurs~\cite{Ashcroft_1976,Kittel_1996}.
In addition, both the transverse and longitudinal $\Gamma_\alpha(\Omega)$ show an $\Omega^2$ dependence, $\Gamma_\alpha \propto \Omega^2$, as seen in Fig.~\ref{fig9b}.
This damping originates from purely anharmonic couplings between phonons through the Umklapp process at finite $T>0$.
The $\omega^2$ dependence can be explained by thermo-elasticity theory or the Boltzmann equation analysis~\cite{Akhieser_1939,Maris_1971}.
At $T=0$, there are no anharmonic effects, and the attenuation becomes exactly zero.

The sound speed $c_\alpha(\Omega)$ of the glass also converges to its macroscopic value, $c_{T0}=\sqrt{G/\rho}$ or $c_{L0}=\sqrt{(K + 4G/3)/\rho}$ (see Eq.~(\ref{fccsoundspeediso})), at low $\Omega$, as shown in Figs.~\ref{fig8a} and~\ref{fig8c}
\footnote{
The longitudinal sound speed does not converge to $c_{L0}$ in the frequency range presented in Fig.~\ref{fig8c}.
However, recent work has shown that it converges as $\Omega$ decreases~\cite{Mizuno_2018}.
}.
As $\Omega$ increases, $c_\alpha(\Omega)$ decreases and reaches a minimum.
This behaviour has been observed in experiments~\cite{Monaco_2009,Baldi_2010} and simulations~\cite{Monaco2_2009,Marruzzo_2013,Mizuno_2014}, and it is called sound softening.
The softening of transverse phonons occurs near the BP frequency $\omega_\text{BP}$.
Above the softening region, $c_\alpha(\Omega)$ increases with increasing $\Omega$; this behaviour is called sound hardening~\cite{Crespo_2016}.

Notably, the attenuation rate $\Gamma_\alpha(\Omega)$ of the glass behaves much differently from that of the crystal, as shown in Fig.~\ref{fig9a}.
First, $\Gamma_\alpha(\Omega)$ takes much larger values than in the crystal.
Second, $\Gamma_\alpha(\Omega)$ shows Rayleigh scattering, $\Gamma_\alpha \propto \Omega^4$, at low $\Omega$ below $\omega_\text{BP}$, and it depends on $\Omega^2$, i.e., $\Gamma_\alpha \propto \Omega^2$, at high $\Omega$ above $\omega_\text{BP}$
\footnote{
The longitudinal phonons do not show Rayleigh scattering behaviour in the frequency range presented in Fig.~\ref{fig9b}.
However, recent work has shown that a Rayleigh scattering regime emerges as $\Omega$ decreases~\cite{Mizuno_2018}.
}.
These anomalous attenuation behaviours have also been observed in experiments~\cite{Masciovecchio_2006,Monaco_2009,Baldi_2010} and simulations~\cite{Monaco2_2009,Marruzzo_2013,Mizuno_2014}.
There are two sources of scattering in glasses: one is anharmonic effects, as in crystals, and the other is the disordered structure.
The latter effects are dominant over the former and induce much stronger sound damping compared to that in crystals; in particular, Rayleigh scattering at low $\Omega$ and an $\Omega^2$ dependence at high $\Omega$ are induced.
Although we cannot observe any evidence of them in Fig.~\ref{fig9a}, the former, anharmonic effects should become visible as some dependence on $\Omega$, possibly a dependence on $\Omega^2$ as in the crystal~\cite{Akhieser_1939,Maris_1971}, in the lower $\Omega$ regime.
Indeed, both experimental works~\cite{Baldi_2014,Ferrante_2013} and theoretical works~\cite{Schirmacher_2010,Tomaras_2010,Marruzzo2_2013} have studied these anharmonic effects, which add to the effects of the disordered structure.
Notably, close to the glass transition, the fractal frequency dependence of the damping, $\Gamma_\alpha \propto \Omega^{3/2}$, has been reported by experiment~\cite{Ferrante_2013} and predicted by theory~\cite{Marruzzo2_2013}.

At low $\Omega$ below the boson peak, we observe a macroscopic sound speed and Rayleigh scattering in the attenuation behaviour.
These phonon transport characteristics can be understood by considering the glass as an elastic medium with point defects.
This analysis indicates that the disordered structure is not uniformly coarse-grained even at macroscopic length scales, but rather, it still plays a defect-type role.
On the other hand, at shorter length scales, phonons are more strongly influenced by the disordered structure, which induces sound softening and an $\Omega^2$ dependence of damping.
$\Gamma_\alpha \propto \Omega^2$ exhibits dynamics characteristic of viscous damping~\cite{shintani_2008}.
In this regime, the scattering is so strong that a phonon does not propagate as a plane wave but rather immediately attenuates to become diffusive.
Refs.~\onlinecite{Allen_1993,Feldman_1993} refer to this vibrational behaviour as diffusion (nonpropagating, delocalized vibration).
We therefore expect the BP frequency $\omega_\text{BP}$ to represent the upper bound on the frequency at which phonons can propagate as plane waves, as will be discussed next.

%%%%%%%%%%%%%%%%%%%%%%%%%%%%%%%%%%%%%%%%%%%%%%%%%%%%%%%%%%%%%%%
\begin{figure}[t]
\centerline{
\subfigure[Glass, $\alpha =T,\ L$.]
{\includegraphics[width=0.5\textwidth]{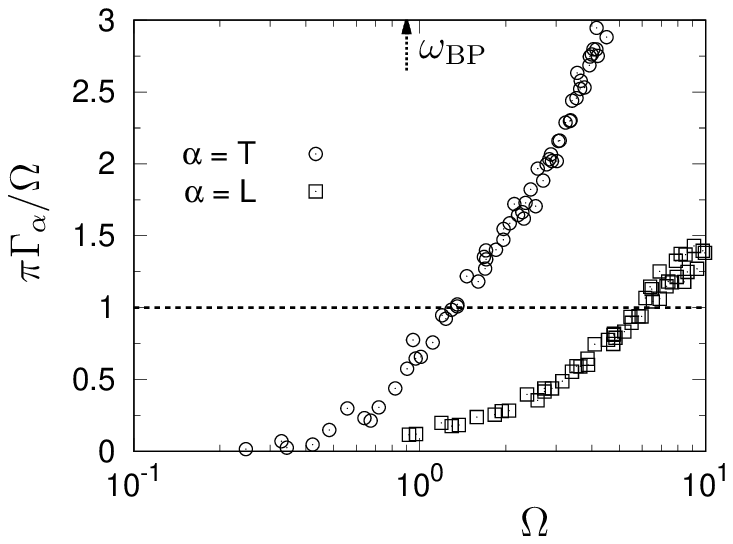}\label{fig10a}}
\hspace*{0mm}
\subfigure[Crystal, $\alpha =T_2,\ L$ in \text{$[110]$}.]
{\includegraphics[width=0.5\textwidth]{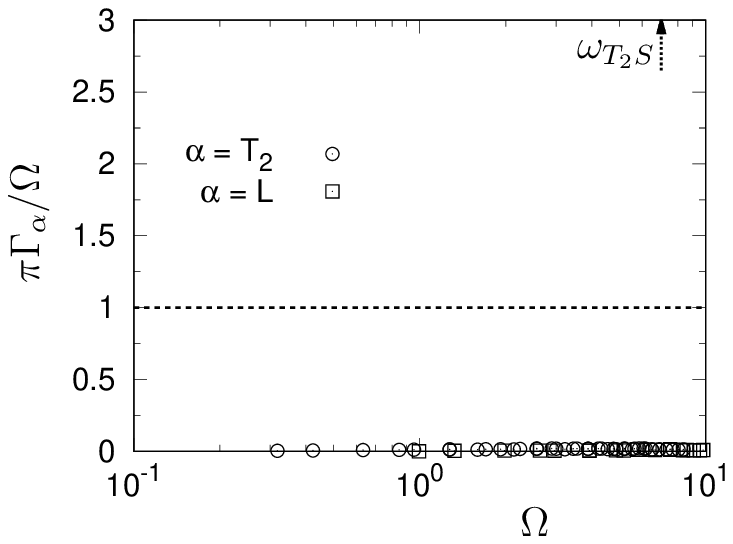}\label{fig10b}}
}
\caption{
Ioffe-Regel~(IR) frequency for an SC system.
(a) Glass, $\alpha =T,\ L$.
(b) FCC crystal, $\alpha =T_2,\ L$ in $[110]$.
We plot $\pi \Gamma_\alpha/\Omega$ versus $\Omega$.
The data for $\Omega$ and $\Gamma_\alpha$ are taken from Figs.~\ref{fig8} and~\ref{fig9}.
The point of intersection with $\pi \Gamma_\alpha / \Omega = 1$ as indicated by the dashed line gives the value of $\Omega_{\alpha \text{IR}}$.
See also the caption of Fig.~\ref{fig8}.
}
\label{fig10}
\end{figure}
%%%%%%%%%%%%%%%%%%%%%%%%%%%%%%%%%%%%%%%%%%%%%%%%%%%%%%%%%%%%%%%

%%%%%%%%%%%%%%%%%%%%%%%%%%%%%%%%%%%%%%%%%%%%%%%%%%%%%%%%%%%%%%%%%%%%%%%%%%%%%%%%%%%%%%%%%%%
\subsection{Ioffe-Regel~(IR) frequency}\label{sec.irfrequency}
We introduce the Ioffe-Regel~(IR) frequency, $\Omega_{\alpha \text{IR}}$, which corresponds to an upper bound on the frequency at which a phonon can propagate as a plane wave~\cite{shintani_2008,Monaco2_2009,Mizuno_2014}.
For $\alpha$ phonons, $\Omega_{\alpha \text{IR}}$ is defined as $\pi \Gamma_\alpha(\Omega=\Omega_{\alpha \text{IR}}) / \Omega_{\alpha \text{IR}} = 1$.
Above $\Omega_{\alpha \text{IR}}$, the phonon decay time ($= \Gamma_\alpha^{-1}$) becomes shorter than half of the vibrational period ($= \pi/\Omega$), i.e., the phonon decays within half of the duration of one period.
Figure~\ref{fig10} plots $\pi \Gamma_\alpha(\Omega) / \Omega$ as a function of $\Omega$.
The point of intersection with $\pi \Gamma_\alpha(\Omega) / \Omega = 1$ gives the value of $\Omega_{\alpha \text{IR}}$.
Note that the fitting functions in Eqs.~(\ref{eq:dho}) and~(\ref{eq:dho2}) (the damped harmonic oscillator model) may not be appropriate for measuring phonon transport properties above $\Omega_{\alpha \text{IR}}$~\cite{Beltukov_2018}.

For the crystal, $\pi \Gamma_\alpha(\Omega) / \Omega \ll 1$ in the whole $\Omega$ region, which means that phonons propagate as plane waves.
In contrast, the glass shows a finite value of $\Omega_{\alpha \text{IR}}$.
In particular, the $\Omega_{T\text{IR}}$ value for transverse phonons nearly coincides with the BP frequency, $\Omega_{T\text{IR}} \approx \omega_\text{BP}$.
This coincidence has been observed in many glasses~\cite{shintani_2008,Monaco2_2009,Mizuno_2014,Beltukov_2016}.
Moreover, this is consistent with the expectation that $\omega_\text{BP}$ should serve as the upper bound on the frequency at which phonons can propagate as plane waves.

However, the $\Omega_{L\text{IR}}$ for longitudinal phonons in the glass is located at a much higher frequency, $\Omega_{L\text{IR}} \gg \Omega_{T\text{IR}} \approx \omega_\text{BP}$~\cite{shintani_2008,Monaco2_2009,Mizuno_2014,Beltukov_2016}, meaning that longitudinal phonons can propagate even above $\omega_\text{BP}$.
This result originates from the fact that the attenuation of longitudinal phonons ($\Gamma_L(\Omega)$) is much lower than that of transverse phonons ($\Gamma_T(\Omega)$) at a fixed $\Omega$, as shown in Fig.~\ref{fig9a}, which can, in turn, be understood in terms of heterogeneous elasticity theory~\cite{schirmacher_2006,schirmacher_2007}.
This theory assumes that shear modulus heterogeneity dominates compared with bulk modulus heterogeneity, which is true in the present SC system~\cite{Mizuno2_2013,Mizuno_2014,Mizuno2_2016}, and predicts that the shear modulus heterogeneity induces anomalous behaviours for both transverse and longitudinal phonons.
In this theoretical framework, we can understand that $\Gamma_T(\Omega)$ becomes larger than $\Gamma_L(\Omega)$ because transverse phonons are more sensitive to the shear modulus heterogeneity.

%%%%%%%%%%%%%%%%%%%%%%%%%%%%%%%%%%%%%%%%%%%%%%%%%%%%%%%%%%%%%%%%%%%%%%%%%%%%%%%%%%%%%%%%%%%
\subsection{Zero-temperature measurement}
As we have discussed so far, the disordered structure has a significant impact on the phonon transport in a glass.
To study these structure-induced effects, it is most straightforward to measure phonon transport at zero temperature $T = 0$, without thermal fluctuations.
We will present two methods of doing so below.

%%%%%%%%%%%%%%%%%%%%%%%%%%%%%%%%%%%%%%%%%%%%%%%%%%%%%%%%%%%%%%%%%%%%%%%%%%%%%%%%%%%%%%%%%%%
\subsubsection{Dynamic structure factor in the zero-temperature limit}
We first take the zero-temperature limit (as $T \rightarrow 0$) of the dynamic structure factor $S_\alpha (\mathbf{q},\omega)$.
As $T \rightarrow 0$ in the harmonic limit, the velocity vector $\mathbf{v}(t)$ converges as follows:
\begin{equation}
\mathbf{v}(t) = \frac{1}{\sqrt{\mathcal{M}}} \frac{d\mathbf{u}(t)}{dt}\ \longrightarrow \ \sum_{k=1}^{3N} \frac{d u^k(t)}{dt} \left( \frac{\mathbf{e}^{k}}{\sqrt{\mathcal{M}}}\right),
\end{equation}
where we apply Eq.~(\ref{mode1b}) for $\mathbf{u}(t)$ and
\begin{equation}
\frac{d u^k(t)}{dt} = - \left( \mathbf{e}^k \cdot \mathbf{u}_0 \right) \omega^k \sin \left( \omega^k t \right) + \left( \mathbf{e}^k \cdot \dot{\mathbf{u}}_0 \right) \cos \left( \omega^k t \right).
\end{equation}
Eqs.~(\ref{eq.sqw}) and~(\ref{projection}) then lead to
\begin{equation}
S_\alpha (\mathbf{q},\omega) \ \longrightarrow \ \frac{T}{2} \left( \frac{q}{\omega} \right)^2 \sum_{k=1}^{3N} \left| \left( \frac{\mathbf{e}^{k}}{\sqrt{\mathcal{M}}} \right) \cdot \mathbf{u}^{\mathbf{q},\alpha}_\text{ph} \right|^2 \left[ \delta \left(\omega-\omega^k \right) + \delta \left(\omega+\omega^k \right) \right], \label{eq.sqw0}
\end{equation}
where we use $\left<  \left( \mathbf{e}^k \cdot \dot{\mathbf{u}}_0 \right)^2 \right>_0 = T$, which is the equipartition law for energy.
By taking the polarization vector to be $\mathbf{s}_T(\mathbf{q}) = \hat{\mathbf{q}}^\perp$ or $\mathbf{s}_L(\mathbf{q}) = \hat{\mathbf{q}}$, we can also write Eq.~(\ref{eq.sqw0}) in a more familiar form:
\begin{equation}
S_\alpha (\mathbf{q},\omega)\ \longrightarrow \ \frac{T}{2N} \left( \frac{q}{\omega} \right)^2 \sum_{k=1}^{3N} F_\alpha^k (\mathbf{q}) \left[ \delta \left(\omega-\omega^k \right) + \delta \left(\omega+\omega^k \right) \right],
\end{equation}
\begin{equation}
\begin{aligned}
F_T^k (\mathbf{q}) &= \left| \sum_{i=1}^N \left( \frac{ \mathbf{e}^k_i }{\sqrt{m_i}} \times  \hat{\mathbf{q}} \right) \exp(\text{i}\mathbf{q}\cdot \mathbf{R}_i) \right|^2, \\
F_L^k (\mathbf{q}) &= \left| \sum_{i=1}^N \left( \frac{ \mathbf{e}^k_i }{\sqrt{m_i}} \cdot  \hat{\mathbf{q}} \right) \exp(\text{i}\mathbf{q}\cdot \mathbf{R}_i) \right|^2.
\end{aligned} 
\end{equation}

For the case of single-component crystals with mass $m_i \equiv m$, $\mathbf{e}^{k} = \mathbf{u}^{\mathbf{q},\alpha}_\text{ph} = \mathbf{e}^{\mathbf{q},\alpha}_\text{ph}$, and we obtain
\begin{equation}
S_\alpha (\mathbf{q},\omega)\ \longrightarrow \ \frac{T}{2m} \left( \frac{q}{\omega} \right)^2 \left[ \delta \left(\omega-\omega_\alpha(\mathbf{q}) \right) + \delta \left(\omega+\omega_\alpha(\mathbf{q}) \right) \right]. \label{eq.sqw0c}
\end{equation}
We thus confirm that $S_\alpha (\mathbf{q},\omega)$ is the delta function as $T \rightarrow 0$; that the propagation frequency is the eigenfrequency, $\Omega(\mathbf{q})=\omega_\alpha(\mathbf{q})$; and that the attenuation rate is exactly zero, $\Gamma_\alpha(\mathbf{q})=0$.
For the case of a glass, it has been numerically verified that $S_\alpha (\mathbf{q},\omega)$ at low $T$ coincides with the value at $T \rightarrow 0$~\cite{Ruocco_2000}.

%%%%%%%%%%%%%%%%%%%%%%%%%%%%%%%%%%%%%%%%%%%%%%%%%%%%%%%%%%%%%%%%%%%%%%%%%%%%%%%%%%%%%%%%%%%
\subsubsection{Direct measurement at zero temperature}\label{sec.ptdm}
The second method is to directly measure phonon transport at $T=0$~\cite{Gelin_2016,Mizuno_2018}.
We start with the inherent structure $\mathbf{R} \equiv \left[ \mathbf{R}_1^T,\mathbf{R}_2^T,\cdots,\mathbf{R}_N^T \right]^T$ and excite a phonon of wavevector $\mathbf{q}$ and polarization $\alpha$ by setting the initial conditions at $t=0$ to ${\mathbf{u}} \equiv \mathbf{u}_0 = \mathbf{0}$ and $d{\mathbf{u}}/dt \equiv \dot{\mathbf{u}}_0 = \mathbf{u}^{\mathbf{q},\alpha}_\text{ph}$ in Eq.~(\ref{phononvector}).
We next solve the linearized equation of motion, Eq.~(\ref{eom2}):
\begin{equation} 
\frac{d^2 \mathbf{u}(t)}{dt^2} = - \mathcal{D} \mathbf{u}(t) + \dot{\mathbf{u}}_0 \delta (t).
\end{equation}
From $\mathbf{u}(t)$, we calculate the (normalized) velocity-velocity time correlation function:
\begin{equation}
C(t) \equiv \left( \frac{\dot{\mathbf{u}}(t)}{\sqrt{\mathcal{M}}} \cdot \frac{\dot{\mathbf{u}}_0 }{\sqrt{\mathcal{M}}} \right) \left( \frac{\dot{\mathbf{u}}_0}{\sqrt{\mathcal{M}}} \cdot \frac{\dot{\mathbf{u}}_0}{\sqrt{\mathcal{M}}} \right)^{-1}.
\end{equation}

The function $C(t)$ represents the propagation and attenuation behaviours of the initially excited phonon $\dot{\mathbf{u}}_0$ with $\mathbf{q}$ and $\alpha$.
$\Omega(\mathbf{q})$ and $\Gamma_\alpha(\mathbf{q})$ are then extracted by fitting the simulated data for $C(t)$ to the damped harmonic oscillator model in Eq.~(\ref{eq:dho2})~\cite{Gelin_2016,Mizuno_2018}: $C(t) \equiv \cos\left( \Omega(\mathbf{q}) t\right) \exp\left(-{\Gamma_\alpha(\mathbf{q})}t/{2} \right)$.

%%%%%%%%%%%%%%%%%%%%%%%%%%%%%%%%%%%%%%%%%%%%%%%%%%%%%%%%%%%%%%%%%%%%%%%%%%%%%%%%%%%%%%%%%%%
\section{Elastic deformation}~\label{sec.moduli}
In this section, we describe how to measure the elastic response, particularly the elastic moduli, of glasses and any solid-state materials.
The elastic moduli are necessary for calculating the Debye vDOS and the macroscopic sound speed, which are useful references for understanding the vibrational properties of a material.
There are two methods of measuring the elastic moduli.
One is the so-called fluctuation formulation.
In this method, we do not apply any external deformation but rather perform an equilibrium, molecular simulation and use the formulation developed based on linear response theory.
The other method is to measure the elastic moduli directly.
We apply an external strain to the system and measure the stress as a function of the strain, i.e., the stress-strain curve.
The slope of this curve gives the value of the elastic modulus.

%%%%%%%%%%%%%%%%%%%%%%%%%%%%%%%%%%%%%%%%%%%%%%%%%%%%%%%%%%%%%%%%%%%%%%%%%%%%%%%%%%%%%%%%%%%
\subsection{General description}
We first provide a general explanation of the elastic response and elastic moduli.

%%%%%%%%%%%%%%%%%%%%%%%%%%%%%%%%%%%%%%%%%%%%%%%%%%%%%%%%%%%%%%%%%%%%%%%%%%%%%%%%%%%%%%%%%%%
\subsubsection{Elastic modulus tensor}
The free energy is given by~\cite{Hansen_2006}
\begin{equation}
F = -T \ln Z = -T \ln \left\{ \text{Tr} \exp\left[ -\frac{1}{T}\left( \sum_{i=1}^N \frac{\mathbf{p}_i^2}{2m_i} + \Phi(\mathbf{r}) \right) \right]  \right\}, \label{freeenergy}
\end{equation}
where we set the Boltzmann constant to $k_B=1$ and $Z \equiv \text{Tr} \exp\left( -{H}/{T}\right)$ is the partition function~($H \equiv \sum_{i=1}^N {\mathbf{p}_i^2}/{m_i} + \Phi$ is the Hamiltonian, and Tr denotes the trace operator).
Here, we consider the Helmholtz free energy and the canonical ensemble; however, the values of the elastic moduli are insensitive to the ensemble at low $T$
\footnote{
The isothermal moduli~(in canonical ensemble) and the adiabatic moduli~(in micro-canonical ensemble) coincide in the low-$T$ harmonic limit~\cite{Ashcroft_1976,Kittel_1996}.
}.
The stress tensor $\sigma_{\alpha \beta}$ is defined as the first derivative of the free energy $F$ with respect to the strain tensor $\epsilon_{\alpha \beta}$ as~\cite{Born_1954,barron_1965,Alexander_1998,lutsko_1989,Lemaitre_2006}
\begin{equation}
\sigma_{\alpha \beta} \equiv \frac{1}{V} \left. \frac{\partial F}{\partial \epsilon_{\alpha \beta}} \right|_{\epsilon_{\alpha \beta} \to 0},
\label{stressformuation0}
\end{equation}
where $\alpha,\beta~(\gamma,\delta) = x, y, z$.
There are two different definitions of $\epsilon_{\alpha \beta}$
\footnote{
The stress tensors defined by $\epsilon_{\alpha \beta} = e_{\alpha \beta}$ and $\eta_{\alpha \beta}$ in Eq.~(\ref{stressformuation0}) coincide with each other.
}:
the linear strain~(infinitesimal strain) tensor, $e_{\alpha \beta}$, and the Green-Lagrange strain~(finite strain) tensor, $\eta_{\alpha \beta}$:
\begin{equation}
\begin{aligned}
e_{\alpha \beta}    & \equiv \frac{1}{2} \left( \frac{\partial u_\alpha}{\partial r_\beta} + \frac{\partial u_\beta}{\partial r_\alpha} \right) \equiv \frac{1}{2} \left( u_{\alpha \beta} + u_{\beta \alpha} \right), \\
\eta_{\alpha \beta} & \equiv \frac{1}{2} \left( \frac{\partial u_\alpha}{\partial r_\beta} + \frac{\partial u_\beta}{\partial r_\alpha} + \sum_{\gamma=x,y,z} \frac{\partial u_\gamma}{\partial r_\alpha} \frac{\partial u_\gamma}{\partial r_\beta} \right) \equiv \frac{1}{2} \left( u_{\alpha \beta} + u_{\beta \alpha} + \sum_{\gamma=x,y,z} u_{\gamma \alpha} u_{\gamma \beta} \right),
\end{aligned}
\end{equation}
where $[r_x,r_y,r_z]^T$ and $[u_x,u_y,u_z]^T$ represent the spatial coordinates and the displacement field, respectively, and $u_{\alpha \beta} \equiv {\partial u_\alpha}/{\partial r_\beta}$ is the displacement gradient tensor.

We then define the elastic modulus tensor in two different ways~\cite{Born_1954,barron_1965,Alexander_1998,lutsko_1989,Lemaitre_2006}.
Firstly the modulus tensor $\tilde{C}_{\alpha \beta \gamma \delta}$ is defined as the second derivative of $F$ with respect to the Green-Lagrange strain tensor $\eta_{\alpha \beta}$:
\begin{equation}
\tilde{C}_{\alpha \beta \gamma \delta} \equiv \frac{1}{V} \left. \frac{ \partial^2 F }{ \partial \eta_{\alpha \beta} \partial \eta_{\gamma\delta} } \right|_{\eta_{\alpha \beta} \to 0,\ \eta_{\gamma \delta} \to 0}. \label{defofmodulus}
\end{equation}
Secondly the modulus tensor ${C}_{\alpha \beta \gamma \delta}$ is defined as the first derivative of $\sigma_{\alpha \beta}$ with respect to the linear strain tensor $e_{\gamma \delta}$:
\begin{equation}
{C}_{\alpha \beta \gamma \delta} \equiv \left. \frac{ \partial \sigma_{\alpha \beta} }{ \partial e_{\gamma\delta} } \right|_{e_{\gamma\delta} \to 0}. \label{defofmodulus2}
\end{equation}
These two modulus tensors do not coincide for the system under initial stress $\sigma^0_{\alpha \beta} \equiv \sigma_{\alpha \beta} (e_{\gamma\delta}=0)$
\footnote{
$\tilde{C}_{\alpha \beta \gamma \delta}$ is defined in a Lagrangian framework, whereas ${C}_{\alpha \beta \gamma \delta}$ is based on an Eulerian framework~\cite{Lemaitre_2006,barron_1965}.
}.
$\tilde{C}_{\alpha \beta \gamma \delta}$ and ${C}_{\alpha \beta \gamma \delta}$ are related as follows~(please see Refs.~\onlinecite{Lemaitre_2006,barron_1965} for details):
\begin{equation}
\begin{aligned}
C_{\alpha \beta \gamma \delta} &= \tilde{C}_{\alpha \beta \gamma \delta} + C^C_{\alpha \beta \gamma \delta}, \\
C^C_{\alpha \beta \gamma \delta} &= -\frac{1}{2} \left( 2 \sigma^0_{\alpha \beta} \delta_{\gamma \delta}-\sigma^0_{\alpha \gamma} \delta_{\beta \delta}-\sigma^0_{\alpha \delta}\delta_{\beta \gamma} - \sigma^0_{\beta \gamma} \delta_{\alpha \delta} -\sigma^0_{\beta \delta}\delta_{\alpha \gamma} \right).
\end{aligned} \label{transmodulus}
\end{equation}
We refer to $C^C_{\alpha \beta \gamma \delta}$ as the correction term in the following.
If the initial stress tensor $\sigma^0_{\alpha \beta}$ is zero, then $C^C_{\alpha \beta \gamma \delta}$ is zero, and $C_{\alpha \beta \gamma \delta}$ and $\tilde{C}_{\alpha \beta \gamma \delta}$ exactly coincide.
Since $C_{\alpha \beta \gamma \delta}$ is necessary for calculations of the Debye vDOS and the macroscopic sound speed, we focus on $C_{\alpha \beta \gamma \delta}$ below.

Here we note that strictly speaking, for calculations of the Debye vDOS and the macroscopic sound speed, we should apply the modulus tensor ${C}'_{\alpha \beta \gamma \delta}$ which is defined as the first derivative of $\sigma_{\alpha \beta}$ with respect to the displacement gradient tensor $u_{\gamma \delta}$~(instead of $e_{\gamma \delta}$):
\begin{equation}
{C}'_{\alpha \beta \gamma \delta} \equiv \left. \frac{ \partial \sigma_{\alpha \beta} }{ \partial u_{\gamma\delta} } \right|_{u_{\gamma\delta} \to 0}. \label{defofmodulus3}
\end{equation}
However, as long as we consider the situation under hydrostatic pressure where the initial stress is $\sigma^0_{\alpha \beta} = -p \delta_{\alpha \beta}$, ${C}_{\alpha \beta \gamma \delta}$ and ${C}'_{\alpha \beta \gamma \delta}$ are identical~\cite{Lemaitre_2006,barron_1965}.

%%%%%%%%%%%%%%%%%%%%%%%%%%%%%%%%%%%%%%%%%%%%%%%%%%%%%%%%%%%%%%%
\begin{figure}[t]
\centerline{
\includegraphics[width=0.95\textwidth]{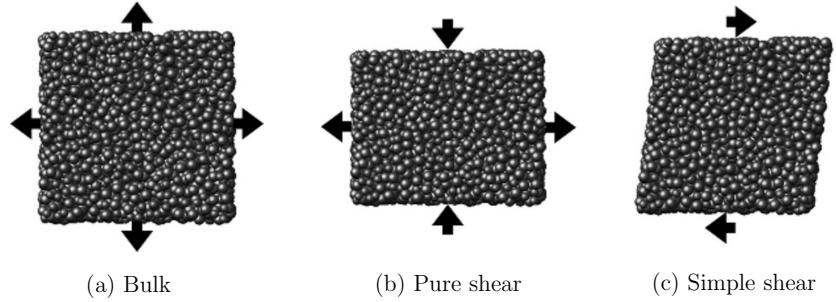}
}
\caption{
Schematic illustrations of (a) bulk, (b) pure shear, and (c) simple shear deformations.
The bulk ($K$), pure shear ($G_p$), and simple shear ($G_s$) moduli correspond to these deformations.
}
\label{fig11}
\end{figure}
%%%%%%%%%%%%%%%%%%%%%%%%%%%%%%%%%%%%%%%%%%%%%%%%%%%%%%%%%%%%%%%

%%%%%%%%%%%%%%%%%%%%%%%%%%%%%%%%%%%%%%%%%%%%%%%%%%%%%%%%%%%%%%%%%%%%%%%%%%%%%%%%%%%%%%%%%%%
\subsubsection{Bulk, pure shear, and simple shear elastic moduli}
We consider three types of deformations in particular, namely, a volume-changing bulk deformation and pure shear~(plane or triaxial strains) and simple shear deformations, which are illustrated in Fig.~\ref{fig11}.
Since there are two independent pure shear deformations and three independent simple shear deformations, we consider six independent deformations in total, including one bulk deformation~\cite{Fung_1977}.
The strain tensors for these six deformations are
\footnote{
We can also consider other pure shear deformations, which are described as superpositions of the present two independent pure shear deformations.
}
\begin{equation}
\epsilon_{xx} = \epsilon_{yy} = \epsilon_{zz} = \epsilon_b \quad \text{(bulk)},
\end{equation}
\begin{equation}
\begin{aligned}
& \epsilon_{xx} = \epsilon_{p1}, \quad \epsilon_{yy} = -\epsilon_{p1} & \text{(pure shear1)}, \\
& \epsilon_{xx} = \epsilon_{p2}, \quad \epsilon_{yy} = \epsilon_{p2}, \quad \epsilon_{zz} = -2 \epsilon_{p2} & \text{(pure shear2)},
\end{aligned}
\end{equation}
\begin{equation}
\begin{aligned}
& \epsilon_{xy} = \epsilon_{yx} = \epsilon_{s1} & \text{(simple shear1)}, \\
& \epsilon_{xz} = \epsilon_{zx} = \epsilon_{s2} & \text{(simple shear2)}, \\
& \epsilon_{yz} = \epsilon_{zy} = \epsilon_{s3} & \text{(simple shear3)},
\end{aligned}
\end{equation}
where no written components in $\epsilon_{\alpha \beta}$ are zero, and $\epsilon_b$, $\epsilon_{p1}$, $\epsilon_{p2}$, $\epsilon_{s1}$, $\epsilon_{s2}$, $\epsilon_{s3}$ represent the applied strains corresponding to the deformations.
$\delta V/V \equiv 3\epsilon_b$ represents the volume change, while $\gamma_{p1} \equiv 2\epsilon_{p1}$, $\gamma_{p2} \equiv 3\epsilon_{p2}$, $\gamma_{s1} \equiv 2\epsilon_{s1}$, $\gamma_{s2} \equiv 2\epsilon_{s2}$, $\gamma_{s3} \equiv 2\epsilon_{s3}$ present the shear strains for the corresponding shear deformations.

When we apply a deformation to the system, the corresponding stress varies with the applied strain. For the bulk deformation, the two pure shear deformations, and the three simple shear deformations, the applied stresses are the pressure, $p \equiv -(\sigma_{xx}+\sigma_{yy}+\sigma_{zz})/3$; two pure shear stresses, $\sigma_{p1} \equiv {(\sigma_{xx} - \sigma_{yy})}/{2}$ and $\sigma_{p2} \equiv {(\sigma_{xx} + \sigma_{yy} -2\sigma_{zz})}/{4}$; and three simple shear stresses, $\sigma_{s1} \equiv \sigma_{xy}$, $\sigma_{s2} \equiv \sigma_{xz}$, and $\sigma_{s3} \equiv \sigma_{yz}$, respectively.
Accordingly, the bulk modulus $K$, the two pure shear moduli $G_{p1}$ and $G_{p2}$, and the three simple shear moduli $G_{s1}$, $G_{s2}$, and $G_{s3}$ are defined as follows:
\begin{equation}
K \equiv \left. -\frac{\partial p}{\partial (\delta V/V)}\right|_{(\delta V/V) \rightarrow 0} = \left. \frac{\partial (\sigma_{xx}+\sigma_{yy}+\sigma_{zz})/3}{\partial (3\epsilon_b)}\right|_{\epsilon_b\rightarrow 0} \quad \text{(bulk)}, \label{modulusslope1}
\end{equation}
\begin{equation}
\begin{aligned}
& G_{p1} \equiv \left. \frac{\partial \sigma_{p1}}{\partial \gamma_{p1}}\right|_{\gamma_{p1} \rightarrow 0} = \left. \frac{\partial (\sigma_{xx} - \sigma_{yy})/2}{\partial (2\epsilon_{p1})}\right|_{\epsilon_{p1} \rightarrow 0} & \text{(pure shear 1)}, \\
& G_{p2} \equiv \left. \frac{\partial \sigma_{p2}}{\partial \gamma_{p2}}\right|_{\gamma_{p2} \rightarrow 0} = \left. \frac{\partial (\sigma_{xx} + \sigma_{yy} -2\sigma_{zz})/4}{\partial (3\epsilon_{p2})}\right|_{\epsilon_{p2} \rightarrow 0} & \text{(pure shear 2)},
\end{aligned} \label{modulusslope2}
\end{equation}
\begin{equation}
\begin{aligned}
& G_{s1} \equiv \left. \frac{\partial \sigma_{s1}}{\partial \gamma_{s1}}\right|_{\gamma_{s1} \rightarrow 0} = \left. \frac{\partial \sigma_{xy}}{\partial (2\epsilon_{s1})}\right|_{\epsilon_{s1} \rightarrow 0} & \text{(simple shear 1)}, \\
& G_{s2} \equiv \left. \frac{\partial \sigma_{s2}}{\partial \gamma_{s2}}\right|_{\gamma_{s2} \rightarrow 0} = \left. \frac{\partial \sigma_{xz}}{\partial (2\epsilon_{s2})}\right|_{\epsilon_{s2} \rightarrow 0} & \text{(simple shear 2)}, \\
& G_{s3} \equiv \left. \frac{\partial \sigma_{s3}}{\partial \gamma_{s3}}\right|_{\gamma_{s3} \rightarrow 0} = \left. \frac{\partial \sigma_{yz}}{\partial (2\epsilon_{s3})}\right|_{\epsilon_{s3} \rightarrow 0} & \text{(simple shear 3)}.
\end{aligned} \label{modulusslope3}
\end{equation}

$K$, $G_{p1}$, $G_{p2}$, $G_{s1}$, $G_{s2}$, and $G_{s3}$ can be formulated in terms of the elastic modulus tensor $C_{\alpha \beta \gamma \delta}$~\cite{Fung_1977}:
\begin{equation}
K = \frac{(C_{xxxx}+C_{yyyy}+C_{zzzz}+C_{xxyy}+C_{yyxx}+C_{xxzz}+C_{zzxx}+C_{yyzz}+C_{zzyy})}{9},
\label{elasticmodulus4a}
\end{equation}
\begin{equation}
\begin{aligned}
G_{p1} & = \frac{(C_{xxxx}+C_{yyyy}-C_{xxyy}-C_{yyxx})}{4}, \\
G_{p2} & = \frac{(C_{xxxx}+C_{yyyy}+4C_{zzzz}+C_{xxyy}+C_{yyxx}-2C_{xxzz}-2C_{zzxx}-2C_{yyzz}-2C_{zzyy})}{12},
\end{aligned}
\label{elasticmodulus4b}
\end{equation}
\begin{equation}
G_{s1} = {C_{xyxy}}, \qquad G_{s2} = {C_{xzxz}}, \qquad G_{s3} = {C_{yzyz}}.
\label{elasticmodulus4c}
\end{equation}
Note that $C_{xyxy}=C_{xyyx}$, $C_{xzxz}=C_{xzzx}$, $C_{yzyz}=C_{yzzy}$.
For cubic crystals such as FCC crystals, the two pure shear moduli coincide, i.e., $G_p \equiv G_{p1} = G_{p2}$, and the three simple shear moduli also coincide, i.e., $G_s \equiv G_{s1} = G_{s2} = G_{s3}$; however, the pure shear and simple shear moduli generally take different values, i.e., $G_p \neq G_s$.
On the other hand, for glasses, which are isotropic systems, all five shear moduli coincide: $G \equiv G_{p1} = G_{p2} = G_{s1} = G_{s2} = G_{s3}$.

%%%%%%%%%%%%%%%%%%%%%%%%%%%%%%%%%%%%%%%%%%%%%%%%%%%%%%%%%%%%%%%%%%%%%%%%%%%%%%%%%%%%%%%%%%%
\subsubsection{Affine and non-affine elastic moduli}
The elastic response can be decomposed into affine and non-affine components~\cite{Tanguy_2002,Lemaitre_2006,Wittmer_2013,Mizuno_2013}.
Affine deformation refers to particles that follow the applied affine strain field and are displaced affinely at all scales.
More specifically, for a linear strain $\epsilon_{\alpha \beta} = e_{\alpha \beta}$, the affine deformation causes the displacement of the position of particle $i$, $\mathbf{r}_i = [r_{ix},r_{iy},r_{iz}]^T$, as follows:
\begin{equation}
r_{i \kappa} \rightarrow r_{i \kappa} + \frac{1}{2} \left( u_{\alpha \beta} r_{i \beta} \delta_{\alpha \kappa} + u_{\beta \alpha} r_{i \alpha} \delta_{\beta \kappa} \right),
\label{affine1}
\end{equation}
where $\kappa = x, y, z$.
Similarly, for the Green-Lagrange strain $\epsilon_{\alpha \beta} = \eta_{\alpha \beta}$, particles are displaced as follows:
\begin{equation}
r_{i \kappa} \rightarrow r_{i \kappa} + \frac{1}{2} \left( u_{\alpha \beta} r_{i \beta} \delta_{\alpha \kappa} + u_{\beta \alpha} r_{i \alpha} \delta_{\beta \kappa} + \sum_{\gamma=x,y,z} u_{\gamma \alpha} u_{\gamma \beta} r_{i \alpha} r_{i \beta} \delta_{\gamma \kappa} \right).
\label{affine2}
\end{equation}
The affine strain tensors $\epsilon^A_{\alpha \beta} = e^A_{\alpha\beta}$ and $\eta^A_{\alpha \beta}$ are defined as displacing particles as expressed in Eqs.~(\ref{affine1}) and~(\ref{affine2}).

By applying the affine strain tensors to Eqs.~(\ref{defofmodulus}) and~(\ref{defofmodulus2}), we can define the affine elastic modulus tensors as
\begin{equation}
\tilde{C}^A_{\alpha \beta \gamma \delta} \equiv \frac{1}{V} \left. \frac{ \partial^2 F }{ \partial \eta^A_{\alpha \beta} \partial \eta^A_{\gamma\delta} } \right|_{\eta^A_{\alpha \beta} \to 0,\ \eta^A_{\gamma \delta} \to 0},
\label{defofaffinemodulus}
\end{equation}
\begin{equation}
{C}^A_{\alpha \beta \gamma \delta} \equiv \left. \frac{ \partial \sigma_{\alpha \beta} }{ \partial e^A_{\gamma\delta} } \right|_{e^A_{\gamma\delta} \to 0}.
\label{defofaffinemodulus2}
\end{equation}
The elastic response of a crystal is characterized mostly by affine deformation and the corresponding affine modulus~\cite{Born_1954,Alexander_1998}.
However, importantly, the elastic response of a glass does not follow solely affine deformation.
We additionally need to consider non-affine deformation, which results in additional particle displacements at the microscopic scale and causes the displacements to deviate from the applied affine field~\cite{Tanguy_2002,Lemaitre_2006,Wittmer_2013,Mizuno_2013}.
Non-affine deformation contributes negatively to the overall modulus; thus, the corresponding modulus is defined as the non-affine modulus $-C_{\alpha \beta \gamma \delta}^{N}$.
We therefore describe the elastic modulus tensor as
\begin{equation}
{C}_{\alpha \beta \gamma \delta} = {C}^A_{\alpha \beta \gamma \delta} - C_{\alpha \beta \gamma \delta}^{N}.
\label{fformuation}
\end{equation}
As we will see below, the non-affine component is important in glasses, being comparable in magnitude to the affine modulus~\cite{Tanguy_2002,Lemaitre_2006,Wittmer_2013,Mizuno_2013}.

%%%%%%%%%%%%%%%%%%%%%%%%%%%%%%%%%%%%%%%%%%%%%%%%%%%%%%%%%%%%%%%%%%%%%%%%%%%%%%%%%%%%%%%%%%%
\subsection{Fluctuation formulation}
We now present how to measure the elastic moduli by means of computational simulations.
We first introduce the fluctuation formulation.
In the following, we consider a system with a pair-wise potential, where particles $i$ and $j$ interact through the potential $\phi_{ij}(r_{ij})$ ($r_{ij}$ is the distance between them) and the total potential is $\Phi = \sum_{i<j} \phi_{ij}(r_{ij})$.
We also employ the linear strain tensor $\epsilon_{\alpha \beta} = e_{\alpha \beta}$.
Please see, e.g., Refs.~\onlinecite{lutsko_1989,Lemaitre_2006,Wittmer_2013}, for detailed derivations of the following formulations.

%%%%%%%%%%%%%%%%%%%%%%%%%%%%%%%%%%%%%%%%%%%%%%%%%%%%%%%%%%%%%%%%%%%%%%%%%%%%%%%%%%%%%%%%%%%
\subsubsection{Finite temperature $T>0$}
By using Eqs.~(\ref{freeenergy}) and (\ref{stressformuation0}), we can obtain the stress tensor in the form of
\begin{equation}
\begin{aligned}
\sigma_{\alpha \beta} &= \left< \hat{\sigma}_{\alpha \beta}\right>, \\
\hat{\sigma}_{\alpha \beta} &= \frac{1}{V} \left[ -\sum_{i=1}^{N} m_i v_{i \alpha}v_{i \beta} + \sum_{i<j} \left( r_{ij} \deri{\phi_{ij}}{r_{ij}} \right) {n_{ij \alpha}n_{ij \beta}} \right], \label{stressformulation}
\end{aligned}
\end{equation}
where $\left< \right>$ denotes the canonical ensemble average, $\hat{\rho} = {N}/{V}$ is the number density, and $\mathbf{n}_{ij} = \left[ n_{ij x},n_{ij y},n_{ij z} \right]^T \equiv (\mathbf{r}_i-\mathbf{r}_j)/r_{ij}$ is the unit vector connecting particle $j$ to particle $i$.
The elastic modulus tensor $C_{\alpha \beta \gamma \delta}$ is formulated using Eqs.~(\ref{defofmodulus}) and~(\ref{transmodulus}), which includes the affine modulus and the non-affine modulus as in Eq.~(\ref{fformuation}).
The affine modulus tensor $C^A_{\alpha \beta \gamma \delta}$~[Eq.~(\ref{defofaffinemodulus2})] is formulated as
\begin{equation}
\begin{aligned}
C^A_{\alpha \beta \gamma \delta} &= C^B_{\alpha \beta \gamma \delta} + C^K_{\alpha \beta \gamma \delta} + C^C_{\alpha \beta \gamma \delta}, \\
C^B_{\alpha \beta \gamma \delta} &= \frac{1}{V} \left< \sum_{i<j} \left( r_{ij}^2 \derri{\phi_{ij}}{{r_{ij}}^2} - {r_{ij}}\deri{\phi_{ij}}{r_{ij}} \right) {n_{ij \alpha} n_{ij \beta} n_{ij \gamma} n_{ij \delta}} \right>, \\
C^K_{\alpha \beta \gamma \delta} &= 2 \hat{\rho} T (\delta_{\alpha \gamma} \delta_{\beta \delta} + \delta_{\alpha \delta}\delta_{\beta \gamma}), \\
C^C_{\alpha \beta \gamma \delta} &= -\frac{1}{2} \left( 2  \left< \hat{\sigma}_{\alpha \beta} \right> \delta_{\gamma \delta} -  \left< \hat{\sigma}_{\alpha \gamma} \right> \delta_{\beta \delta}- \left< \hat{\sigma}_{\alpha \delta} \right> \delta_{\beta \gamma} -  \left< \hat{\sigma}_{\beta \gamma} \right> \delta_{\alpha \delta} - \left< \hat{\sigma}_{\beta \delta} \right> \delta_{\alpha \gamma} \right),
\end{aligned}
\label{afformuation}
\end{equation}
where $C^B_{\alpha \beta \gamma \delta}$ is the so-called Born term and $C^K_{\alpha \beta \gamma \delta}$ is the kinetic contribution.
Since these two terms are formulated by Eq.~(\ref{defofaffinemodulus}) based on the Green-Lagrange strain, we need the correction term $C^C_{\alpha \beta \gamma \delta}$ to obtain the modulus tensor based on the linear strain, as shown in Eq.~(\ref{transmodulus})~\cite{Lemaitre_2006,barron_1965}.
In addition, the non-affine modulus is formulated as
\begin{equation}
C^N_{\alpha \beta \gamma \delta} = \frac{V}{T} \left[ \left< \hat{\sigma}_{\alpha \beta} \hat{\sigma}_{\gamma \delta} \right>-\left< \hat{\sigma}_{\alpha \beta} \right>\left< \hat{\sigma}_{\gamma \delta} \right> \right] = \frac{V}{T} \left< \delta \hat{\sigma}_{\alpha \beta} \delta \hat{\sigma}_{\gamma \delta} \right>,
\label{nfformuation}
\end{equation}
where $\delta \hat{\sigma}_{\alpha \beta} \equiv \hat{\sigma}_{\alpha \beta}- \left< \hat{\sigma}_{\alpha \beta} \right>$ is the stress fluctuation.
$C^N_{\alpha \beta \gamma \delta}$ is therefore formulated as the correlation function for stress fluctuations.

%%%%%%%%%%%%%%%%%%%%%%%%%%%%%%%%%%%%%%%%%%%%%%%%%%%%%%%%%%%%%%%
\begin{table}[t]
\tbl{The elastic moduli, $K$, $G_p$, and $G_s$, of SC systems: a glass and an FCC crystal.
We present values obtained using two methods: the fluctuation formulation~(FF) at a finite temperature of $T=10^{-2}$~(almost at the zero-temperature limit) and direct measurement~(DM) at zero temperature, $T=0$.
We present the average values of pure shear modulus, $G_{p} = (G_{p1} + G_{p2})/2$, and simple shear modulus, $G_{s} = (G_{s1} + G_{s2} + G_{s3})/3$.
The affine and non-affine moduli are also presented.}
{\centerline{
\renewcommand{\arraystretch}{1.1}
\begin{tabular}{c|c|ccc|ccc|ccc}
\hline
\hline
            & \text{Method} & $K$    & $K^A$  & $K^N$ & $G_p$ & $G_p^A$ & $G_p^N$ & $G_s$  & $G_s^A$ & $G_s^N$ \\
\hline
\hline
Glass       & \text{FF}     & $40.9$ & $41.0$ & $0.1$ & $6.2$ & $14.7$  & $8.5$   & $6.3$  & $14.8$  & $8.5$   \\
\cline{2-11}
            & \text{DM}     & $40.5$ & $40.5$ & $0.0$ & $6.3$ & $14.4$  & $8.1$   & $6.4$  & $14.7$  & $8.3$   \\
\hline
Crystal & \text{FF}     & $32.9$ & $33.0$ & $0.1$ & $5.1$ & $5.2$   & $0.1$   & $16.1$ & $16.3$  & $0.2$   \\
\cline{2-11}
            & \text{DM}     & $32.6$ & $32.6$ & $0.0$ & $5.4$ & $5.4$   & $0.0$   & $16.2$ & $16.2$  & $0.0$   \\
\hline
\hline
\end{tabular}
}} \label{table1}
\end{table}
%%%%%%%%%%%%%%%%%%%%%%%%%%%%%%%%%%%%%%%%%%%%%%%%%%%%%%%%%%%%%%%

In our computer simulation, we first perform an equilibrium ($NVT$) MD or MC simulation at a finite $T>0$ and generate configurations $\mathbf{r} \equiv \left[ \mathbf{r}_1^T, \mathbf{r}_2^T,\cdots,\mathbf{r}_N^T \right]^T$.
From these configurations, we evaluate  $C_{\alpha \beta \gamma \delta}$ by using Eqs.~(\ref{fformuation}) to (\ref{nfformuation}).
From $C_{\alpha \beta \gamma \delta}$, we then calculate the elastic moduli, $K$, $G_p$, and $G_s$, using Eqs.~(\ref{elasticmodulus4a}) to~(\ref{elasticmodulus4c}).
Table~\ref{table1} presents the values of $K$, $G_p$, and $G_s$ for a glass and an FCC crystal with the SC potential.
Here, we also present the values of the affine moduli $K^A$, $G_p^A$, and $G_s^A$ and the non-affine moduli $K^N$, $G_p^N$, and $G_s^N$.
The temperature is very low, $T=10^{-2}$, almost at the zero-temperature limit.
We confirm that the crystal shows negligible values of the non-affine components in both the bulk and shear moduli.
The elastic response of the crystal is therefore characterized by affine deformation, which is due to the symmetry of the lattice structure~\cite{Born_1954,Alexander_1998}.

In contrast to the crystal, the glass shows large values of the non-affine components in the shear moduli, comparable to those of the affine components.
Therefore, non-affine deformation plays an important role in the elastic response of the glass, making it distinct from the elastic response of the crystal~\cite{Tanguy_2002,Lemaitre_2006,Wittmer_2013,Mizuno_2013}.
Note that $G_p$ and $G_s$ coincide in the glass due to its isotropic structure.
On the other hand, for bulk deformation, the glass also shows a very small non-affine component, which is due to the isotropic nature of bulk deformation.
In particular, we can demonstrate that a mono-disperse system with an inverse-power-law potential, such as the present SC potential~[Eq.~(\ref{sc})], shows a non-affine modulus of zero under bulk deformation at zero temperature.

%%%%%%%%%%%%%%%%%%%%%%%%%%%%%%%%%%%%%%%%%%%%%%%%%%%%%%%%%%%%%%%%%%%%%%%%%%%%%%%%%%%%%%%%%%%
\subsubsection{Zero-temperature limit $T \rightarrow 0$}
It is useful to take the limit as $T\rightarrow 0$ in the fluctuation formulation given in Eqs.~(\ref{fformuation}) to (\ref{nfformuation})~\cite{lutsko_1989,Lemaitre_2006}.
As $T\rightarrow 0$, the configuration converges to the inherent structure, $\mathbf{r} \rightarrow \mathbf{R} = \left[ \mathbf{R}_1^T, \mathbf{R}_2^T, \cdots, \mathbf{R}_N^T \right]^T$, and we obtain $C^B_{\alpha \beta \gamma \delta}$, $C^K_{\alpha \beta \gamma \delta}$, and $C^C_{\alpha \beta \gamma \delta}$ as follows:
\begin{equation}
\begin{aligned}
C^B_{\alpha \beta \gamma \delta} &\ \longrightarrow \  \frac{1}{V} \sum_{i<j} \left. \left( r_{ij}^2 \derri{\phi_{ij}}{{r_{ij}}^2} - {r_{ij}}\deri{\phi_{ij}}{r_{ij}} \right) {n_{ij \alpha} n_{ij \beta} n_{ij \gamma} n_{ij \delta}} \right|_{\mathbf{r} = \mathbf{R}}, \\
C^K_{\alpha \beta \gamma \delta} &\ \longrightarrow \ 0,\\
C^C_{\alpha \beta \gamma \delta} &\ \longrightarrow \ -\frac{1}{2} \left( 2 \sigma_{\alpha \beta 0} \delta_{\gamma \delta} - \sigma_{\alpha \gamma 0} \delta_{\beta \delta}- \sigma_{\alpha \delta 0} \delta_{\beta \gamma} - \sigma_{\beta \gamma 0} \delta_{\alpha \delta} - \sigma_{\beta \delta 0} \delta_{\alpha \gamma} \right),
\end{aligned}
\label{fformuation2}
\end{equation}
where $\sigma_{\alpha \beta 0}$ is the stress tensor $\sigma_{\alpha \beta}$ at $T \rightarrow 0$,
\begin{equation}
\sigma_{\alpha \beta}\ \longrightarrow \ \sigma_{\alpha \beta 0} \equiv \frac{1}{V} \left. \sum_{i<j} \left( r_{ij} \deri{\phi_{ij}}{r_{ij}} \right) {n_{ij \alpha}n_{ij \beta}} \right|_{\mathbf{r} = \mathbf{R}}.
\label{stressformulation2}
\end{equation}

The non-affine term $C_{\alpha \beta \gamma \delta}^{N}$ is formulated in terms of the eigenfrequencies $\omega^k$ and the eigenvectors $\mathbf{e}^k$ as follows.
The stress fluctuation is
\begin{equation}
\begin{aligned}
\delta \hat{\sigma}_{\alpha \beta} \ \longrightarrow \ \deri{\sigma_{\alpha \beta 0}}{\mathbf{R}} \cdot \frac{\mathbf{u}(t)}{\sqrt{\mathcal{M}}}
= \sum_{k=1}^{3N} u^k(t) \left( \deri{\sigma_{\alpha \beta 0}}{\mathbf{R}} \cdot \frac{\mathbf{e}^{k}}{ \sqrt{\mathcal{M}} }  \right),
\end{aligned}
\label{fformuation3}
\end{equation}
where we use $\mathbf{u}(t)$ in Eq.~(\ref{mode1b}).
We then obtain
\begin{equation}
C_{\alpha \beta \gamma \delta}^{N} \ \longrightarrow \ V \sum_{k=1}^{3N} \frac{1}{{\omega^k}^2} \left( \deri{\sigma_{\alpha \beta 0}}{\mathbf{R}} \cdot \frac{ \mathbf{e}^{k} }{ \sqrt{\mathcal{M}} } \right) \left( \deri{\sigma_{\gamma \delta 0}}{\mathbf{R}} \cdot \frac{ \mathbf{e}^{k} }{ \sqrt{\mathcal{M}} } \right),
\label{nonaffineterm1}
\end{equation}
where we use $\left< \left( \mathbf{e}^k \cdot \dot{\mathbf{u}}_0 \right)^2 \right> = {\omega^k}^2 \left< \left( \mathbf{e}^k \cdot {\mathbf{u}}_0 \right)^2 \right> = T$, which is the equipartition law for energy.
It is worth noting that the affine component, $C^A_{\alpha \beta \gamma \delta} \equiv C^B_{\alpha \beta \gamma \delta} + C^K_{\alpha \beta \gamma \delta} + C^C_{\alpha \beta \gamma \delta}$, is determined by static structural properties, whereas the non-affine term, $C^N_{\alpha \beta \gamma \delta}$, reflects the vibrational properties.
We also note that the non-affine term $C_{\alpha \beta \gamma \delta}^{N}$ in Eq.~(\ref{nonaffineterm1}) can be written as
\begin{equation}
C_{\alpha \beta \gamma \delta}^{N} \ \longrightarrow \ V \deri{\sigma_{\alpha \beta 0}}{\mathbf{R}^T} \left[ \left. \frac{\partial^2 \Phi}{\partial \mathbf{r} \partial \mathbf{r}^T} \right|_{\mathbf{r}=\mathbf{R}} \right]^{-1} \deri{\sigma_{\alpha \beta 0}}{\mathbf{R}}.
\label{nonaffineterm2}
\end{equation}
From Eqs.~(\ref{fformuation2}), (\ref{stressformulation2}), and~(\ref{nonaffineterm2}), we can confirm that elastic modulus at zero temperature does not depend on mass of particles but rather it depends on only the potential.

%%%%%%%%%%%%%%%%%%%%%%%%%%%%%%%%%%%%%%%%%%%%%%%%%%%%%%%%%%%%%%%
\begin{figure}[t]
\centerline{
\subfigure[Glass~(bulk).]
{\includegraphics[width=0.4\textwidth]{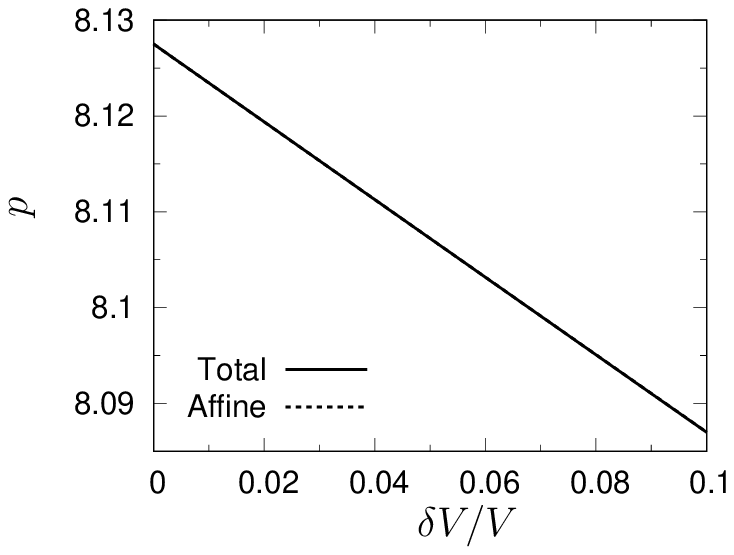}\label{fig12a}}
\hspace*{4mm}
\subfigure[Crystal~(bulk).]
{\includegraphics[width=0.4\textwidth]{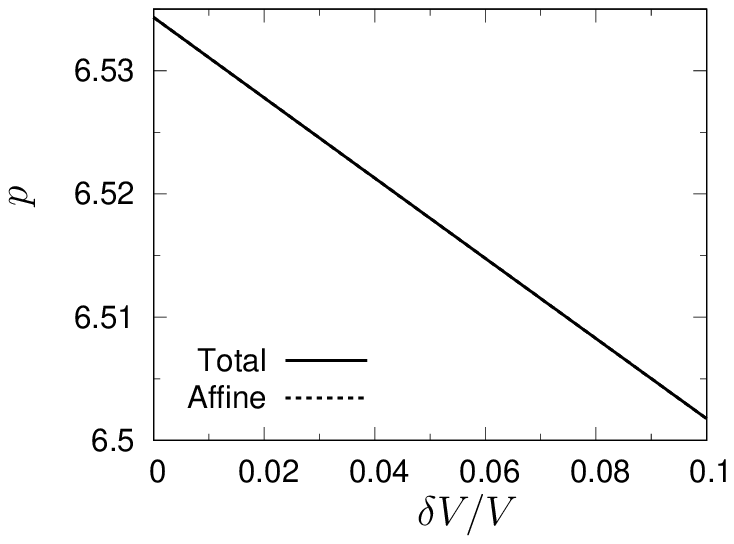}\label{fig12b}}
}

\vspace*{1.5mm}
\centerline{
\subfigure[Glass~(pure shear$1$).]
{\includegraphics[width=0.4\textwidth]{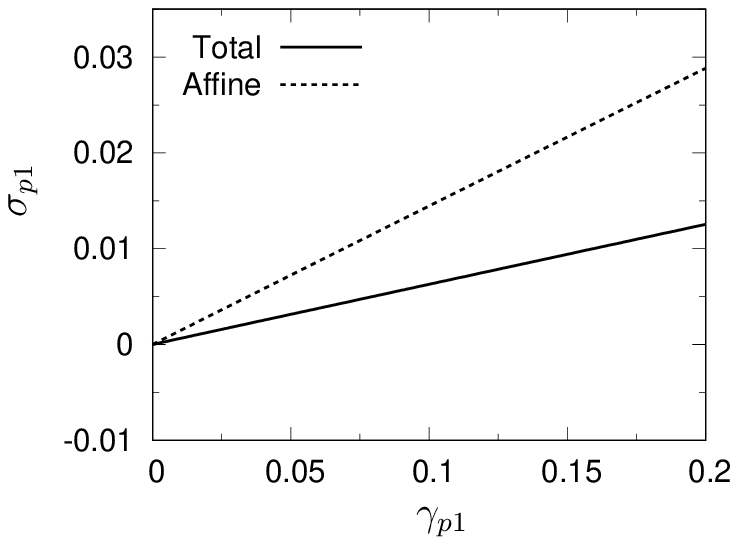}\label{fig12c}}
\hspace*{4mm}
\subfigure[Crystal~(pure shear$1$).]
{\includegraphics[width=0.4\textwidth]{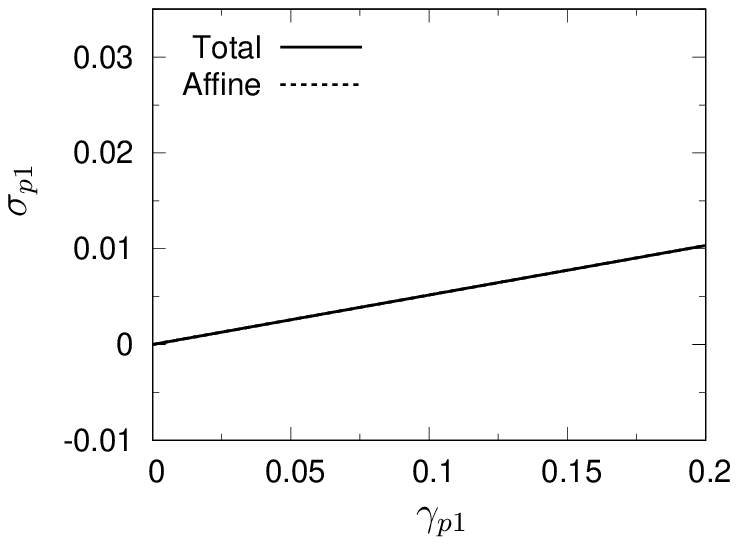}\label{fig12d}}
}

\vspace*{1.5mm}
\centerline{
\subfigure[Glass~(simple shear$1$).]
{\includegraphics[width=0.4\textwidth]{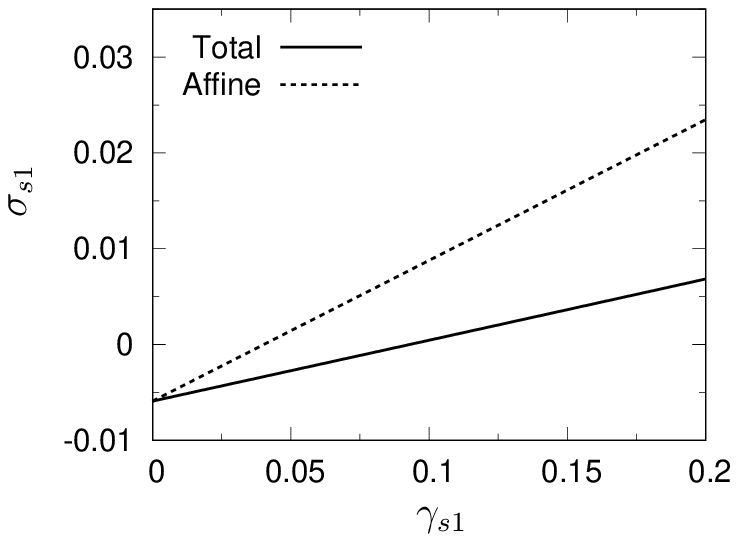}\label{fig12e}}
\hspace*{4mm}
\subfigure[Crystal~(simple shear$1$).]
{\includegraphics[width=0.4\textwidth]{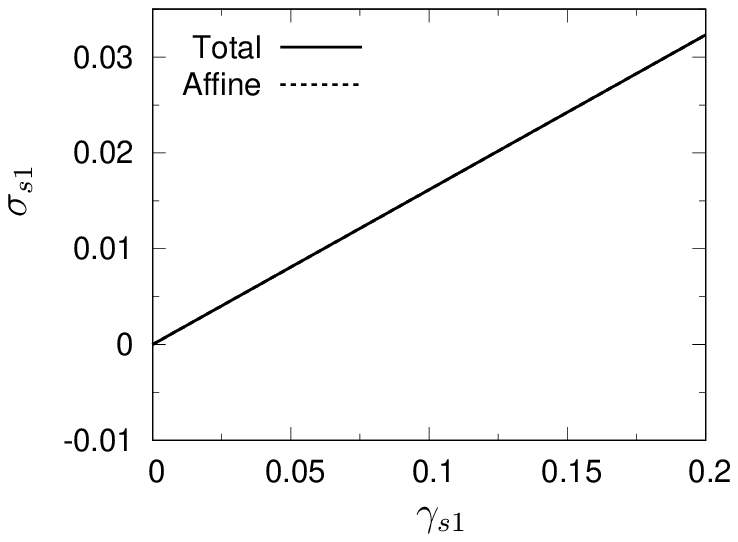}\label{fig12f}}
}
\caption{
The elastic response of an SC system.
(a),(c),(e) Glass.
(b),(d),(f) FCC crystal.
We plot $p$ versus $\delta V / V$ for bulk deformation in (a),(b), $\sigma_{p1}$ versus $\gamma_{p1}$ for pure shear $1$ deformation in (c),(d), and $\sigma_{s1}$ versus $\gamma_{s1}$ for simple shear $1$ deformation in (e),(f).
The dashed line represents affine deformation, while the solid line represents the overall response, including non-affine deformation.
The slopes of the stress-strain curves yield the values of the elastic moduli, $K$, $G_{p1}$, and $G_{s1}$, as in Eqs.~(\ref{modulusslope1}) to~(\ref{modulusslope3}).
The obtained values are presented in Table~\ref{table1}.
}
\label{fig12}
\end{figure}
%%%%%%%%%%%%%%%%%%%%%%%%%%%%%%%%%%%%%%%%%%%%%%%%%%%%%%%%%%%%%%%

%%%%%%%%%%%%%%%%%%%%%%%%%%%%%%%%%%%%%%%%%%%%%%%%%%%%%%%%%%%%%%%%%%%%%%%%%%%%%%%%%%%%%%%%%%%
\subsection{Direct measurement}
We can apply an external strain to the system to directly measure the elastic moduli.
By increasing the strain, we obtain the stress-strain curve, and the slope of this curve yields the corresponding elastic modulus.
Specifically, the bulk, pure shear~$1$, and simple shear~$1$ deformations, which are illustrated in Fig.~\ref{fig11}, yield the $p-\delta V/V$, $\sigma_{p1}-\gamma_{p1}$, and $\sigma_{s1}-\gamma_{s1}$ curves, respectively.
Figure~\ref{fig12} (solid lines) plots these stress-strain curves, which we obtained by applying the linear strain $\epsilon_{\alpha \beta}=e_{\alpha \beta}$.
The slopes of these curves give the bulk modulus $K$, the pure shear modulus $G_{p1}$, and the simple shear modulus $G_{s1}$, as shown in Eqs.~(\ref{modulusslope1}) to~(\ref{modulusslope3}).
We note that other pure shear, $G_{p2}$, and simple shear, $G_{s2}$ and $G_{s3}$, moduli can be also obtained in the same manner.

The obtained values of the elastic moduli are presented in Table~\ref{table1}.
From Table~\ref{table1}, we can confirm that these values coincide well with those obtained from the fluctuation formulation.
We note that the fluctuation formulation and direct measurement method were implemented at a finite temperature of $T=10^{-2}$ and at zero temperature ($T=0$), respectively.
However, almost negligible differences are observed at these different $T$s, since $T=10^{-2}$ is almost at the zero-temperature limit and the elastic moduli are insensitive to $T$ in the low-$T$ regime.

To obtain the response under affine deformation, we can apply an affine strain, $\epsilon^A_{\alpha \beta}=e^A_{\alpha \beta}$, to displace the particles according to Eq.~(\ref{affine1}).
After the application of an affine strain, the particles are generally not in mechanical equilibrium; however, we do not permit the particles to relax to the equilibrium state.
By doing so, we obtain the stress-strain curve under affine deformation, which is shown by the dashed lines in Fig.~\ref{fig12}.
On the other hand, when we allow the particles to relax, we obtain the overall stress-strain curve, as shown by the solid lines in Fig.~\ref{fig12}.
This relaxation process, i.e., non-affine deformation, generally reduces the stress and causes a reduction in the elastic modulus.

As shown in Figs.~\ref{fig12b},~\ref{fig12d}, and~\ref{fig12f}, for the crystal, the stress-strain curves of the overall response~(solid line) and the affine response~(dashed line) coincide.
The elastic response of the crystal is therefore determined by the affine deformation without relaxation, as discussed above.
In contrast to the crystal, the glass shows a large non-affine contribution in its shear deformations, as shown in Figs.~\ref{fig12c} and~\ref{fig12e}.
For the bulk deformation of the glass, the non-affine relaxation is negligible, as shown in Fig.~\ref{fig12a}; this is due to the isotropic nature of bulk deformation, as also discussed above.

%%%%%%%%%%%%%%%%%%%%%%%%%%%%%%%%%%%%%%%%%%%%%%%%%%%%%%%%%%%%%%%%%%%%%%%%%%%%%%%%%%%%%%%%%%%
\section{Recent advances concerning the vibrational properties of glasses}\label{sec.recent}
At the end of this chapter, we will introduce recent advances in the understanding of the vibrational properties of glasses.
Currently, the capability of computers is rapidly growing.
This enables us to perform large-scale computational simulations and to understand the vibrations of the particles in glasses in greater detail.
In particular, recent simulations~\cite{Lerner_2016,Mizuno_2017,Shimada_2017,Lerner_2017,Lerner2_2017,Angelani_2018,Bouchbinder_2018,Mizuno_2018,Shimada_2018,Wang_2019} have revealed the vibrational properties in the low-frequency continuum limit.
In this section, we present simulation data for vibrational eigenmodes and phonon transport, giving particular attention to the low-frequency regime.
The data presented below were obtained in Refs.~\onlinecite{Mizuno_2017,Mizuno_2018} by simulating an atomic glass~(packed glass) with a harmonic potential, $\Phi_\text{HA}$, as expressed~in Eq.~(\ref{ha}).

As described in Section~\ref{sec.modeanalysis}, the eigenmodes in a crystal, i.e., phonons, smoothly converge to elastic waves at low frequencies, and their vDOS converges to the Debye vDOS~\cite{Ashcroft_1976,Kittel_1996}.
This means that crystals behave as elastic media at long length scales, where the microscopic lattice structures are uniformly coarse-grained.
Similarly, we might expect that the disordered structures of glasses are also uniformly coarse-grained at long length scales and that they would therefore also behave as elastic media.
The eigenmodes and vDOS of a glass would then be expected to converge to elastic waves and the Debye vDOS at low frequencies.
However, in contrast to this expectation, we will show below that the disordered structures of glasses are not completely uniformly coarse-grained but rather act as defects even at macroscopic scales.
Glasses therefore behave not as uniform elastic media but rather as elastic media with defects.

%%%%%%%%%%%%%%%%%%%%%%%%%%%%%%%%%%%%%%%%%%%%%%%%%%%%%%%%%%%%%%%
\begin{figure}[t]
\centerline{
\includegraphics[width=0.995\textwidth]{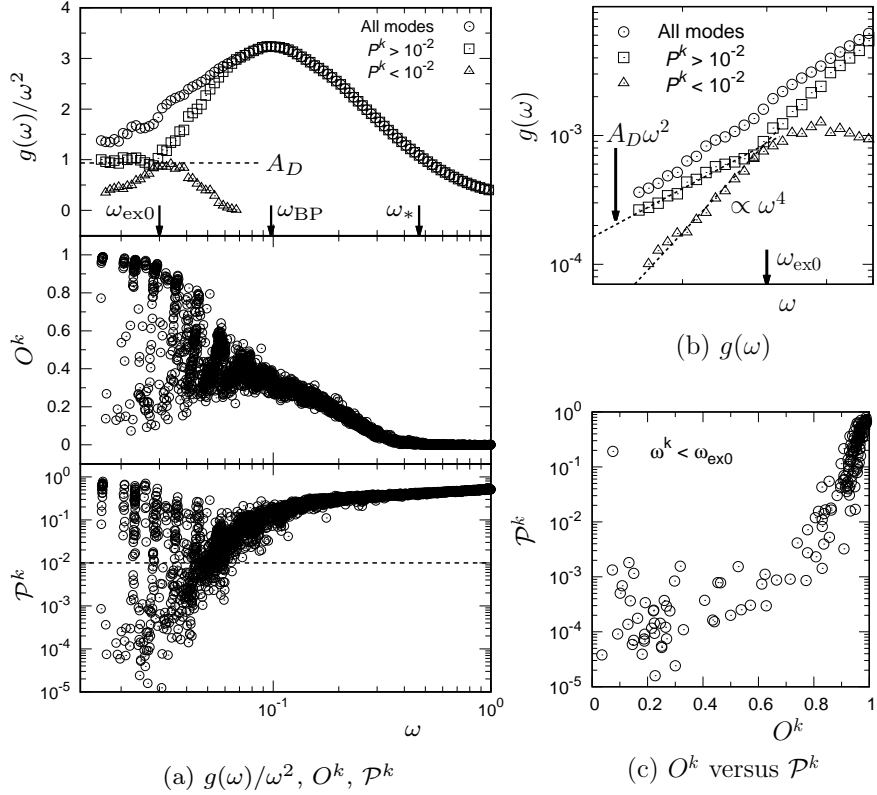}\label{fig13}
}
\caption{
Vibrational eigenmodes in a glass with a harmonic potential.
(a) Plots of the reduced vDOS $g(\omega)/\omega^2$ and of the phonon order parameter $O^k$ and participation ratio $\mathcal{P}^k$ for each eigenmode $k$ as functions of the frequency $\omega$.
(b) The vDOS $g(\omega)$.
(c) $O^k$ versus $\mathcal{P}^k$ for the low-frequency eigenmodes with $\omega^k < \omega_\text{ex0}$.
Regarding the vDOS, data are also presented for $g_\text{ex}(\omega)$, the vDOS of the extended modes ($\mathcal{P}^k > 10^{-2}$), and for $g_\text{loc}(\omega)$, the vDOS of the localized modes ($\mathcal{P}^k < 10^{-2}$).
The horizontal lines at the top and bottom of (a) represent the Debye level $A_D$ and $\mathcal{P}=10^{-2}$, respectively.
In the top panel of (a), we use arrows to indicate the characteristic frequencies $\omega_\text{ex0}$, $\omega_\text{BP}$, and $\omega_\ast$, as discussed in the main text.
}
\label{fig13}
\end{figure}
%%%%%%%%%%%%%%%%%%%%%%%%%%%%%%%%%%%%%%%%%%%%%%%%%%%%%%%%%%%%%%%

%%%%%%%%%%%%%%%%%%%%%%%%%%%%%%%%%%%%%%%%%%%%%%%%%%%%%%%%%%%%%%%%%%%%%%%%%%%%%%%%%%%%%%%%%%%
\subsection{Vibrational eigenmodes in glasses}
Figure~\ref{fig13} presents the results for the vibrational eigenmodes in a glass with a harmonic potential.
The top panel of Fig.~\ref{fig13}(a)~(circles) shows the reduced vDOS $g(\omega)/\omega^2$ together with the Debye level $A_0$.
The reduced vDOS clearly exhibits a maximum, i.e., the boson peak (the arrow indicates the BP frequency, $\omega_\text{BP}$).
As the frequency decreases below $\omega_\text{BP}$, $g(\omega)/\omega^2$ decreases toward but does not reach $A_0$ in the present frequency region.
We will carefully discuss this point below.

To enable characterization of the eigenmodes, Figure~\ref{fig13}(a) also presents data on the phonon order parameter $O^k$ defined in Eq.~(\ref{equofok})~(middle panel) and the participation ratio $\mathcal{P}^k$ defined in Eq.~(\ref{participation})~(bottom panel).
Let us first consider $O^k$, which measures the extent to which the eigenmode $k$ exhibits phonon-like vibrations~\cite{Mizuno_2017,Shimada_2017}: $O^k$ takes values from $1$ (phonon) to $0$ (non-phonon).
At high $\omega$, $O^k$ is nearly zero, which confirms that these eigenmodes are considerably different to phonons.
We may define the frequency, $\omega_\ast$, at which $O^k$ converges to zero
\footnote{
The present glass system shows a characteristic plateau in the vDOS~\cite{Silbert_2005,Wyart_2006,Silbert_2009}.
The value of $\omega_\ast$ is usually defined as the onset frequency of this plateau, which is plotted in Fig.~\ref{fig13}(a).
However, we can confirm in Fig.~\ref{fig13}(a) that the frequency at which $O^k$ converges to zero is consistent with the onset frequency $\omega_\ast$.
}.
It has been reported that at the high $\omega > \omega_\ast$, the eigenmodes are disordered and extended~\cite{Silbert_2005,Wyart_2006,Silbert_2009}; this report is consistent with the present result of $O^k \approx 0$.
As $\omega$ decreases from $\omega_\ast$ to $\omega_\text{BP}$, $O^k$ smoothly increases to $\approx 0.3$.
This result indicates that the eigenmodes near $\omega_\text{BP}$ show phonon-like vibrations to some extent.
Notably, as $\omega$ further decreases below $\omega_\text{BP}$, the modes can be divided into two groups: $O^k$ increases with decreasing $\omega$ in one group, whereas $O^k$ decreases in the other group.
In the former group, $O^k$ converges to almost $1$ at $\omega_\text{ex0}$ (we will provide the precise definition of $\omega_\text{ex0}$ later), indicating that these modes are phonon modes
\footnote{
The values of $O^k$ for these phonon modes are close to but not exactly $1$, which indicates that they are weakly perturbed.
An exact value of $O^k = 1$ may be realized only in the limit of $\omega \to 0$.
}.

We next consider $\mathcal{P}^k$, which evaluates the extent of spatial localization of eigenmode $k$~\cite{Schober_1991,mazzacurati_1996,Taraskin_1999}.
$\mathcal{P}^k$ takes values from 1 (extended over all particles equally) to $1/N \ll 1$ (localized to one particle).
$\mathcal{P}^k$ more clearly exhibits the division of the modes into two groups: one group approaches $\mathcal{P}^k = \mathcal{O}(1)$ with decreasing $\omega$, and the other approaches $\mathcal{P}^k = \mathcal{O}(1/N)$.
Figure~\ref{fig13}(c) shows that the non-phonon modes (small $O^k$) are localized (small $\mathcal{P}^k$) at low $\omega < \omega_\text{ex0}$, whereas the phonon modes (large $O^k$) are extended (large $\mathcal{P}^k$).
Therefore, the data for $O^k$ and $\mathcal{P}^k$ unambiguously demonstrate that phonon modes and non-phonon localized modes coexist at the low $\omega < \omega_\text{ex0}$.

The distinction between phonon modes and localized modes enables us to separately consider the vDOSs for these two types of modes.
We define $g_\text{ex}(\omega)$ as the vDOS for modes with $\mathcal{P}^k > \mathcal{P}_c$, and we define $g_\text{loc}(\omega)$ as the vDOS for modes with $\mathcal{P}^k < \mathcal{P}_c$.
Here, we set a reasonable threshold value of $\mathcal{P}_c=10^{-2}$, as in the data for $\mathcal{P}^k$ in Fig.~\ref{fig13}(a)~(bottom panel)
\footnote{
However, the results are insensitive to the choice of $\mathcal{P}_c$ for $5\times 10^{-3} < \mathcal{P}_c < 2\times 10^{-2}$~\cite{Mizuno_2017}.
}.
We plot the reduced versions of $g_\text{ex}(\omega)$ and $g_\text{loc}(\omega)$ in Fig.~\ref{fig13}(a)~(top panel).
We also plot the vDOS itself in Fig.~\ref{fig13}(b).
$g_\text{ex}(\omega)$ converges exactly to the Debye vDOS $A_D \omega^2$ at a finite value of $\omega$, which we define as $\omega_{\text{ex0}}$.
On the other hand, $g_\text{loc}(\omega)$ follows a different scaling law, $g_\text{loc}(\omega) \propto \omega^4$.
Thus, we conclude that the phonon modes that follow the Debye law ($g_\text{ex}(\omega) = A_D \omega^2$) and the localized modes that follow the other, non-Debye law ($g_\text{loc}(\omega) \propto \omega^4$) coexist at $\omega < \omega_\text{ex0}$.
Note that the $\omega^4$ scaling is the same law proposed in the soft-potential model~\cite{Karpov_1983,Buchenau_1991,Buchenau_1992,Gurevich_2003,Gurevich_2005,Gurevich_2007}.
The total vDOS $g(\omega)$ can therefore be described as
\begin{equation}
g(\omega) = g_\text{ex}(\omega) + g_\text{loc}(\omega) = A_D \omega^2 + A_\text{loc} \omega^4. \label{doscon}
\end{equation}
Since $g_\text{loc}(\omega)$ always takes finite values at $\omega < \omega_\text{ex0}$, $g(\omega)$ coincides with the Debye vDOS only at zero frequency.
This means that the disordered structures of glasses are not uniformly coarse-grained even at macroscopic scales but rather continue to produce localized vibrations.

Based on the vDOS in Eq.~(\ref{doscon}), the heat capacity $C(T)$ can be predicted within the harmonic approximation as follows~\cite{Kittel_1996,Ashcroft_1976}:
\begin{equation}
C(T) = 3 \int \hbar \omega \left( \frac{\partial f(\omega,T)}{\partial T} \right) g(\omega) d\omega, \label{spheat}
\end{equation}
where $f(\omega,T) = \left[ \exp(\hbar\omega/k_B T)-1 \right]^{-1}$ is the Bose-Einstein distribution, $k_B$ is the Boltzmann constant, and $\hbar = h/2\pi$ with $h$ being the Plank constant.
At the low temperature, $k_B T \lesssim \hbar \omega_\text{ex0}$, $C(T)$ consists of two terms: the Debye term $C_\text{ex}(T) \propto T^3$ and the non-Debye term $C_\text{loc}(T) \propto T^5$.
This result for $C(T)$ in the harmonic approximation cannot correctly capture experimental observations~\cite{Zeller_1971,Phillips_1981,Elliott_1990}, in which a linear $T$ dependence appears at low temperatures.
The present results therefore demonstrate that anharmonicities should play an important role in the low-$T$ properties of glasses, e.g., a two-level system might give rise to the linear $T$ dependence~\cite{Anderson_1972,Perez-Castaneda_2014,Perez-Castaneda2_2014}.
This is completely different to the case of crystals, which are well described within the harmonic approximation~\cite{Kittel_1996,Ashcroft_1976}.

%%%%%%%%%%%%%%%%%%%%%%%%%%%%%%%%%%%%%%%%%%%%%%%%%%%%%%%%%%%%%%%
\begin{figure}[t]
\centerline{
\subfigure[Transverse phonon transport.]
{\includegraphics[width=0.5\textwidth]{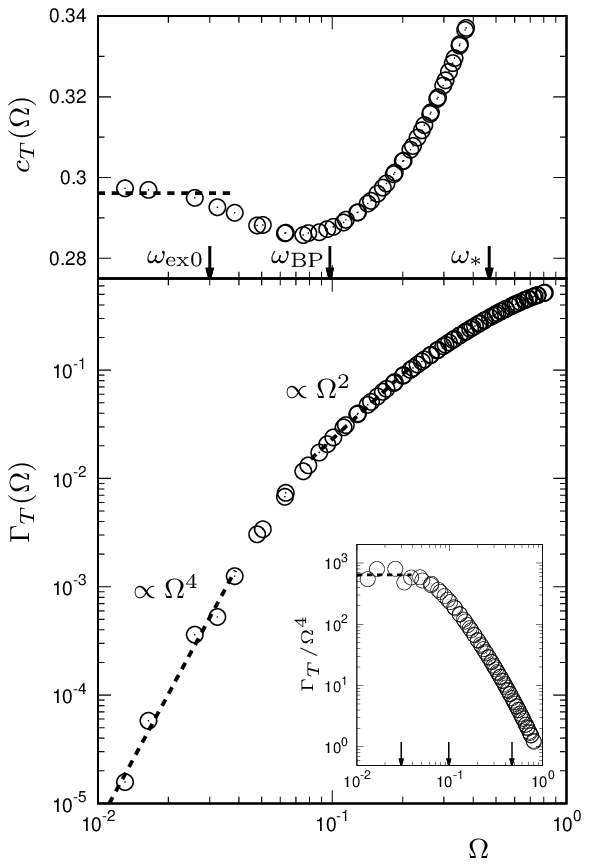}\label{fig14a}}
\hspace*{0mm}
\subfigure[Longitudinal phonon transport.]
{\includegraphics[width=0.5\textwidth]{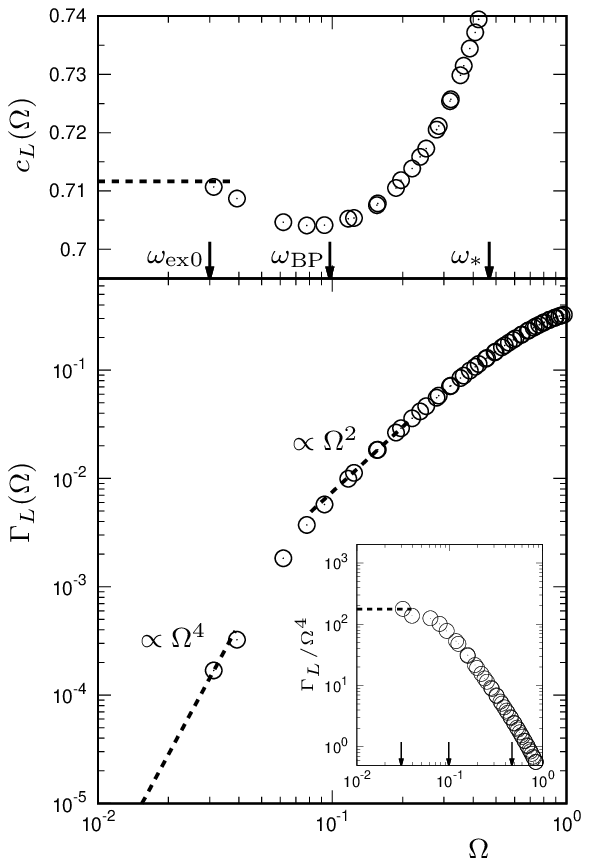}\label{fig14b}}
}
\caption{
Phonon transport in a glass with a harmonic potential.
(a) Transverse phonon transport.
(b) Longitudinal phonon transport.
We plot the sound speed $c_\alpha(\Omega)$ (top panel) and the attenuation rate $\Gamma_\alpha(\Omega)$ (bottom) as functions of the frequency $\Omega$.
The dashed line indicates the macroscopic sound speed (top panel) or the power-law scaling with $\Omega$ (bottom).
The inset in the bottom panel plots $\Gamma_\alpha/\Omega^4$ versus $\Omega$ to clearly demonstrate the Rayleigh scattering law.
We use arrows to indicate the characteristic frequencies $\omega_\text{ex0}$, $\omega_\text{BP}$, and $\omega_\ast$ that are obtained from the vibrational eigenmodes as shown in Fig.~\ref{fig13}.
}
\label{fig14}
\end{figure}
%%%%%%%%%%%%%%%%%%%%%%%%%%%%%%%%%%%%%%%%%%%%%%%%%%%%%%%%%%%%%%%

%%%%%%%%%%%%%%%%%%%%%%%%%%%%%%%%%%%%%%%%%%%%%%%%%%%%%%%%%%%%%%%%%%%%%%%%%%%%%%%%%%%%%%%%%%%
\subsection{Phonon transport in glasses}
We next present the results for phonon transport in a glass with a harmonic potential.
Figure~\ref{fig14} shows the sound speeds, $c_\alpha(\Omega)$, and the attenuation rates, $\Gamma_\alpha(\Omega)$, for transverse ($\alpha = T$, Fig.~\ref{fig14a}) and longitudinal ($\alpha = L$, Fig.~\ref{fig14b}) phonon transport.
These results were obtained through direct measurement at zero temperature (see Section~\ref{sec.ptdm}).
In this figure, arrows indicate the characteristic frequencies, $\omega_\ast$, $\omega_\text{BP}$, and $\omega_\text{ex0}$, all of which are obtained from the eigenmode results in Fig.~\ref{fig13}.
In the boson peak regime, $\Omega \sim \omega_\text{BP}$, $c_\alpha$ takes its minimum value, corresponding to sound softening.
$\Gamma_\alpha$ shows an $\Omega^2$ dependence, $\Gamma_\alpha \propto \Omega^2$.
We also find that the IR limit for transverse phonons is $\Omega_{T\text{IR}} \approx \omega_\text{BP}$.
These results indicate that such a phonon does not propagate as a plane wave but rather exhibits dynamics characteristic of viscous damping~\cite{shintani_2008}.
As $\Omega$ decreases to $\Omega \lesssim \omega_\text{ex0}$, we observe a clear crossover to Rayleigh scattering behaviour.
$c_\alpha$ converges to its macroscopic value: $c_{T0} = \sqrt{G/\rho}$ or $c_{L0} = \sqrt{(K + 4G/3)/\rho}$.
Additionally, $\Gamma_\alpha$ shows an $\Omega^4$ dependence, $\Gamma_\alpha \propto \Omega^4$.

All of the above observations were previously discussed in Section~\ref{sec.phonon}.
The crossover in the nature of the phonon transport has been observed in both experiments~\cite{Masciovecchio_2006,Monaco_2009,Baldi_2010} and simulations~\cite{Monaco2_2009,Marruzzo_2013,Mizuno_2014,Beltukov_2018,Beltukov_2016}.
This crossover can be predicted by mean-field theories~\cite{schirmacher_2006,schirmacher_2007,Wyart_2010,DeGiuli_2014}.
In particular, both transverse and longitudinal phonons show similar transport properties: we indeed observe the same crossover frequency $\omega_\text{ex0}$ for both of them.
This result is consistent with the prediction of heterogeneous elasticity theory~\cite{schirmacher_2006,schirmacher_2007}: the shear modulus heterogeneity induces anomalous behaviours for both transverse and longitudinal phonon transport.

We can understand the crossover in phonon transport in terms of the underlying vibrational eigenmodes as follows.
Because the eigenmodes at $\Omega \sim \omega_\text{BP}$ correspond to disordered and extended vibrations and an initially excited phonon is decomposed into these eigenmodes, it immediately attenuates to become diffusive.
On the other hand, the eigenmodes at $\Omega \lesssim \omega_\text{ex0} $ consist of phonon modes and localized modes.
Here, we measure the overlap $O_\text{loc}$ between the initially excited phonon, $\dot{\mathbf{u}}_0 = \mathbf{u}^{\mathbf{q},\alpha}_\text{ph}$~[Eq.~(\ref{phononvector})],
and the localized modes, $\mathbf{e}^k$, with participation ratio $\mathcal{P}^k < 10^{-2}$, by calculating
\begin{equation}
O_\text{loc} = \sum_{k;\ \mathcal{P}^k < 10^{-2}} \left| \mathbf{u}^{\mathbf{q},\alpha}_\text{ph} \cdot \mathbf{e}^k \right|^2,
\end{equation}
and we find that only a few~$\%$ (at most) of the initially excited phonon is made up of localized modes.
From this observation, we can understand that the initially excited phonon is decomposed mainly into phonon modes, so it attenuates only slowly.
Therefore, the present results unambiguously link these two types of phonon transport to the eigenmodes in the corresponding frequency regimes.
We are able to identify the crossover frequency as $\omega_\text{ex0}$, at which the nature of the underlying eigenmodes changes.

As was already discussed in Section~\ref{sec.irfrequency}, we note that the IR limit for longitudinal phonons, $\Omega_{L\text{IR}}$, is much higher than that for transverse phonons: $\Omega_{L\text{IR}} \gg \Omega_{T\text{IR}} \approx \omega_\text{BP}$.
Therefore, a longitudinal phonon can propagate even at $\Omega > \omega_\text{BP}$.
This result indicates that although the disordered eigenmodes are dominant at $\Omega > \omega_\text{BP}$, longitudinal phonon modes also exist in this $\Omega$ regime, which support the propagation of longitudinal phonons.

In summary, we have shown that even in the low-frequency limit, the glass exhibits localized eigenmodes and Rayleigh scattering in phonon transport.
The same conclusion has also been obtained for an LJ glass~\cite{Monaco2_2009,Shimada_2017}.
These results indicate that the disordered structures of glasses are not uniformly coarse-grained even at macroscopic scales but rather play the role of defects, influencing the vibrational properties of the material.
Below the frequency $\omega_\text{ex0}$, which corresponds to the continuum limit frequency, glasses behave as elastic media with defects.

As a final remark, the above results are $T=0$ properties, in the harmonic approximation limit.
Here we mention anharmonic properties of vibrational excitations.
At finite temperatures $T>0$, systems undergo anharmonic processes.
In crystals phonons excited by thermal fluctuations couple through phonon-phonon interactions~\cite{Ashcroft_1976,Kittel_1996}.
For example, in three-phonon processes, one phonon splits into two different ones, or conversely, two phonons combine into a single excitation.
Similarly, in glasses, vibrational eigenmodes also show anharmonic processes due to mode-mode interactions~\cite{Fabian_1996,Fabian_2003}.

In contrast to crystals, however, it has been demonstrated that an additional anharmonic channel of different origin emerges in glasses, which induces unconventional intermittent rearrangements of particles~\cite{Heuer_2008,Xu_2010,Mizuno_2020,Mizuno2_2020}.
The intermittent rearrangements can be considered as relics of the liquid state which survive the complete dynamic arrest taking place at the glass transition temperature~\cite{Berthier_2016,Scalliet_2017,Scalliet_2019}.
It was also demonstrated that glasses exhibit anomalous temperature dependence of phonon attenuation at finite temperatures~\cite{Ferrante_2013,Baldi_2014,mizuno2019impact,mizuno2019impact2,Wang_2020}, which can be attributed to this additional anharmonic channel.

%%%%%%%%%%%%%%%%%%%%%%%%%%%%%%%%%%%%%%%%%%%%%%%%%%%%%%%%%%%%%%%%%%%%%%%%%%%%%%%%%%%%%%%%%%%
\bibliographystyle{ws-rv-van}
\bibliography{reference}

%\blankpage
%\printindex[aindx]                 % to print author index
%\printindex                         % to print subject index

\end{document}